\documentclass[10pt,aps,twocolumn,superscriptaddress,prl]{revtex4-2}

\usepackage{amsmath,amssymb,amsfonts}
\usepackage{xcolor}
\usepackage{graphicx}
\definecolor{myred}{rgb}{1,0.,0.3}
\definecolor{myblue}{rgb}{0.2,0.2,0.8}
\usepackage[colorlinks=true,citecolor=myblue,linkcolor=myred]{hyperref}
\usepackage{bm}%
\newcommand{\ket}[1]{\left|#1\right\rangle}
\newcommand{\bra}[1]{\left\langle#1\right|}

\newcommand{\ba}{\begin{eqnarray}}
\newcommand{\ea}{\end{eqnarray}}
\newcommand{\be}{\begin{equation}}
\newcommand{\ee}{\end{equation}}

\allowdisplaybreaks

\begin{document}
\title{All-to-All Connectivity in Superconducting Qubits via Metamaterial-Mediated Interactions}%

\author{Or Golan}
\affiliation{%
AWS Center for Quantum Computing, Pasadena, California 91125, USA
}%
\affiliation{Racah Institute of Physics, The Hebrew University of Jerusalem, Jerusalem 91904, Givat Ram, Israel}
\email{or.golan@mail.huji.ac.il}
\author{Alexey V. Gorshkov}
\affiliation{%
AWS Center for Quantum Computing, Pasadena, California 91125, USA
}%
\author{Andreas Butler}
\affiliation{%
AWS Center for Quantum Computing, Pasadena, California 91125, USA
}%
\author{Gihwan Kim}
\affiliation{%
AWS Center for Quantum Computing, Pasadena, California 91125, USA
}%
\author{Fernando G.S.L. Brandão}
\affiliation{%
AWS Center for Quantum Computing, Pasadena, California 91125, USA
}%
\author{Oskar Painter}
\affiliation{%
AWS Center for Quantum Computing, Pasadena, California 91125, USA
}%
\author{Alex Retzker}
\affiliation{%
AWS Center for Quantum Computing, Pasadena, California 91125, USA
}%
\affiliation{Racah Institute of Physics, The Hebrew University of Jerusalem, Jerusalem 91904, Givat Ram, Israel}

\date{\today}

\begin{abstract}
We show that encoding superconducting qubits as dual-rail or spin-locked transmons lets a metamaterial waveguide couple them all-to-all while suppressing the spurious entanglement that burdens shared-mode architectures, a step toward the non-local connectivity that quantum low-density parity-check codes require. In a bare-transmon architecture, the dispersive coupling between each populated mode and every idle qubit attached to it entangles spectators with the gate, and cancelling it by dynamical decoupling lengthens every gate. The encoding suppresses this coupling at the hardware level, from second to fourth order in the inverse qubit--mode detuning. A two-tone cross-resonance drive then generates a M{\o}lmer-S{\o}rensen interaction between encoded qubits, which we use for an entangling gate and a joint parity measurement, both selective and parallelizable through different modes. Simulating two dual-rail transmons sharing three resonators with a mode lifetime of $30\,\mu$s, we find that heralding transmon relaxation as erasure leaves a gate infidelity of $1.4\times10^{-3}$ at $152$~ns, dominated by photon loss. Photon loss and mode crowding together bound the array size. The gate duration must exceed the inverse mode spacing, which shrinks as the array lengthens, while photon loss grows with gate duration. This bound grows linearly with the mode lifetime and with the inter-resonator coupling; at a lifetime of $150\,\mu$s, five times the simulated value, and a coupling of $80$~MHz, a $10^{-3}$ error budget admits $13$ qubits on $25$ resonators.
\end{abstract}

\maketitle

Superconducting qubits are a leading platform for fault-tolerant quantum information processing, as recent demonstrations of quantum error correction show~\cite{krinner2022realizing, google2023suppressing, acharya2024quantum, wang_demonstration_2025, putterman_hardware-efficient_2025, brock_quantum_2025}, particularly of the surface code~\cite{fowler2012surface}. The surface code suits this platform because it requires only the nearest-neighbor couplings that superconducting qubits provide natively. Its error threshold already exceeds the error rates achieved in some devices~\cite{acharya2024quantum}. Its cost is overhead. The number of physical qubits per logical qubit grows as the square of the code distance, so the encoding rate vanishes as the code is scaled up~\cite{fowler2012surface}.

Quantum low-density parity-check (qLDPC) codes remove this cost. Families with constant encoding rate and growing distance exist~\cite{panteleev2022asymptotically, dinur_good_2022}, and finite-size instances already encode far more logical qubits than surface codes of the same distance and physical size, at comparable thresholds~\cite{bravyi2024high}. The price is connectivity. By the Bravyi-Poulin-Terhal bound, a code with constant rate and growing distance cannot be realized with geometrically local checks in two dimensions~\cite{bravyi_no-go_2009, bravyi_tradeoffs_2010}, so qLDPC codes require non-local connectivity that a planar layout of nearest-neighbor couplings cannot provide~\cite{baspin_quantifying_2022, berthusen_toward_2025}.

Superconducting qubits offer three routes to non-local connectivity. The first is a direct link, a fixed or tunable coupler between two distant qubits~\cite{wang_demonstration_2025, marxer_long-distance_2023}; it is simple and selective, but its reach is limited and each link consumes much of a qubit's capacitance budget, so a qubit can afford only a few. The second is a one-to-many bus, a resonator shared by many qubits. Driving the bus, as in the resonator-induced phase gate~\cite{cross2015optimized, paik2016experimental, puri2016high, malekakhlagh2022optimization, kumph_demonstration_2024}, reaches every qubit on it but is not selective, since every pair entangles at once. Alternatively, one qubit's state is swapped into the bus and the gate performed there~\cite{rengerSuperconductingQubitResonatorQuantum2025}; this is selective but serial, because the bus holds one state at a time and the pairs must take turns. The third is shuttling, moving a qubit state through a programmable network of oscillators to its partner~\cite{liuEfficientRoutingQuantum2026}; it is selective and needs few couplers, but every gate pays a control overhead for routing and a compilation overhead for scheduling. No route yet offers a bus that is both selective and parallel.

A metamaterial waveguide, a chain of coupled resonators whose modes extend over the whole chain, is such a bus. In earlier work~\cite{gorshkov25}, we and our collaborators presented the cavity-mediated cross-cross-resonance gate, in which each of the two transmons is driven near the frequency of a shared mode, a cross-resonance interaction~\cite{magesan2020effective} with that mode. Only the driven pair couples, and different pairs can be driven through different modes at the same time. The mode plays the role of the phonon bus of a trapped-ion system, mediating the qubit--qubit interaction through a state-dependent force, the mechanism behind geometric-phase and M{\o}lmer-S{\o}rensen gates in ion traps~\cite{sorensen1999quantum}. Unlike the phonon bus, however, the superconducting realization carries an unwanted dispersive coupling between the qubits and the modes. A populated mode shifts the frequency of every idle ``spectator'' qubit coupled to it, entangling the spectators with the gate and limiting the scalability of the approach. In Ref.~\cite{gorshkov25} we proposed cancelling this shift with dynamical decoupling pulses, the `flowers' sequence. The burden on the sequence is the total dispersive error it must cancel. Each qubit--mode shift is fixed, but the number of such terms grows linearly with the number of qubits and with the number of driven modes, so the approach becomes harder to scale. The goal of this work is therefore to suppress this dispersive coupling at its source.

We suppress it at the qubit rather than in the pulse sequence, by encoding each qubit as a dual-rail transmon (DRT)~\cite{levine2023demonstrating, kubica2023erasure, huang_logical_2025, hung2026fasthighfidelityerasuredetection} [Fig.~\ref{fig:overview}(a)] or, as shown in the Supplemental Material~\cite{SM}, a spin-locked transmon (SLT)~\cite{caiRobustDynamicalDecoupling2012a, zuk2024robust}. Both encodings protect the logical qubit from dephasing, DRTs through strong capacitive coupling between two resonant transmons and SLTs through continuous driving. On the bare transmons, the dispersive coupling has the same operator structure as dephasing noise, so protection against one is protection against the other, passive and built into the hardware. The encoded gate no longer relies on active decoupling, and it reaches a lower infidelity than the bare-transmon gate with or without its decoupling sequence (Fig.~\ref{fig:gate_fidelity}).

Both encodings protect against dephasing in the same way. A strong, noiseless term in the Hamiltonian, orthogonal to the noise, rotates the qubit's energy axis, so that noise couples the new eigenstates only off-diagonally and the energy gap between them suppresses the transitions it would induce. The same rotation suppresses the dispersive coupling through the same gap, but it also transforms the cross-resonance interaction~\cite{magesan2020effective} into a transverse interaction $\tilde{X}(c+c^\dagger)$ between a qubit and a mode, where $\tilde{X}$, $\tilde{Y}$, $\tilde{Z}$ are the logical Pauli operators on the qubit and $c$ ($c^\dagger$) is the annihilation (creation) operator on the mode. The interaction is transformed, not suppressed. The rotation has two consequences for the drive. First, activating the interaction now requires two tones to convert it from transverse to longitudinal, one driving the flip-flop process ($\tilde{\sigma}^+c + \text{h.c.}$) and one the flip-flip process ($\tilde{\sigma}^+c^\dagger + \text{h.c.}$) [Fig.~\ref{fig:overview}(b,c)]. Second, because the desired interaction and the residual dispersive coupling ($\tilde Z c^\dagger c$) now act on orthogonal qubit axes, adding $\tilde{X}$ echoes suppresses the dispersive coupling further. With ideal echo pulses, this removes the need for the `flowers' sequence of Ref.~\cite{gorshkov25}; with imperfect pulses, flowers may still be required. Because the suppression is built into the hardware, the spectator-qubit dispersive coupling no longer limits the number of qubits coupled to the metamaterial. The remaining costs of scale are the erasure of spectator qubits and the competition between resonator photon loss and the crowding of the mode spectrum, which we quantify below.

The drive itself provides two operations. It displaces the shared mode along a trajectory in phase space set by the joint $\tilde{X}\tilde{X}$ eigenvalue of the two qubits. Reading out the resulting mode population realizes a joint parity measurement of $\tilde{X}\tilde{X}$, the two-qubit check measurement from which Floquet codes are built~\cite{hastingsDynamicallyGeneratedLogical2021}, including their non-local versions~\cite{higgottConstructionsPerformanceHyperbolic2024, jacobyStairwayCodesFloquetifying2026}. Adding a small detuning instead closes the trajectory into a loop, imparting a conditional geometric phase for an $\tilde{X}\tilde{X}$ entangling gate. Both operations inherit the selectivity and parallelism of the drive, so any pairing of qubits can be addressed.

\begin{figure}[tb!]
\includegraphics[width=\linewidth]{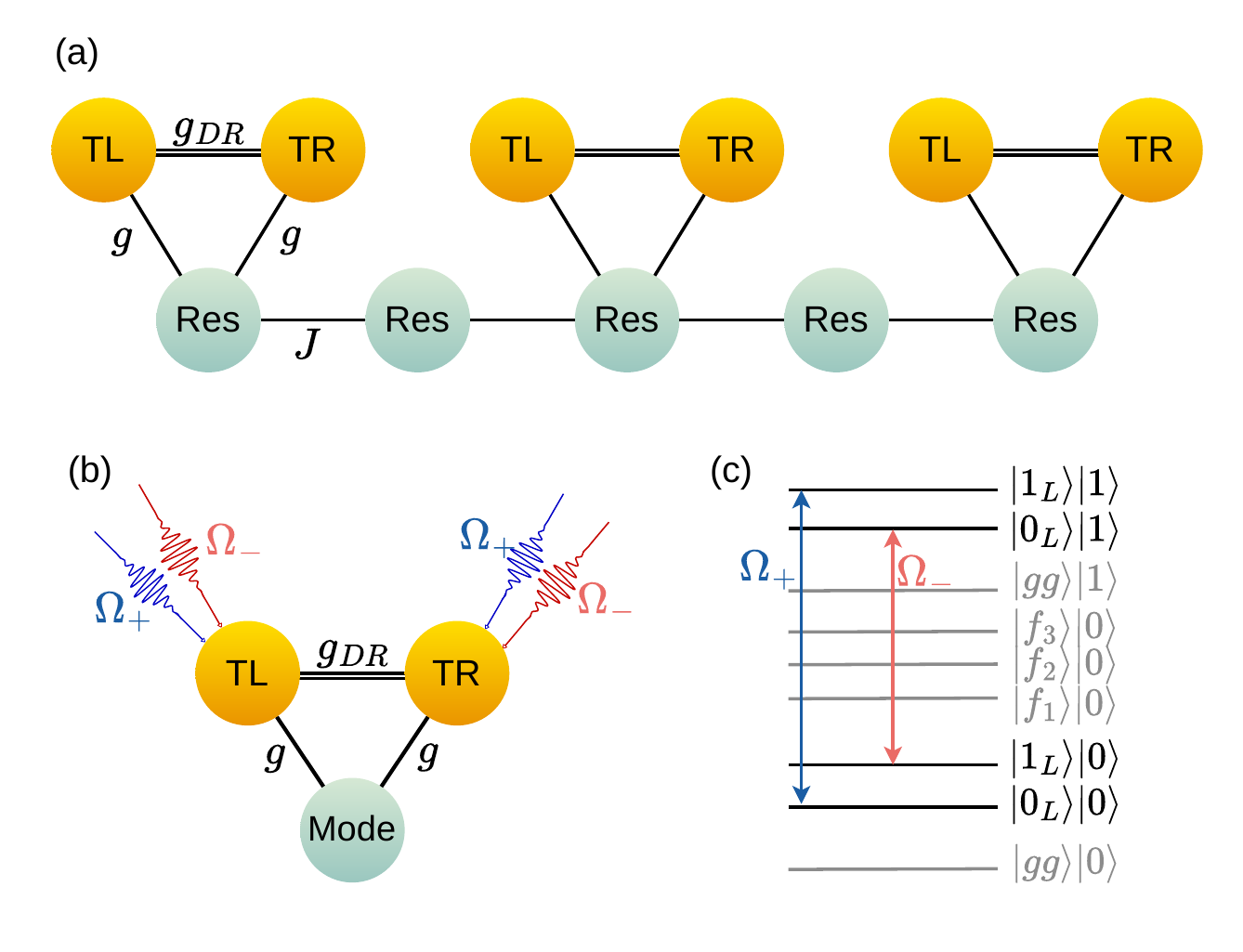}
\caption{A metamaterial waveguide couples dual-rail transmons (DRTs), and a two-tone cross-resonance drive activates their interaction with its modes. (a) Each DRT is a pair of transmons (TL, TR) coupled by $g_{DR}$, which opens the dual-rail gap $2g_{DR}$, and coupled with equal strength $g$ to one resonator of a chain with inter-resonator coupling $J$, one DRT per two resonators. (b) A single DRT coupled to a single mode. The two transmons are driven with opposite phase, in cross-resonance with the mode, by two tones, one blue-detuned from the mode frequency by the dual-rail gap ($\omega_+ \approx \nu + 2g_{DR}$, amplitude $\Omega_+$) and one red-detuned ($\omega_- \approx \nu - 2g_{DR}$, amplitude $\Omega_-$). (c) Energy levels of the DRT--mode system. The blue tone drives the flip-flip transition $|0_L0\rangle \to |1_L1\rangle$ and the red tone the flip-flop transition $|1_L0\rangle \to |0_L1\rangle$. Grey levels lie outside the logical subspace. Erasure drops the DRT to $|gg\rangle$, and $|f_i\rangle$ are its two-excitation states.}
\label{fig:overview}
\end{figure}

\textit{DRT--Mode Coupling.}---Our full system is shown in Fig.~\ref{fig:overview}(a). For a single DRT coupled to one mode [Fig.~\ref{fig:overview}(b)], the Hamiltonian in units with $\hbar=1$, $H=H_0+H_\text{c}+H_\text{d}$, is
\begin{subequations} \label{eq:drt_ham} \begin{align} H_0 &= \omega(a^\dagger a + b^\dagger b) - \frac{\eta}{2}(a^\dagger a^\dagger aa + b^\dagger b^\dagger bb) \nonumber \\
&+ g_{DR}(a^\dagger b + b^\dagger a) + \nu c^\dagger c, \label{eq:h0} \end{align} \begin{equation} H_\text{c} = g((a + b)^\dagger c + (a + b) c^\dagger), \label{eq:hcoupling} \end{equation}
\begin{equation}
H_\text{d} = \big(\Omega _+\cos(\omega_+ t) + \Omega_- \cos(\omega_- t)\big)(a^\dagger - b^\dagger + \text{h.c.}).\label{eq:hdrive}
\end{equation}
\end{subequations}
Here, $a, b$ ($a^\dagger, b^\dagger$) are the annihilation (creation) operators of the left and right transmons (TL, TR) of the DRT, its two rails. In $H_0$, the rails are taken to be identical, with frequency $\omega$ and anharmonicity $\eta>0$, $\nu$ is the mode frequency, and $g_{DR}$ is the coupling strength between the rails. In $H_\text{c}$, $g$ is the coupling strength between each rail and the mode; in $H_\text{d}$, $\Omega_{\pm}$ are the nonnegative drive amplitudes and $\omega_\pm$ the drive frequencies.

In terms of the symmetric and anti-symmetric combinations $\tilde{a}_\pm=\frac{a\pm b}{\sqrt{2}}$, any operator acting on the DRT splits into a part that is symmetric under exchange of the two rails and a part that is anti-symmetric. The logical states, $|1_L\rangle \equiv \tilde{a}_+^\dagger|gg\rangle$ and $|0_L\rangle \equiv \tilde{a}_-^\dagger|gg\rangle$, are respectively symmetric and anti-symmetric. Each drive tone must connect $|0_L\rangle$ and $|1_L\rangle$ [Fig.~\ref{fig:overview}(c)], states of opposite symmetry, so the engineered matrix element must be anti-symmetric. We obtain it by combining the symmetric coupling $H_\text{c}$ with the anti-symmetric drive $H_\text{d}$ of Eq.~\eqref{eq:hdrive}. Deviations from exact symmetry are suppressed by the dual-rail gap; coupling, frequency, and drive asymmetries between the rails are quantified in the Supplemental Material~\cite{SM}.
\nocite{clerkIntroductionQuantumNoise2010, majerCouplingSuperconductingQubits2007, nielsen_simple_2002, mercurioQuantumToolboxjlEfficientJulia2025a, pedersenFidelityQuantumOperations2007, astafievResonanceFluorescenceSingle2010}

The drive and the coupling to the mode induce dynamics within the logical subspace of the DRT. To isolate them, we block-diagonalize the full Hamiltonian with a second-order time-dependent Schrieffer-Wolff transformation~\cite{bravyi2011schrieffer, malekakhlaghFirstprinciplesAnalysisCrossresonance2020a, petrescuAccurateMethodsAnalysis2023}, which yields an effective Hamiltonian projected onto the DRT logical subspace while retaining the mode explicitly; the full derivation is given in the Supplemental Material~\cite{SM}. We assume $\Delta = \nu - \omega \gg g, \Omega_\pm, \eta$, with $\Delta >0$ from here on, and set $\Omega_- = \frac{(\Delta - g_{DR})(\Delta - g_{DR} + \eta)}{(\Delta + g_{DR})(\Delta + g_{DR} + \eta) - 4g_{DR}^2}\,\Omega_+$ so that the two tones contribute equally, writing $\Omega\equiv\Omega_+$. Cross-resonance driving here uses two tones centered on the mode frequency, one red-detuned and one blue-detuned from it by the dual-rail gap, $\omega_\pm \approx \nu \pm 2g_{DR}$. In the interaction picture relative to the dual-rail gap, we then obtain an effective cross-resonance interaction between the DRT logical subspace and the mode:
\begin{equation}
\label{eq:effham}
\begin{gathered}
H_\text{eff}=-g_{CR}\tilde{X}(c^\dagger+c),\\
g_{CR}\equiv\frac{\eta g\Omega}{(\Delta + g_{DR})(\Delta + g_{DR} + \eta) - 4g_{DR}^2}.
\end{gathered}
\end{equation}
To leading order in $1/\Delta$, $g_{CR}\approx \frac{\eta g\Omega}{\Delta^2}$.
This interaction is analogous to the conventional $ZX$ cross-resonance interaction~\cite{magesan2020effective}, with the control-qubit $Z$ replaced by the logical operator $\tilde{X}$ and the target-qubit $X$ replaced by the mode quadrature $c^\dagger+c$. SLTs yield the same effective interaction through an analogous derivation, given in the Supplemental Material~\cite{SM}.

Beyond enabling the two-qubit operations below, this interaction allows longitudinal readout of the DRT logical state, by conditionally displacing a readout mode and measuring its output-field quadratures. For an initially empty mode, Eq.~\eqref{eq:effham} gives $\langle c(t)\rangle = i g_{CR}t\langle \tilde{X}\rangle$, so the displacement is conditioned on the eigenvalue of the logical operator $\tilde{X}$. Unlike the conditional-displacement readout of bare transmons~\cite{touzard_gated_2019}, this measurement carries negligible cross-Kerr interaction between the qubit and the readout mode. It may also be faster than current DRT readout, which requires a preparatory adiabatic mapping of the logical state~\cite{campbell_universal_2020}.

\begin{figure}[t]
\includegraphics[width=\linewidth]{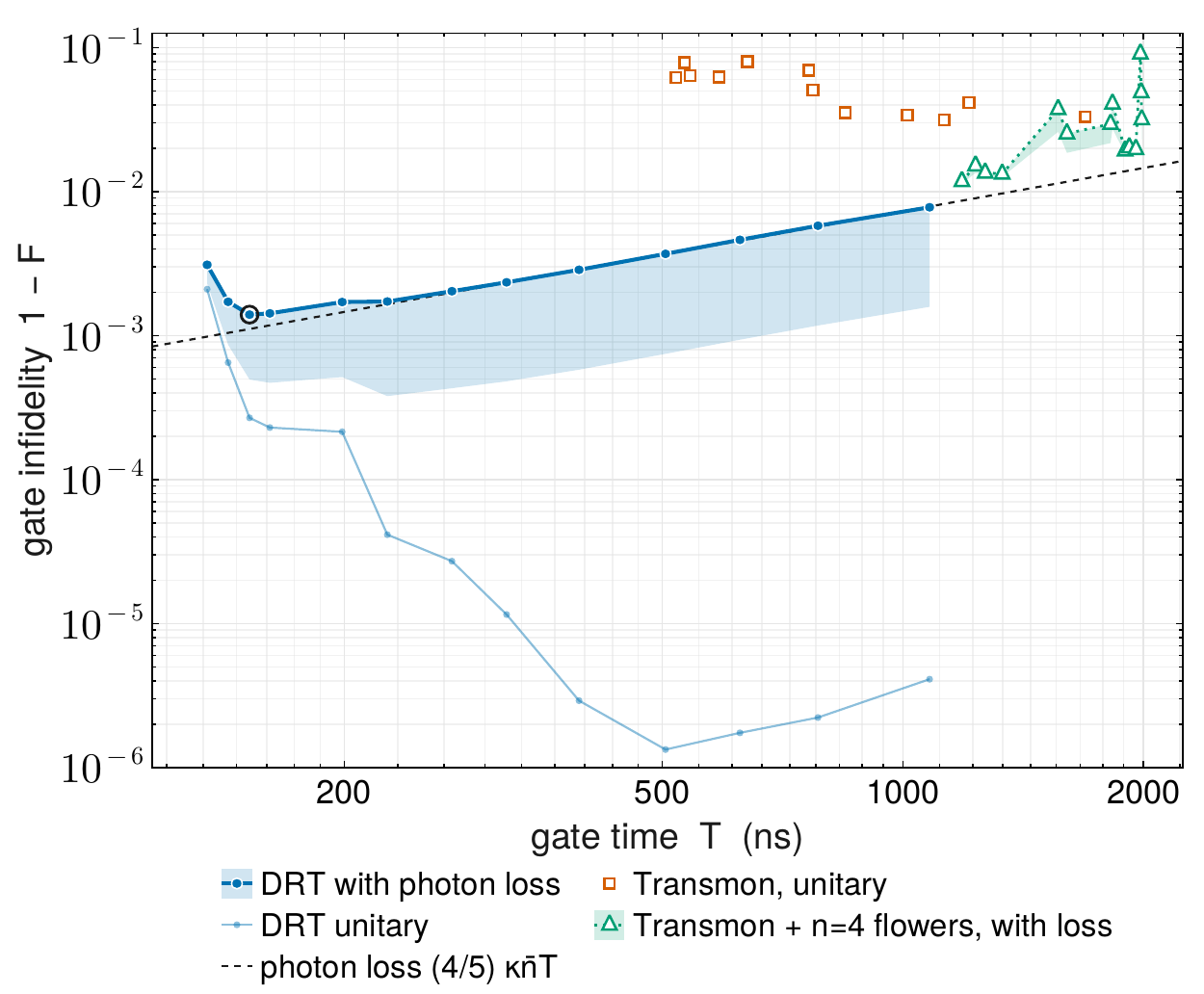}
\caption{Encoding the qubits in dual-rail transmons (DRTs) reaches a lower two-qubit gate infidelity than the bare-transmon gate of Ref.~\cite{gorshkov25}, with or without its dynamical-decoupling sequence. Simulated infidelity $1-F$ versus gate duration $T$, set by the amplitude $\Omega \equiv \Omega_{+, 1} = \Omega_{+, 2}$ of the blue-detuned tone in Eq.~\eqref{eq:hdrivemm}. Blue circles show the $\tilde{X}\tilde{X}$ gate between two DRTs coupled to a metamaterial of three identical resonators (Fig.~\ref{fig:overview}), with the drive frequencies and amplitude ratios optimized numerically at each $\Omega$. The light-blue trace is the coherent error alone, the band adds metamaterial photon loss for mode $T_1 = 30$--$150\,\mu$s, and the solid line is its $30\,\mu$s edge. The dashed line is the photon-loss infidelity $\frac{4}{5}\kappa\langle\bar{n}\rangle T$ of the End Matter with the simulated $\langle\bar{n}\rangle = 0.273$ at mode $T_1 = 30\,\mu$s. The coherent error rises fivefold between $T = 226$ and $199$~ns and stays at $2$--$3\times10^{-4}$ down to $152$~ns, where an off-resonant three-photon process leaks one DRT into a two-excitation state, and at shorter $T$ the effective description breaks down. The black circle marks the minimum of the solid line, $1-F = 1.4\times10^{-3}$ at $T = 152$~ns, of which $81\%$ is photon loss. Orange squares show the bare-transmon gate without decoupling, coherent error only. Green triangles show the same gate with its four-petal flowers sequence, including coherent error, transmon relaxation at $T_1 = 100\,\mu$s, and photon loss at mode $T_1 = 30\,\mu$s, with the band spanning the same mode-lifetime range. Transmon relaxation is excluded from the blue traces because on a DRT it leaves the logical subspace and is heralded as an erasure, with probability $3.2\times10^{-3}$ at the blue minimum for transmon $T_1 = 100\,\mu$s. On a bare transmon it is unheralded. Simulation details are in the Supplemental Material~\cite{SM}.}
\label{fig:gate_fidelity}
\end{figure}

\textit{Parity Measurement.}---Direct two-body measurements have been proposed and demonstrated in circuit QED~\cite{Chow2010JointReadout, Ciani2017ThreeQubitParity, Criger2016MultiQubitJoint, DiCarlo2009TwoQubitAlgorithms, DiVincenzo2013MultiQubitParity, Filipp2009JointDispersiveReadout, Govia2015ParityPhotonCounter, Huembeli2017HeraldedParity, Kerckhoff2009ContinuousParity, Lalumiere2010TunableJoint, Livingston2021ContinuousQEC, Mohseninia2020AlwaysOn, Riste2012InitializationByMeasurement, Royer2018ParityParametricDriving, Tornberg2014ThreeQubitParity}. Here, the same interaction that drives the gate provides one. The effective Hamiltonian of Eq.~\eqref{eq:effham}, extended to two DRTs, allows direct measurement of the logical operator $\tilde{X}\tilde{X}$, whose eigenvalue is the parity of the pair, $+1$ (even) for $|\tilde{+}\tilde{+}\rangle$ and $|\tilde{-}\tilde{-}\rangle$ and $-1$ (odd) otherwise, with $|\tilde{\pm}\rangle$ the $\tilde{X}$ eigenstates. The mode's phase-space trajectory is $\langle c(t)\rangle = i \left\langle \sum\limits_ig_{CR,i}\tilde{X}_i \right\rangle t$. With equal amplitudes, $g_{CR,1}=g_{CR,2}\equiv g_{CR}$, obtained by fixing the ratio of $\Omega_1$ to $\Omega_2$, the displacement of a joint eigenstate with eigenvalues $x_{1,2}=\pm1$ is $|\alpha|^2 = g_{CR}^2(x_1+x_2)^2t^2$, which depends only on the parity $x_1x_2$. The mode is displaced for even parity and stays empty for odd.

We read out the photon number with an auxiliary transmon dispersively coupled to the mode and driven by a frequency comb~\cite{essig_multiplexed_2021}; a populated mode signals even parity, an empty one odd, and neither outcome reveals the individual $\tilde{X}$ eigenvalues. Within the even sector, the two eigenstates acquire a relative phase $(-1)^n$, correctable once the photon number $n$ is known. The measurement takes several tens of $1/\chi$, where $\chi$ is the dispersive shift between the auxiliary transmon and the mode, because the auxiliary transmon saturates~\cite{SM}.
This native parity measurement enables a qubit-efficient implementation of Floquet codes~\cite{hastingsDynamicallyGeneratedLogical2021, higgottConstructionsPerformanceHyperbolic2024, jacobyStairwayCodesFloquetifying2026}.

\textit{Two-qubit Gate.}---Adding a detuning to the drives gives, in the frame rotating at the drive-center frequency and with the dual-rail gap, the effective Hamiltonian
$H_\text{eff} = \epsilon c^\dagger c - \sum_{i=1}^2 g_{CR,i} \tilde{X}_i (c^\dagger + c),$
where $\epsilon = \nu - \omega_\text{center}$ is the detuning of the mode from the drive-center frequency. This is the M{\o}lmer-S{\o}rensen Hamiltonian~\cite{sorensen1999quantum, sorensen_entanglement_2000}, and its time evolution is given exactly by the propagator
\begin{equation} \begin{gathered}
\label{eq:2qprop}
U(t) = \exp\left(\sum\limits_i g_{CR,i} \frac{1 - e^{i\epsilon t}}{\epsilon} \tilde{X}_i c - \text{h.c.} \right) \\
\times \exp\left(2i \tilde{X}_1 \tilde{X}_2 g_{CR,1}g_{CR,2} \frac{\epsilon t - \sin \epsilon t}{\epsilon^2} \right)\\
\times\exp\left(2i\left(g_{CR,1}^2 + g_{CR,2}^2\right)\frac{t}{\epsilon}\right)\exp\left(i\epsilon t c^\dagger c\right). \end{gathered} \end{equation}
The propagator describes closed loops in the mode phase space whose enclosed areas depend on the logical states of the DRTs, entangling the DRTs through a state-dependent geometric phase. A maximally entangling gate requires $\epsilon t = 2\pi n$ and $\frac{2 t}{\epsilon} g_{CR,1}g_{CR,2} = \frac{\pi}{4} + 2\pi m$, with the shortest gate time at $n = 1$ and $m = 0$. For these parameters, the propagator is $U(t)\propto e^{i\frac{\pi}{4}\tilde{X}_1\tilde{X}_2}$, which is equivalent to a CZ gate up to single-qubit rotations.

The encoding suppresses the dispersive coupling between each DRT and the mode. For a bare transmon, this coupling appears at second order in the inverse transmon--mode detuning, $\propto g^2\eta/\Delta^2$; here it is pushed to fourth order, $\frac{2\eta g^2(g^2-\eta g_{DR})}{\Delta^4}\tilde Z c^\dagger c$, as derived in the Supplemental Material~\cite{SM}. The suppression sets how strong the qubit--qubit coupling can be made while the dispersive coupling stays within the fidelity budget, and so allows faster gates (Fig.~\ref{fig:gate_fidelity}). Without it, the dispersive coupling dominates the coherent error of the bare-transmon gate [orange squares in Fig.~\ref{fig:gate_fidelity}]. Active decoupling~\cite{gorshkov25} can reduce the residual further.

\textit{Coupling to the Metamaterial.}---The metamaterial consists of $N$ identical resonators with nearest-neighbor coupling $J$, forming an ideally uniform one-dimensional chain. After circuit quantization, its Hamiltonian~\cite{zhang2023superconducting} is $H = \sum\limits_{j=1}^{N} \nu c^\dagger_jc_j + \sum\limits_{j=1}^{N-1}J (c^\dagger_jc_{j+1} + c_{j+1}^\dagger c_j)$, which the momentum-space modes $c_k = \sqrt{\frac{2}{N+1}}\sum\limits_j \sin \left(\frac{\pi kj}{N+1}\right) c_j$ diagonalize to $H=\sum_{k=1}^{N}\nu_k c^\dagger_k c_k$ with $\nu_k = \nu + 2J\cos\left(\frac{\pi k}{N+1}\right)$. Coupling $N_q$ DRTs to the metamaterial, with DRT $n$ attached to resonator $j_n$, extends the Hamiltonian of Eq.~\eqref{eq:drt_ham} to
\begin{subequations}
\label{eq:drt_mm_ham}
\begin{align} H_0 =& \sum\limits_{n=1}^{N_q}\Big(\omega_n (a^\dagger_n a_n + b^\dagger_n b_n) + g_{DR,n}(a^\dagger_n b_n + b^\dagger_n a_n)\nonumber\\
&- \frac{\eta_n}{2}(a^\dagger_na^\dagger_na_na_n + b^\dagger_nb^\dagger_nb_nb_n)\Big)
 + \sum\limits_{k=1}^{N}\nu_k c^\dagger_kc_k, \label{eq:h0mm}
\end{align}
\begin{equation}
H_\text{c} = \sqrt{\frac{2}{N+1}}\sum_{kn}\sin \frac{j_nk\pi}{N+1}\left((g_{na} a_n^\dagger + g_{nb} b_n^\dagger) c_k + \text{h.c.}\right), \label{eq:hcouplingmm}
\end{equation}
\begin{align} H_\text{d} = \sum\limits_n\Big(&\Omega_{ +,n}\cos(\omega_{+,n} t) + \Omega_{-, n}\cos(\omega_{-,n} t)\Big)\\\times(&a^\dagger_n - b^\dagger_n + \text{h.c.}). \label{eq:hdrivemm}
\end{align}
\end{subequations}

Compared with the Hamiltonian of Eq.~\eqref{eq:drt_ham}, the coupling here is diluted by the factor $\sqrt{\frac{2}{N+1}}$. A second difference is a cross-Kerr coupling between DRTs, mediated by the modes, which is strongest when two DRTs share a resonator. Separating the attachment points by several resonators suppresses it exponentially in the site distance between them~\cite{kumph_demonstration_2024}~\footnote{This holds when the intra-metamaterial coupling is weak compared to the qubit-mode detuning, and the detuning is large compared to the free spectral range.}. Exact diagonalization at our simulation parameters gives a suppression of $\sim(J/\Delta)^2 \approx 1/100$ per unit of site distance, independent of the total chain length. A site distance of $2$, one resonator between the attachment sites, already brings the mediated cross-Kerr below $1$~kHz, and a distance of $3$ leaves only a few Hz~\cite{SM}. We therefore place one DRT per two resonators [Fig.~\ref{fig:overview}(a)] and generate the DRT--DRT interaction by driving cross-resonantly with one of the modes. Coupling many qubits to a shared metamaterial introduces three error channels absent in the single-mode picture, control crosstalk between drive lines, spectral crosstalk between mode sidebands, and the erasure of spectator DRTs, which we treat in the End Matter.

As the metamaterial grows, the shrinking free spectral range crowds the sideband spectrum; this crowding and resonator photon loss together set the number of modes the array can support. We budget $10^{-3}$ for the two-qubit gate error, an order of magnitude below the surface-code threshold~\cite{fowler2012surface} and the physical error rate at which bivariate-bicycle qLDPC codes are projected to outperform the surface code with a tenth of the physical qubits~\cite{bravyi2024high}. Photon loss grows linearly with the gate duration (End Matter), so this budget bounds the duration from above by $T_\text{max}\propto T_1$. Resolving one mode from its neighbors, spaced by $\simeq 2\pi J/(N+1)$ near mid-band, bounds it from below by $T_\text{min}\propto (N+1)/J$. For flat-top pulses with smooth ramps the two bounds meet at $N+1 \approx 2.6\times10^{-3}\,(J/2\pi)\,T_1$~\cite{SM}. This mode-count ceiling grows linearly with both $T_1$ and $J$, and is independent of the qubit--mode coupling, the drive, and the anharmonicity, which only set where the gate duration falls between the bounds. Spectator erasure adds a third term, which grows as $(N+1)^2$ and depends on the coupling through the bare dispersive shift (End Matter). At $J/2\pi = 80$~MHz, mode $T_1 = 150\,\mu$s, and transmon $T_1 = 100\,\mu$s, the budget admits $N = 25$ modes ($N_q = 13$ DRTs), with spectator erasure taking a seventh of it. Reaching it requires a drive strong enough to give $g_{CR}/2\pi \approx 1.3$~MHz at $N = 25$ despite the coupling dilution $\sqrt{2/(N+1)}$; at weaker drives the array is drive-limited, and its capacity grows as $(\Omega T_1)^2$. At $T_1 = 30\,\mu$s the requirement rises to $g_{CR}/2\pi \approx 5.2$~MHz, above the $g_{CR}$ of the simulation in Fig.~\ref{fig:gate_fidelity}, so the simulated array is drive-limited, and photon loss already exceeds the $10^{-3}$ budget at $N = 3$. Amplitude-, phase-, and multi-tone-modulated gates, developed for the same crowding in large trapped-ion crystals~\cite{zhuArbitraryspeedQuantumGates2006, roosIonTrapQuantum2008, haddadfarshiHighFidelityQuantum2016, shapiraRobustEntanglementGates2018, webbResilientEntanglingGates2018, milnePhaseModulatedEntanglingGates2020}, apply directly to our sideband drives and can raise the prefactor of the ceiling. We avoid the band-edge channels, where the mode spacing falls as $1/(N+1)^2$ rather than $1/(N+1)$ (End Matter) and the sideband crowding is worst.

\textit{Parallel Parity Measurements.}---Parity measurements on different pairs of DRTs can run at once, by reading out the photon numbers of the populated modes. Each auxiliary transmon, coupled to the metamaterial and to its own readout line, is probed with a frequency comb as in Ref.~\cite{essig_multiplexed_2021}. Every populated mode shifts every auxiliary transmon, so we decode the photon numbers of all populated modes jointly, by maximum a posteriori estimation from the reflected comb signals. The modes $k$ and $N+1-k$ have the same weight at every site, so an auxiliary transmon cannot tell them apart, leaving at most $\lfloor(N+1)/2\rfloor$ channels. Two effects set the cost. The auxiliary transmons saturate, which caps their readout rate at their linewidth, itself bounded by the diluted dispersive shifts. When excited, they also mix photon configurations of different populated modes, which limits how many modes they read at strong coupling. Optimizing the auxiliary transmons and the number of pairs per round brings the cost to $3$ to $5\,\mu$s per pair for chains of $N = 5$ to $20$ resonators at unit detection efficiency, with three or four pairs per round optimal for $N \ge 10$~\cite{SM}.

\textit{Conclusions.}---We have presented an architecture for all-to-all connectivity in superconducting processors. Coupling many qubits through shared modes generates a dispersive coupling between every qubit and every mode, which has so far demanded active dynamical decoupling~\cite{gorshkov25}. Encoding the qubits as DRTs or SLTs suppresses this coupling passively, at the hardware level, and leaves a two-qubit gate and a joint parity measurement that are selective and run in parallel through the modes of one metamaterial. In simulation the gate reaches an infidelity of $1.4\times10^{-3}$, dominated by photon loss, and the number of qubits one metamaterial can host grows linearly with the mode lifetime and the inter-resonator coupling.
\begin{acknowledgments}
We thank the staff from across the AWS Center for Quantum Computing that enabled this project. We also thank James Hamilton, Nafea Bshara, and Peter DeSantis at AWS for their involvement and support of the research activities at the AWS Center for Quantum Computing.
During the preparation of this manuscript, the authors used Claude (Anthropic) to assist with phrasing and editing of the text, and to debug and extend the simulation code. All AI-assisted code changes and numerical results were verified by the authors.
\end{acknowledgments}
\bibliography{bibliography}

@PREAMBLE{
 "\providecommand{\noopsort}[1]{}" 
 # "\providecommand{\singleletter}[1]{#1}%" 
}

@article{kubica2023erasure,
  title={Erasure qubits: Overcoming the t 1 limit in superconducting circuits},
  author={Kubica, Aleksander and Haim, Arbel and Vaknin, Yotam and Levine, Harry and Brand{\~a}o, Fernando and Retzker, Alex},
  journal={Physical Review X},
  volume={13},
  number={4},
  pages={041022},
  year={2023},
  publisher={APS},
  doi={10.1103/PhysRevX.13.041022},
  url={https://doi.org/10.1103/PhysRevX.13.041022}
}

@article{bravyi2011schrieffer,
  title={Schrieffer--Wolff transformation for quantum many-body systems},
  author={Bravyi, Sergey and DiVincenzo, David P and Loss, Daniel},
  journal={Annals of physics},
  volume={326},
  number={10},
  pages={2793--2826},
  year={2011},
  publisher={Elsevier},
  doi={10.1016/j.aop.2011.06.004},
  url={https://doi.org/10.1016/j.aop.2011.06.004}
}

@article{levine2023demonstrating,
  title={Demonstrating a long-coherence dual-rail erasure qubit using tunable transmons},
  author={Levine, Harry and Haim, Arbel and Hung, Jimmy SC and Alidoust, Nasser and Kalaee, Mahmoud and DeLorenzo, Laura and Wollack, E Alex and Arriola, Patricio Arrangoiz and Khalajhedayati, Amirhossein and Vaknin, Yotam and others},
  journal={arXiv preprint arXiv:2307.08737},
  year={2023},
  doi={10.48550/arXiv.2307.08737},
  url={https://arxiv.org/abs/2307.08737}
}

@article{magesan2020effective,
  title={Effective Hamiltonian models of the cross-resonance gate},
  author={Magesan, Easwar and Gambetta, Jay M},
  journal={Physical Review A},
  volume={101},
  number={5},
  pages={052308},
  year={2020},
  publisher={APS},
  doi={10.1103/PhysRevA.101.052308},
  url={https://doi.org/10.1103/PhysRevA.101.052308}
}

@article{paik2016experimental,
  title={Experimental demonstration of a resonator-induced phase gate in a multiqubit circuit-qed system},
  author={Paik, Hanhee and Mezzacapo, Antonio and Sandberg, Martin and McClure, Doug T and Abdo, Baleegh and C{\'o}rcoles, Antonio D and Dial, O and Bogorin, Daniela F and Plourde, BLT and Steffen, Matthias and others},
  journal={Physical review letters},
  volume={117},
  number={25},
  pages={250502},
  year={2016},
  publisher={APS},
  doi={10.1103/PhysRevLett.117.250502},
  url={https://doi.org/10.1103/PhysRevLett.117.250502}
}

@article{puri2016high,
  title={High-fidelity resonator-induced phase gate with single-mode squeezing},
  author={Puri, Shruti and Blais, Alexandre},
  journal={Physical review letters},
  volume={116},
  number={18},
  pages={180501},
  year={2016},
  publisher={APS},
  doi={10.1103/PhysRevLett.116.180501},
  url={https://doi.org/10.1103/PhysRevLett.116.180501}
}

@article{cross2015optimized,
  title={Optimized pulse shapes for a resonator-induced phase gate},
  author={Cross, Andrew W and Gambetta, Jay M},
  journal={Physical Review A},
  volume={91},
  number={3},
  pages={032325},
  year={2015},
  publisher={APS},
  doi={10.1103/PhysRevA.91.032325},
  url={https://doi.org/10.1103/PhysRevA.91.032325}
}

@article{malekakhlagh2022optimization,
  title={Optimization of the resonator-induced phase gate for superconducting qubits},
  author={Malekakhlagh, Moein and Shanks, William and Paik, Hanhee},
  journal={Physical Review A},
  volume={105},
  number={2},
  pages={022607},
  year={2022},
  publisher={APS},
  doi={10.1103/PhysRevA.105.022607},
  url={https://doi.org/10.1103/PhysRevA.105.022607}
}

@article{zuk2024robust,
  title={Robust gates with spin-locked superconducting qubits},
  author={Zuk, Ido and Cohen, Daniel and Gorshkov, Alexey V and Retzker, Alex},
  journal={Physical Review Research},
  volume={6},
  number={1},
  pages={013217},
  year={2024},
  publisher={APS},
  doi={10.1103/PhysRevResearch.6.013217},
  url={https://doi.org/10.1103/PhysRevResearch.6.013217}
}

@article{acharya2024quantum,
  title={Quantum error correction below the surface code threshold},
  author={Acharya, Rajeev and Aghababaie-Beni, Laleh and Aleiner, Igor and Andersen, Trond I and Ansmann, Markus and Arute, Frank and Arya, Kunal and Asfaw, Abraham and Astrakhantsev, Nikita and Atalaya, Juan and others},
  journal={arXiv preprint arXiv:2408.13687},
  year={2024},
  doi={10.48550/arXiv.2408.13687},
  url={https://arxiv.org/abs/2408.13687}
}

@article{krinner2022realizing,
  title={Realizing repeated quantum error correction in a distance-three surface code},
  author={Krinner, Sebastian and Lacroix, Nathan and Remm, Ants and Di Paolo, Agustin and Genois, Elie and Leroux, Catherine and Hellings, Christoph and Lazar, Stefania and Swiadek, Francois and Herrmann, Johannes and others},
  journal={Nature},
  volume={605},
  number={7911},
  pages={669--674},
  year={2022},
  publisher={Nature Publishing Group UK London},
  doi={10.1038/s41586-022-04566-8},
  url={https://doi.org/10.1038/s41586-022-04566-8}
}

@article{google2023suppressing,
  title={Suppressing quantum errors by scaling a surface code logical qubit},
  author={Abe Asfaw and Anthony Megrant and Cody Jones and Craig Gidney and Dave Bacon and Dripto Debroy and Dvir Kafri and Erik Lucero and Hartmut Neven and Jeremy Hilton and Jimmy Chen and John Platt and Jonathan Gross and Juan Atalaya and Julian Kelly and Kenny Lee and Kevin Satzinger and Michael Newman and Sergio Boixo and Vadim Smelyanskiy and Yu Chen and Catherine Vollgraff Heidweiller},
  journal={Nature},
  volume={614},
  number={7949},
  pages={676--681},
  year={2023},
  publisher={Nature Publishing Group UK London},
  doi={10.1038/s41586-022-05434-1},
  url={https://doi.org/10.1038/s41586-022-05434-1}
}

@article{fowler2012surface,
  title={Surface codes: Towards practical large-scale quantum computation},
  author={Fowler, Austin G and Mariantoni, Matteo and Martinis, John M and Cleland, Andrew N},
  journal={Physical Review A—Atomic, Molecular, and Optical Physics},
  volume={86},
  number={3},
  pages={032324},
  year={2012},
  publisher={APS},
  doi={10.1103/PhysRevA.86.032324},
  url={https://doi.org/10.1103/PhysRevA.86.032324}
}

@inproceedings{panteleev2022asymptotically,
  title={Asymptotically good quantum and locally testable classical LDPC codes},
  author={Panteleev, Pavel and Kalachev, Gleb},
  booktitle={Proceedings of the 54th Annual ACM SIGACT Symposium on Theory of Computing},
  pages={375--388},
  year={2022},
  doi={10.1145/3519935.3520017},
  url={https://doi.org/10.1145/3519935.3520017}
}

@article{bravyi2024high,
  title={High-threshold and low-overhead fault-tolerant quantum memory},
  author={Bravyi, Sergey and Cross, Andrew W and Gambetta, Jay M and Maslov, Dmitri and Rall, Patrick and Yoder, Theodore J},
  journal={Nature},
  volume={627},
  number={8005},
  pages={778--782},
  year={2024},
  publisher={Nature Publishing Group UK London},
  doi={10.1038/s41586-024-07107-7},
  url={https://doi.org/10.1038/s41586-024-07107-7}
}

@article{sorensen1999quantum,
  title={Quantum computation with ions in thermal motion},
  author={S{\o}rensen, Anders and M{\o}lmer, Klaus},
  journal={Physical review letters},
  volume={82},
  number={9},
  pages={1971},
  year={1999},
  publisher={APS},
  doi={10.1103/PhysRevLett.82.1971},
  url={https://doi.org/10.1103/PhysRevLett.82.1971}
}

@article{zhang2023superconducting,
  title={A superconducting quantum simulator based on a photonic-bandgap metamaterial},
  author={Zhang, Xueyue and Kim, Eunjong and Mark, Daniel K and Choi, Soonwon and Painter, Oskar},
  journal={Science},
  volume={379},
  number={6629},
  pages={278--283},
  year={2023},
  publisher={American Association for the Advancement of Science},
  doi={10.1126/science.ade7651},
  url={https://doi.org/10.1126/science.ade7651}
}

@article{caiRobustDynamicalDecoupling2012a,
  title = {Robust Dynamical Decoupling with Concatenated Continuous Driving},
  author = {Cai, J.-M. and Naydenov, B. and Pfeiffer, R. and McGuinness, L. P. and Jahnke, K. D. and Jelezko, F. and Plenio, M. B. and Retzker, A.},
  year = {2012},
  month = nov,
  journal = {New Journal of Physics},
  volume = {14},
  number = {11},
  pages = {113023},
  publisher = {IOP Publishing},
  issn = {1367-2630},
  doi = {10.1088/1367-2630/14/11/113023},
  urldate = {2022-03-02},
  langid = {english}
}

@article{bravyi_no-go_2009,
    title = {A no-go theorem for a two-dimensional self-correcting quantum memory based on stabilizer codes},
    volume = {11},
    issn = {1367-2630},
    url = {http://arxiv.org/abs/0810.1983},
    doi = {10.1088/1367-2630/11/4/043029},
    number = {4},
    urldate = {2025-03-12},
    journal = {New Journal of Physics},
    author = {Bravyi, Sergey and Terhal, Barbara},
    month = apr,
    year = {2009},
    note = {arXiv:0810.1983 [quant-ph]},
    pages = {043029},
}

@article{bravyi_tradeoffs_2010,
    title = {Tradeoffs for reliable quantum information storage in {2D} systems},
    volume = {104},
    issn = {0031-9007, 1079-7114},
    url = {http://arxiv.org/abs/0909.5200},
    doi = {10.1103/PhysRevLett.104.050503},
    number = {5},
    urldate = {2025-03-04},
    journal = {Physical Review Letters},
    author = {Bravyi, Sergey and Poulin, David and Terhal, Barbara},
    month = feb,
    year = {2010},
    note = {arXiv:0909.5200 [quant-ph]},
    pages = {050503},
}

@article{baspin_quantifying_2022,
    title = {Quantifying {Nonlocality}: {How} {Outperforming} {Local} {Quantum} {Codes} {Is} {Expensive}},
    volume = {129},
    shorttitle = {Quantifying {Nonlocality}},
    url = {https://link.aps.org/doi/10.1103/PhysRevLett.129.050505},
    doi = {10.1103/PhysRevLett.129.050505},
    number = {5},
    urldate = {2025-04-10},
    journal = {Physical Review Letters},
    author = {Baspin, Nouédyn and Krishna, Anirudh},
    month = jul,
    year = {2022},
    note = {Publisher: American Physical Society},
    pages = {050505},
}

@misc{huang_logical_2025,
    title = {Logical multi-qubit entanglement with dual-rail superconducting qubits},
    url = {http://arxiv.org/abs/2504.12099},
    doi = {10.48550/arXiv.2504.12099},
    urldate = {2025-04-21},
    publisher = {arXiv},
    author = {Huang, Wenhui and Sun, Xuandong and Zhang, Jiawei and Guo, Zechen and Huang, Peisheng and Liang, Yongqi and Liu, Yiting and Sun, Daxiong and Wang, Zilin and Xiong, Yuzhe and Yang, Xiaohan and Zhang, Jiajian and Zhang, Libo and Chu, Ji and Guo, Weijie and Jiang, Ji and Liu, Song and Niu, Jingjing and Qiu, Jiawei and Tao, Ziyu and Zhou, Yuxuan and Linpeng, Xiayu and Zhong, Youpeng and Yu, Dapeng},
    month = apr,
    year = {2025},
    note = {arXiv:2504.12099 [quant-ph]},
}

@misc{kumph_demonstration_2024,
    title = {Demonstration of {RIP} gates in a quantum processor with negligible transverse coupling},
    url = {http://arxiv.org/abs/2406.11770},
    doi = {10.48550/arXiv.2406.11770},
    urldate = {2025-04-29},
    publisher = {arXiv},
    author = {Kumph, Muir and Raftery, James and Finck, Aaron and Blair, John and Carniol, April and Carnevale, Santino and Keefe, George A. and Arena, Vincent and Hall, Shawn and McKay, David and Stehlik, George},
    month = jun,
    year = {2024},
    note = {arXiv:2406.11770 [quant-ph]},
}

@article{gorshkov25,
  title = {Cavity-Mediated Cross-Cross-Resonance Gate},
  author = {Gorshkov, Alexey V. and Cohen, Daniel and Haim, Arbel and Rotem, Amit and Golan, Or and Kim, Gihwan and Butler, Andreas and Hann, Connor T. and Painter, Oskar and Brand{\~a}o, Fernando G. S. L. and Retzker, Alex},
  year = {2025},
  journal = {arXiv:2506.03239},
  doi = {10.48550/arXiv.2506.03239},
  url = {https://arxiv.org/abs/2506.03239}
}

@article{essig_multiplexed_2021,
    title = {Multiplexed {Photon} {Number} {Measurement}},
    volume = {11},
    issn = {2160-3308},
    url = {https://link.aps.org/doi/10.1103/PhysRevX.11.031045},
    doi = {10.1103/PhysRevX.11.031045},
    number = {3},
    urldate = {2025-06-24},
    journal = {Physical Review X},
    author = {Essig, Antoine and Ficheux, Quentin and Peronnin, Théau and Cottet, Nathanaël and Lescanne, Raphaël and Sarlette, Alain and Rouchon, Pierre and Leghtas, Zaki and Huard, Benjamin},
    month = aug,
    year = {2021},
    pages = {031045},
}

@article{nielsen_simple_2002,
    title = {A simple formula for the average gate fidelity of a quantum dynamical operation},
    volume = {303},
    issn = {03759601},
    url = {http://arxiv.org/abs/quant-ph/0205035},
    doi = {10.1016/S0375-9601(02)01272-0},
    number = {4},
    urldate = {2025-05-16},
    journal = {Physics Letters A},
    author = {Nielsen, Michael A.},
    month = oct,
    year = {2002},
    note = {arXiv:quant-ph/0205035},
    pages = {249--252},
}

@article{sorensen_entanglement_2000,
    title = {Entanglement and quantum computation with ions in thermal motion},
    volume = {62},
    copyright = {http://link.aps.org/licenses/aps-default-license},
    issn = {1050-2947, 1094-1622},
    url = {https://link.aps.org/doi/10.1103/PhysRevA.62.022311},
    doi = {10.1103/PhysRevA.62.022311},
    number = {2},
    urldate = {2025-10-21},
    journal = {Physical Review A},
    author = {Sørensen, Anders and Mølmer, Klaus},
    month = jul,
    year = {2000},
    pages = {022311},
}

@article{touzard_gated_2019,
    title = {Gated conditional displacement readout of superconducting qubits},
    volume = {122},
    issn = {0031-9007, 1079-7114},
    url = {http://arxiv.org/abs/1809.06964},
    doi = {10.1103/PhysRevLett.122.080502},
    number = {8},
    urldate = {2025-04-15},
    journal = {Physical Review Letters},
    author = {Touzard, S. and Kou, A. and Frattini, N. E. and Sivak, V. V. and Puri, S. and Grimm, A. and Frunzio, L. and Shankar, S. and Devoret, M. H.},
    month = feb,
    year = {2019},
    note = {arXiv:1809.06964 [quant-ph]},
    pages = {080502},
}

@article{campbell_universal_2020,
    title = {Universal {Nonadiabatic} {Control} of {Small}-{Gap} {Superconducting} {Qubits}},
    volume = {10},
    issn = {2160-3308},
    url = {https://link.aps.org/doi/10.1103/PhysRevX.10.041051},
    doi = {10.1103/PhysRevX.10.041051},
    number = {4},
    urldate = {2025-10-22},
    journal = {Physical Review X},
    author = {Campbell, Daniel L. and Shim, Yun-Pil and Kannan, Bharath and Winik, Roni and Kim, David K. and Melville, Alexander and Niedzielski, Bethany M. and Yoder, Jonilyn L. and Tahan, Charles and Gustavsson, Simon and Oliver, William D.},
    month = dec,
    year = {2020},
    pages = {041051},
}

@misc{wang_demonstration_2025,
    title = {Demonstration of low-overhead quantum error correction codes},
    copyright = {Creative Commons Attribution 4.0 International},
    url = {https://arxiv.org/abs/2505.09684},
    doi = {10.48550/ARXIV.2505.09684},
    urldate = {2025-12-03},
    publisher = {arXiv},
    author = {Wang, Ke and Lu, Zhide and Zhang, Chuanyu and Liu, Gongyu and Chen, Jiachen and Wang, Yanzhe and Wu, Yaozu and Xu, Shibo and Zhu, Xuhao and Jin, Feitong and Gao, Yu and Tan, Ziqi and Cui, Zhengyi and Wang, Ning and Zou, Yiren and Zhang, Aosai and Li, Tingting and Shen, Fanhao and Zhong, Jiarun and Bao, Zehang and Zhu, Zitian and Han, Yihang and He, Yiyang and Shen, Jiayuan and Wang, Han and Yang, Jia-Nan and Song, Zixuan and Deng, Jinfeng and Dong, Hang and Sun, Zheng-Zhi and Li, Weikang and Ye, Qi and Jiang, Si and Ma, Yixuan and Shen, Pei-Xin and Zhang, Pengfei and Li, Hekang and Guo, Qiujiang and Wang, Zhen and Song, Chao and Wang, H. and Deng, Dong-Ling},
    year = {2025},
    note = {Version Number: 1},
}

@article{putterman_hardware-efficient_2025,
    title = {Hardware-efficient quantum error correction via concatenated bosonic qubits},
    volume = {638},
    copyright = {2025 The Author(s)},
    issn = {1476-4687},
    url = {https://www.nature.com/articles/s41586-025-08642-7},
    doi = {10.1038/s41586-025-08642-7},
    number = {8052},
    urldate = {2025-12-03},
    journal = {Nature},
    author = {Putterman, Harald and Noh, Kyungjoo and Hann, Connor T. and MacCabe, Gregory S. and Aghaeimeibodi, Shahriar and Patel, Rishi N. and Lee, Menyoung and Jones, William M. and Moradinejad, Hesam and Rodriguez, Roberto and Mahuli, Neha and Rose, Jefferson and Owens, John Clai and Levine, Harry and Rosenfeld, Emma and Reinhold, Philip and Moncelsi, Lorenzo and Alcid, Joshua Ari and Alidoust, Nasser and Arrangoiz-Arriola, Patricio and Barnett, James and Bienias, Przemyslaw and Carson, Hugh A. and Chen, Cliff and Chen, Li and Chinkezian, Harutiun and Chisholm, Eric M. and Chou, Ming-Han and Clerk, Aashish and Clifford, Andrew and Cosmic, R. and Curiel, Ana Valdes and Davis, Erik and DeLorenzo, Laura and D’Ewart, J. Mitchell and Diky, Art and D’Souza, Nathan and Dumitrescu, Philipp T. and Eisenmann, Shmuel and Elkhouly, Essam and Evenbly, Glen and Fang, Michael T. and Fang, Yawen and Fling, Matthew J. and Fon, Warren and Garcia, Gabriel and Gorshkov, Alexey V. and Grant, Julia A. and Gray, Mason J. and Grimberg, Sebastian and Grimsmo, Arne L. and Haim, Arbel and Hand, Justin and He, Yuan and Hernandez, Mike and Hover, David and Hung, Jimmy S. C. and Hunt, Matthew and Iverson, Joe and Jarrige, Ignace and Jaskula, Jean-Christophe and Jiang, Liang and Kalaee, Mahmoud and Karabalin, Rassul and Karalekas, Peter J. and Keller, Andrew J. and Khalajhedayati, Amirhossein and Kubica, Aleksander and Lee, Hanho and Leroux, Catherine and Lieu, Simon and Ly, Victor and Madrigal, Keven Villegas and Marcaud, Guillaume and McCabe, Gavin and Miles, Cody and Milsted, Ashley and Minguzzi, Joaquin and Mishra, Anurag and Mukherjee, Biswaroop and Naghiloo, Mahdi and Oblepias, Eric and Ortuno, Gerson and Pagdilao, Jason and Pancotti, Nicola and Panduro, Ashley and Paquette, J. P. and Park, Minje and Peairs, Gregory A. and Perello, David and Peterson, Eric C. and Ponte, Sophia and Preskill, John and Qiao, Johnson and Refael, Gil and Resnick, Rachel and Retzker, Alex and Reyna, Omar A. and Runyan, Marc and Ryan, Colm A. and Sahmoud, Abdulrahman and Sanchez, Ernesto and Sanil, Rohan and Sankar, Krishanu and Sato, Yuki and Scaffidi, Thomas and Siavoshi, Salome and Sivarajah, Prasahnt and Skogland, Trenton and Su, Chun-Ju and Swenson, Loren J. and Teo, Stephanie M. and Tomada, Astrid and Torlai, Giacomo and Wollack, E. Alex and Ye, Yufeng and Zerrudo, Jessica A. and Zhang, Kailing and Brandão, Fernando G. S. L. and Matheny, Matthew H. and Painter, Oskar},
    month = feb,
    year = {2025},
    note = {Publisher: Nature Publishing Group},
    pages = {927--934},
}

@article{brock_quantum_2025,
    title = {Quantum error correction of qudits beyond break-even},
    volume = {641},
    copyright = {2025 The Author(s)},
    issn = {1476-4687},
    url = {https://www.nature.com/articles/s41586-025-08899-y},
    doi = {10.1038/s41586-025-08899-y},
    number = {8063},
    urldate = {2025-12-03},
    journal = {Nature},
    author = {Brock, Benjamin L. and Singh, Shraddha and Eickbusch, Alec and Sivak, Volodymyr V. and Ding, Andy Z. and Frunzio, Luigi and Girvin, Steven M. and Devoret, Michel H.},
    month = may,
    year = {2025},
    note = {Publisher: Nature Publishing Group},
    pages = {612--618},
}

@misc{dinur_good_2022,
    title = {Good {Quantum} {LDPC} {Codes} with {Linear} {Time} {Decoders}},
    url = {http://arxiv.org/abs/2206.07750},
    doi = {10.48550/arXiv.2206.07750},
    urldate = {2025-03-04},
    publisher = {arXiv},
    author = {Dinur, Irit and Hsieh, Min-Hsiu and Lin, Ting-Chun and Vidick, Thomas},
    month = jun,
    year = {2022},
    note = {arXiv:2206.07750 [quant-ph]},
}

@article{berthusen_toward_2025,
    title = {Toward a {2D} {Local} {Implementation} of {Quantum} {Low}-{Density} {Parity}-{Check} {Codes}},
    volume = {6},
    url = {https://link.aps.org/doi/10.1103/PRXQuantum.6.010306},
    doi = {10.1103/PRXQuantum.6.010306},
    number = {1},
    urldate = {2025-12-03},
    journal = {PRX Quantum},
    author = {Berthusen, Noah and Devulapalli, Dhruv and Schoute, Eddie and Childs, Andrew M. and Gullans, Michael J. and Gorshkov, Alexey V. and Gottesman, Daniel},
    month = jan,
    year = {2025},
    note = {Publisher: American Physical Society},
    pages = {010306},
}

@article{marxer_long-distance_2023,
    title = {Long-{Distance} {Transmon} {Coupler} with cz-{Gate} {Fidelity} above \$99.8{\textbackslash}mathrm\{\%\}\$},
    volume = {4},
    url = {https://link.aps.org/doi/10.1103/PRXQuantum.4.010314},
    doi = {10.1103/PRXQuantum.4.010314},
    number = {1},
    urldate = {2025-12-03},
    journal = {PRX Quantum},
    author = {Marxer, Fabian and Vepsäläinen, Antti and Jolin, Shan W. and Tuorila, Jani and Landra, Alessandro and Ockeloen-Korppi, Caspar and Liu, Wei and Ahonen, Olli and Auer, Adrian and Belzane, Lucien and Bergholm, Ville and Chan, Chun Fai and Chan, Kok Wai and Hiltunen, Tuukka and Hotari, Juho and Hyyppä, Eric and Ikonen, Joni and Janzso, David and Koistinen, Miikka and Kotilahti, Janne and Li, Tianyi and Luus, Jyrgen and Papic, Miha and Partanen, Matti and Räbinä, Jukka and Rosti, Jari and Savytskyi, Mykhailo and Seppälä, Marko and Sevriuk, Vasilii and Takala, Eelis and Tarasinski, Brian and Thapa, Manish J. and Tosto, Francesca and Vorobeva, Natalia and Yu, Liuqi and Tan, Kuan Yen and Hassel, Juha and Möttönen, Mikko and Heinsoo, Johannes},
    month = feb,
    year = {2023},
    note = {Publisher: American Physical Society},
    pages = {010314},
}

@article{clerkIntroductionQuantumNoise2010,
    title = {Introduction to quantum noise, measurement, and amplification},
    volume = {82},
    copyright = {http://link.aps.org/licenses/aps-default-license},
    issn = {0034-6861, 1539-0756},
    url = {https://link.aps.org/doi/10.1103/RevModPhys.82.1155},
    doi = {10.1103/RevModPhys.82.1155},
    number = {2},
    urldate = {2026-01-30},
    journal = {Reviews of Modern Physics},
    author = {Clerk, A. A. and Devoret, M. H. and Girvin, S. M. and Marquardt, Florian and Schoelkopf, R. J.},
    month = apr,
    year = {2010},
    pages = {1155--1208},
}

@article{majerCouplingSuperconductingQubits2007,
    title = {Coupling {Superconducting} {Qubits} via a {Cavity} {Bus}},
    volume = {449},
    issn = {0028-0836, 1476-4687},
    url = {http://arxiv.org/abs/0709.2135},
    doi = {10.1038/nature06184},
    number = {7161},
    urldate = {2026-04-14},
    journal = {Nature},
    author = {Majer, J. and Chow, J. M. and Gambetta, J. M. and Koch, Jens and Johnson, B. R. and Schreier, J. A. and Frunzio, L. and Schuster, D. I. and Houck, A. A. and Wallraff, A. and Blais, A. and Devoret, M. H. and Girvin, S. M. and Schoelkopf, R. J.},
    month = sep,
    year = {2007},
    note = {arXiv:0709.2135 [cond-mat]},
    pages = {443--447},
}

@article{bermudezDissipationAssistedQuantumInformation2013a,
    title = {Dissipation-{Assisted} {Quantum} {Information} {Processing} with {Trapped} {Ions}},
    volume = {110},
    url = {https://link.aps.org/doi/10.1103/PhysRevLett.110.110502},
    doi = {10.1103/PhysRevLett.110.110502},
    number = {11},
    urldate = {2026-04-12},
    journal = {Physical Review Letters},
    publisher = {American Physical Society},
    author = {Bermudez, A. and Schaetz, T. and Plenio, M. B.},
    month = mar,
    year = {2013},
    pages = {110502},
}

@article{hastingsDynamicallyGeneratedLogical2021,
    title = {Dynamically {Generated} {Logical} {Qubits}},
    volume = {5},
    url = {https://quantum-journal.org/papers/q-2021-10-19-564/},
    doi = {10.22331/q-2021-10-19-564},
    urldate = {2025-11-18},
    journal = {Quantum},
    publisher = {Verein zur Förderung des Open Access Publizierens in den Quantenwissenschaften},
    author = {Hastings, Matthew B. and Haah, Jeongwan},
    month = oct,
    year = {2021},
    pages = {564},
}

@article{higgottConstructionsPerformanceHyperbolic2024,
    title = {Constructions and {Performance} of {Hyperbolic} and {Semi}-{Hyperbolic} {Floquet} {Codes}},
    volume = {5},
    url = {https://link.aps.org/doi/10.1103/PRXQuantum.5.040327},
    doi = {10.1103/PRXQuantum.5.040327},
    number = {4},
    urldate = {2026-07-19},
    journal = {PRX Quantum},
    publisher = {American Physical Society},
    author = {Higgott, Oscar and Breuckmann, Nikolas P.},
    month = nov,
    year = {2024},
    pages = {040327},
}

@misc{jacobyStairwayCodesFloquetifying2026,
    title = {Stairway {Codes}: {Floquetifying} {Bivariate} {Bicycle} {Codes} and {Beyond}},
    shorttitle = {Stairway {Codes}},
    url = {http://arxiv.org/abs/2603.00228},
    doi = {10.48550/arXiv.2603.00228},
    urldate = {2026-07-19},
    publisher = {arXiv},
    author = {Jacoby, Shoham and Retzker, Alex and Pastawski, Fernando},
    month = feb,
    year = {2026},
    note = {arXiv:2603.00228 [quant-ph]},
}

@article{mercurioQuantumToolboxjlEfficientJulia2025a,
  author  = {Mercurio, Alberto and Huang, Yi-Te and Cai, Li-Xun and
             Chen, Yueh-Nan and Savona, Vincenzo and Nori, Franco},
  title   = {{QuantumToolbox}.jl: {An} efficient {Julia} framework for
             simulating open quantum systems},
  journal = {Quantum},
  volume  = {9},
  pages   = {1866},
  year    = {2025},
  doi     = {10.22331/q-2025-09-29-1866}
}

@misc{SM,
    note = {See the Supplemental Material below for the derivation of the effective Hamiltonian, including the drive-induced frequency shifts, the residual and mediated cross-Kerr interactions, and robustness to rail asymmetry; the spin-locked transmon analysis; the backaction, decoding, and measurement-time analysis of the single and parallel parity measurements; and the average gate fidelity under spectator erasure},
}

@article{haddadfarshiHighFidelityQuantum2016,
  title = {High Fidelity Quantum Gates of Trapped Ions in the Presence of Motional Heating},
  author = {Haddadfarshi, Farhang and Mintert, Florian},
  year = 2016,
  month = dec,
  journal = {New Journal of Physics},
  volume = {18},
  number = {12},
  pages = {123007},
  publisher = {IOP Publishing},
  issn = {1367-2630},
  doi = {10.1088/1367-2630/18/12/123007},
  urldate = {2026-08-12},
  langid = {english},
}

@article{milnePhaseModulatedEntanglingGates2020,
  title = {Phase-{{Modulated Entangling Gates Robust}} to {{Static}} and {{Time-Varying Errors}}},
  author = {Milne, Alistair R. and Edmunds, Claire L. and Hempel, Cornelius and Roy, Federico and Mavadia, Sandeep and Biercuk, Michael J.},
  year = 2020,
  month = feb,
  journal = {Physical Review Applied},
  volume = {13},
  number = {2},
  pages = {024022},
  publisher = {American Physical Society},
  doi = {10.1103/PhysRevApplied.13.024022},
  urldate = {2026-08-12},
}

@article{roosIonTrapQuantum2008,
  title = {Ion Trap Quantum Gates with Amplitude-Modulated Laser Beams},
  author = {Roos, Christian F},
  year = 2008,
  month = jan,
  journal = {New Journal of Physics},
  volume = {10},
  number = {1},
  pages = {013002},
  issn = {1367-2630},
  doi = {10.1088/1367-2630/10/1/013002},
  urldate = {2026-08-12},
  langid = {english},
}

@article{shapiraRobustEntanglementGates2018,
  title = {Robust {{Entanglement Gates}} for {{Trapped-Ion Qubits}}},
  author = {Shapira, Yotam and Shaniv, Ravid and Manovitz, Tom and Akerman, Nitzan and Ozeri, Roee},
  year = 2018,
  month = nov,
  journal = {Physical Review Letters},
  volume = {121},
  number = {18},
  pages = {180502},
  publisher = {American Physical Society},
  doi = {10.1103/PhysRevLett.121.180502},
  urldate = {2026-08-12},
}

@article{webbResilientEntanglingGates2018,
  title = {Resilient {{Entangling Gates}} for {{Trapped Ions}}},
  author = {Webb, A. E. and Webster, S. C. and Collingbourne, S. and Bretaud, D. and Lawrence, A. M. and Weidt, S. and Mintert, F. and Hensinger, W. K.},
  year = 2018,
  month = nov,
  journal = {Physical Review Letters},
  volume = {121},
  number = {18},
  pages = {180501},
  publisher = {American Physical Society},
  doi = {10.1103/PhysRevLett.121.180501},
  urldate = {2026-08-12},
}

@article{zhuArbitraryspeedQuantumGates2006,
  title = {Arbitrary-Speed Quantum Gates within Large Ion Crystals through Minimum Control of Laser Beams},
  author = {Zhu, Shi-Liang and Monroe, C. and Duan, L.-M.},
  year = 2006,
  month = jan,
  journal = {Europhysics Letters},
  volume = {73},
  number = {4},
  pages = {485},
  publisher = {IOP Publishing},
  issn = {0295-5075},
  doi = {10.1209/epl/i2005-10424-4},
  urldate = {2026-08-12},
  langid = {english},
}

@article{malekakhlaghFirstprinciplesAnalysisCrossresonance2020a,
  title = {First-principles analysis of cross-resonance gate operation},
  author = {Malekakhlagh, Moein and Magesan, Easwar and McKay, David C.},
  journal = {Physical Review A},
  volume = {102},
  number = {4},
  pages = {042605},
  year = {2020},
  publisher = {American Physical Society},
  doi = {10.1103/PhysRevA.102.042605},
  url = {https://link.aps.org/doi/10.1103/PhysRevA.102.042605}
}

@article{petrescuAccurateMethodsAnalysis2023,
  title = {Accurate Methods for the Analysis of Strong-Drive Effects in Parametric Gates},
  author = {Petrescu, Alexandru and Le Calonnec, Camille and Leroux, Catherine and Di Paolo, Agustin and Mundada, Pranav and Sussman, Sara and Vrajitoarea, Andrei and Houck, Andrew A. and Blais, Alexandre},
  journal = {Physical Review Applied},
  volume = {19},
  number = {4},
  pages = {044003},
  year = {2023},
  publisher = {American Physical Society},
  doi = {10.1103/PhysRevApplied.19.044003},
  url = {https://link.aps.org/doi/10.1103/PhysRevApplied.19.044003}
}

@misc{liuEfficientRoutingQuantum2026,
  title = {Efficient {{Routing}} of {{Quantum LDPC Codes}} on {{Programmable 2D Toric Architectures}}},
  author = {Liu, Kun and Tsunoda, Takahiro and Xue, Sophia H. and McKinney, Evan and Zhou, Zeyuan and Xu, Shifan and Schoelkopf, Robert J. and Ding, Yongshan},
  year = 2026,
  month = apr,
  number = {arXiv:2604.18714},
  eprint = {2604.18714},
  primaryclass = {quant-ph},
  publisher = {arXiv},
  doi = {10.48550/arXiv.2604.18714},
  urldate = {2026-07-23},
  archiveprefix = {arXiv},
}

@misc{hung2026fasthighfidelityerasuredetection,
      title={Fast, High-Fidelity Erasure Detection of Dual-Rail Qubits with Symmetrically Coupled Readout}, 
      author={Jimmy Shih-Chun Hung and Arbel Haim and Mouktik Raha and Gihwan Kim and Ziwen Huang and Ming-Han Chou and Mitch D'Ewart and Erik Davis and Anurag Mishra and Patricio Arrangoiz Arriola and Amirhossein Khalajhedayati and David Hover and Fernando G. S. L. Brandão and Aashish A. Clerk and Alex Retzker and Harry Levine and Oskar Painter},
      year={2026},
      eprint={2604.16292},
      archivePrefix={arXiv},
      primaryClass={quant-ph},
      url={https://arxiv.org/abs/2604.16292}, 
}

@article{Chow2010JointReadout,
  author  = {Chow, J. M. and DiCarlo, L. and Gambetta, J. M. and Nunnenkamp, A. and Bishop, Lev S. and Frunzio, L. and Devoret, M. H. and Girvin, S. M. and Schoelkopf, R. J.},
  title   = {Detecting highly entangled states with a joint qubit readout},
  journal = {Physical Review A},
  volume  = {81},
  number  = {6},
  pages   = {062325},
  year    = {2010},
  doi     = {10.1103/PhysRevA.81.062325}
}

@article{Ciani2017ThreeQubitParity,
  author  = {Ciani, Alessandro and DiVincenzo, David P.},
  title   = {Three-qubit direct dispersive parity measurement with tunable coupling qubits},
  journal = {Physical Review B},
  volume  = {96},
  number  = {21},
  pages   = {214511},
  year    = {2017},
  doi     = {10.1103/PhysRevB.96.214511}
}

@article{Criger2016MultiQubitJoint,
  author  = {Criger, Ben and Ciani, Alessandro and DiVincenzo, David P.},
  title   = {Multi-qubit joint measurements in circuit {QED}: stochastic master equation analysis},
  journal = {EPJ Quantum Technology},
  volume  = {3},
  number  = {1},
  pages   = {1--21},
  year    = {2016},
  doi     = {10.1140/epjqt/s40507-016-0044-6}
}

@article{DiCarlo2009TwoQubitAlgorithms,
  author  = {DiCarlo, L. and Chow, J. M. and Gambetta, J. M. and Bishop, Lev S. and Johnson, B. R. and Schuster, D. I. and Majer, J. and Blais, A. and Frunzio, L. and Girvin, S. M. and Schoelkopf, R. J.},
  title   = {Demonstration of two-qubit algorithms with a superconducting quantum processor},
  journal = {Nature},
  volume  = {460},
  number  = {7252},
  pages   = {240--244},
  year    = {2009},
  doi     = {10.1038/nature08121}
}

@article{DiVincenzo2013MultiQubitParity,
  author  = {DiVincenzo, David P. and Solgun, Firat},
  title   = {Multi-qubit parity measurement in circuit quantum electrodynamics},
  journal = {New Journal of Physics},
  volume  = {15},
  number  = {7},
  pages   = {075001},
  year    = {2013},
  doi     = {10.1088/1367-2630/15/7/075001}
}

@article{Filipp2009JointDispersiveReadout,
  author  = {Filipp, S. and Maurer, P. and Leek, P. J. and Baur, M. and Bianchetti, R. and Fink, J. M. and G{\"o}ppl, M. and Steffen, L. and Gambetta, J. M. and Blais, A. and Wallraff, A.},
  title   = {Two-qubit state tomography using a joint dispersive readout},
  journal = {Physical Review Letters},
  volume  = {102},
  pages   = {200402},
  year    = {2009},
  doi     = {10.1103/PhysRevLett.102.200402}
}

@article{Govia2015ParityPhotonCounter,
  author  = {Govia, Luke C. G. and Pritchett, Emily J. and Plourde, B. L. T. and Vavilov, Maxim G. and McDermott, R. and Wilhelm, Frank K.},
  title   = {Scalable two- and four-qubit parity measurement with a threshold photon counter},
  journal = {Physical Review A},
  volume  = {92},
  number  = {2},
  pages   = {022335},
  year    = {2015},
  doi     = {10.1103/PhysRevA.92.022335}
}

@article{Huembeli2017HeraldedParity,
  author  = {Huembeli, Patrick and Nigg, Simon E.},
  title   = {Towards a heralded eigenstate-preserving measurement of multi-qubit parity in circuit {QED}},
  journal = {Physical Review A},
  volume  = {96},
  number  = {1},
  pages   = {012313},
  year    = {2017},
  doi     = {10.1103/PhysRevA.96.012313}
}

@article{Kerckhoff2009ContinuousParity,
  author  = {Kerckhoff, Joseph and Bouten, Luc and Silberfarb, Andrew and Mabuchi, Hideo},
  title   = {Physical model of continuous two-qubit parity measurement in a cavity-{QED} network},
  journal = {Physical Review A},
  volume  = {79},
  number  = {2},
  pages   = {024305},
  year    = {2009},
  doi     = {10.1103/PhysRevA.79.024305}
}

@article{Lalumiere2010TunableJoint,
  author  = {Lalumi{\`e}re, Kevin and Gambetta, J. M. and Blais, Alexandre},
  title   = {Tunable joint measurements in the dispersive regime of cavity {QED}},
  journal = {Physical Review A},
  volume  = {81},
  number  = {4},
  pages   = {040301},
  year    = {2010},
  doi     = {10.1103/PhysRevA.81.040301}
}

@article{Livingston2021ContinuousQEC,
  author  = {Livingston, William P. and Blok, Machiel S. and Flurin, Emmanuel and Dressel, Justin and Jordan, Andrew N. and Siddiqi, Irfan},
  title   = {Experimental demonstration of continuous quantum error correction},
  journal = {arXiv preprint arXiv:2107.11398},
  year    = {2021},
  eprint  = {2107.11398},
  archivePrefix = {arXiv},
  url     = {https://arxiv.org/abs/2107.11398}
}

@article{Mohseninia2020AlwaysOn,
  author  = {Mohseninia, Razieh and Yang, Jing and Siddiqi, Irfan and Jordan, Andrew N. and Dressel, Justin},
  title   = {Always-on quantum error tracking with continuous parity measurements},
  journal = {Quantum},
  volume  = {4},
  pages   = {358},
  year    = {2020},
  doi     = {10.22331/q-2020-11-04-358}
}

@article{Riste2012InitializationByMeasurement,
  author  = {Rist{\`e}, D. and van Leeuwen, J. G. and Ku, H.-S. and Lehnert, K. W. and DiCarlo, L.},
  title   = {Initialization by measurement of a superconducting quantum bit circuit},
  journal = {Physical Review Letters},
  volume  = {109},
  number  = {5},
  pages   = {050507},
  year    = {2012},
  doi     = {10.1103/PhysRevLett.109.050507}
}

@article{Royer2018ParityParametricDriving,
  author  = {Royer, Baptiste and Puri, Shruti and Blais, Alexandre},
  title   = {Qubit parity measurement by parametric driving in circuit {QED}},
  journal = {Science Advances},
  volume  = {4},
  number  = {11},
  pages   = {eaau1695},
  year    = {2018},
  doi     = {10.1126/sciadv.aau1695}
}

@article{Tornberg2014ThreeQubitParity,
  author  = {Tornberg, Lars and Barzanjeh, Sh. and DiVincenzo, David P.},
  title   = {Stochastic-master-equation analysis of optimized three-qubit nondemolition parity measurements},
  journal = {Physical Review A},
  volume  = {89},
  number  = {3},
  pages   = {032314},
  year    = {2014},
  doi     = {10.1103/PhysRevA.89.032314}
}

@misc{rengerSuperconductingQubitResonatorQuantum2025,
  title = {A {{Superconducting Qubit-Resonator Quantum Processor}} with {{Effective All}}-to-{{All Connectivity}}},
  author = {Renger, Michael and Verjauw, Jeroen and Wurz, Nicola and Hosseinkhani, Amin and {Ockeloen-Korppi}, Caspar and Liu, Wei and Rath, Aniket and Thapa, Manish J. and Vigneau, Florian and Wybo, Elisabeth and Bergholm, Ville and Chan, Chun Fai and Csat{\'a}ri, B{\'a}lint and Dahl, Saga and Davletkaliyev, Rakhim and Giri, Rakshyakar and Gusenkova, Daria and Heimonen, Hermanni and Hiltunen, Tuukka and Hsu, Hao and Hyypp{\"a}, Eric and Ikonen, Joni and Jones, Tyler and Khalid, Shabeeb and Kim, Seung-Goo and Koistinen, Miikka and Komlev, Anton and Kotilahti, Janne and Kukushkin, Vladimir and Lamprich, Julia and Landra, Alessandro and Lee, Lan-Hsuan and Li, Tianyi and Liebermann, Per and Majumder, Sourav and M{\"a}ntyl{\"a}, Janne and Marxer, Fabian and de Griend, Arianne Meijer-van and Milchakov, Vladimir and Mro{\.z}ek, Jakub and Nath, Jayshankar and Orell, Tuure and Papi{\v c}, Miha and Partanen, Matti and Plyushch, Alexander and Pogorzalek, Stefan and Ritvas, Jussi and Romero, Pedro Figueroa and Sampo, Ville and Sepp{\"a}l{\"a}, Marko and Selinmaa, Ville and Sundstr{\"o}m, Linus and Takmakov, Ivan and Tarasinski, Brian and Tuorila, Jani and Tyrkk{\"o}, Olli and V{\"a}limaa, Alpo and Wesdorp, Jaap and Yang, Ping and Yu, Liuqi and Heinsoo, Johannes and Veps{\"a}l{\"a}inen, Antti and Kindel, William and Ku, Hsiang-Sheng and Deppe, Frank},
  year = 2025,
  month = dec,
  number = {arXiv:2503.10903},
  eprint = {2503.10903},
  primaryclass = {quant-ph},
  publisher = {arXiv},
  doi = {10.48550/arXiv.2503.10903},
  urldate = {2026-10-13},
  archiveprefix = {arXiv},
}

@article{pedersenFidelityQuantumOperations2007,
  title = {Fidelity of Quantum Operations},
  author = {Pedersen, Line Hjortsh{\o}j and M{\o}ller, Niels Martin and M{\o}lmer, Klaus},
  year = 2007,
  month = jul,
  journal = {Physics Letters A},
  volume = {367},
  number = {1},
  pages = {47--51},
  issn = {0375-9601},
  doi = {10.1016/j.physleta.2007.02.069},
  urldate = {2026-10-20},
}

@article{astafievResonanceFluorescenceSingle2010,
  title = {Resonance {{Fluorescence}} of a {{Single Artificial Atom}}},
  author = {Astafiev, O. and Zagoskin, A. M. and Abdumalikov, A. A. and Pashkin, {\relax Yu}. A. and Yamamoto, T. and Inomata, K. and Nakamura, Y. and Tsai, J. S.},
  year = 2010,
  month = feb,
  journal = {Science},
  volume = {327},
  number = {5967},
  pages = {840--843},
  issn = {0036-8075, 1095-9203},
  doi = {10.1126/science.1181918},
  urldate = {2026-10-23},
  langid = {english},
}
\makeatletter\let\SMorig@hangfromsection\@hangfrom@section\let\SMorig@sectioncntformat\@sectioncntformat\makeatother
\appendix
\section{Crosstalk}
Coupling many DRTs to a shared metamaterial introduces two crosstalk channels, leakage between control lines that drives spectator DRTs and finite mode spacing that lets a sideband excite spectator modes.

If both sideband tones leak onto a spectator DRT with the same amplitude ratio $\xi$ and phase, control crosstalk is described by $H_\text{eff}=\epsilon c^\dagger c+\Delta_{DR}\tilde Z_s/2+g_{CR}(\tilde X_t+\xi\tilde X_s)(c+c^\dagger)$, where $\Delta_{DR}=2(g_{DR,t}-g_{DR,s})$, with $t$ and $s$ denoting the target and spectator qubits. For a square pulse of duration $T=2\pi/\epsilon$, setting $\epsilon = \Delta_{DR}/m$ with $m\in\mathbb Z\setminus\{0,\pm1\}$ cancels the spectator error; we choose the dual-rail gaps so that neighboring and next-nearest DRTs have $|m|=2$ and $4$. A fractional mistargeting $\delta m=(\Delta_{DR}-m\epsilon)/\epsilon$ leaves a leading infidelity $1-F\propto(\delta m)^2\xi^2$.

Spectral crosstalk arises when a sideband tuned to mode $k_0$, detuned by $\pm\epsilon$ from its resonance, also displaces another mode $k$. Its smallest detuning from any other mode, $\Delta_\text{spec}=\min_{k\neq k_0,\,\sigma=\pm1}|\nu_k-\nu_{k_0}+\sigma\epsilon|$, must exceed both $g_{CR}$ and the pulse bandwidth. The condition is most restrictive near a band edge, where the nearest-mode spacing is $\nu_2-\nu_1\simeq3\pi^2|J|/(N+1)^2$ for $k_0=1$; near mid-band it is $\simeq2\pi|J|/(N+1)$, but still falls as $1/(N+1)$. Longer chains therefore require longer, more selective pulses, and this requirement, together with the photon loss below, caps the number of modes~\cite{SM}.
\section{Dephasing Errors}
At its idling point, the DRT is protected against static dephasing~\cite{kubica2023erasure, levine2023demonstrating}. Frequency noise on the two rails enters as $V(t)=\delta_a(t)\,a^\dagger a+\delta_b(t)\,b^\dagger b=\frac{1}{2}\left[\delta_+(t)L_c+\delta_-(t)L_t\right]$, with $\delta_\pm=\delta_a\pm\delta_b$, a common channel $L_c=\tilde{a}_+^\dagger\tilde{a}_++\tilde{a}_-^\dagger\tilde{a}_-$, and a transverse channel $L_t=\tilde{a}_+^\dagger\tilde{a}_-+\tilde{a}_-^\dagger\tilde{a}_+$. Both logical states hold exactly one excitation, so the common channel writes only a global phase, and the dual-rail gap $2g_{DR}$ suppresses the off-diagonal transitions driven by $L_t$. The cross-resonance drive dresses the logical states with states outside the logical subspace. The transverse protection survives, because the anti-symmetric drive [Eq.~\eqref{eq:hdrive}] and the anharmonicity change the $\tilde{a}_+$ occupation only by even amounts while $L_t$ changes it by one. The common channel is no longer protected by symmetry alone, since the dressed logical states can hold different excitation numbers, $\ell_0\neq\ell_1$ with $\ell_i=\langle\tilde{i}_L|L_c|\tilde{i}_L\rangle$; for low-frequency noise the logical qubit then dephases at the bare-transmon rate scaled by $(\ell_0-\ell_1)^2/2$.

The difference is small. A drive tone admixes each logical state with the vacuum and with two-excitation states. For harmonic transmons the admixtures add the same excitation number to every state, and for anharmonic ones the two-excitation state reached from $|0_L\rangle$ is hybridized with a second doubly excited state $4g_{DR}$ away, the same hybridization that generates the cross-resonance coupling~\cite{SM}. The resulting asymmetry first appears at fifth order in $1/\Delta$ (Supplemental Material~\cite{SM}),
\begin{equation}
\begin{gathered}
\ell_0-\ell_1 \approx \frac{4g_{DR}\,\eta^2\,\Omega^2}{\Delta^5}\left(1+r^2\right),\\
\Gamma_\phi^\text{dressed} \approx 8\,\Gamma^\text{transmon}_\phi\, \frac{g_{DR}^2\,\eta^4\,\Omega^4}{\Delta^{10}}\left(1+r^2\right)^2,
\end{gathered}
\end{equation}
with $r=\Omega_-/\Omega_+$ the tone ratio and $\Gamma_\phi^\text{transmon}$ the dephasing rate of a single bare transmon; we confirmed the expansion against exact diagonalization of the driven dual-rail Hamiltonian.

The simulations of Fig.~\ref{fig:gate_fidelity} include quasi-static frequency noise on each rail with $T_\phi=10\,\mu$s. It adds $7\times10^{-9}$ to $4\times10^{-7}$ to the gate infidelity across the scan, and $\sim\!1\times10^{-7}$ at the operating point, where the drive-induced term above contributes $\sim\!3\times10^{-10}$. Transmon frequency noise lies three to four orders of magnitude below the coherent error, and we omit it from the plotted budget.

\section{Resonator Photon Loss\label{sec:photon_loss}}
Photon loss from the mode during the gate dephases the DRTs. A lost photon carries which-sector information about $\tilde{X}_1\tilde{X}_2$ into the environment, as in the M{\o}lmer-S{\o}rensen gate~\cite{sorensen_entanglement_2000, bermudezDissipationAssistedQuantumInformation2013a}. The coherence between two sectors with mode trajectories $\alpha(t)$ and $\alpha'(t)$ decays with the exponent $\frac{\kappa}{2}\int_0^T|\alpha-\alpha'|^2dt$, where $\kappa$ is the mode decay rate and $\alpha(t)$ follows the circular trajectory of Eq.~\eqref{eq:2qprop}. Averaged over the computational space, the gate infidelity is $\varepsilon_{\text{loss}} = \frac{4}{5}\,\kappa\langle\bar{n}\rangle T$, where $\langle\bar{n}\rangle$ is the mode population averaged over the gate and over the computational basis, and $4/5 = d/(d+1)$ is the average-fidelity weight of a dephasing error in $d=4$ dimensions.

This loss grows linearly with the gate duration, and driving harder does not reduce it. The entangling phase and the integrated population $\int_0^T|\alpha|^2dt$ are the same phase-space integral, because each sector's geometric phase is twice the area its trajectory encloses, so a fixed phase costs a fixed integrated population. The conditions below Eq.~\eqref{eq:2qprop} set $g_{CR}/\epsilon = 1/4$, so a stronger drive traverses the same circle faster. The circle also sets the population. The $x_1 x_2 = +1$ sectors are circles of radius $2g_{CR}/\epsilon = 1/2$ with mean population $1/2$, the $x_1 x_2 = -1$ sectors are undisplaced, and the computational-basis average is $\langle\bar{n}\rangle = 1/4$.

In the simulations of Fig.~\ref{fig:gate_fidelity}, $\langle\bar{n}\rangle = 0.273 \pm 0.003$ across the full scan, with $94\%$ of the integrated population in the driven mode and the rest in its neighbors. At the optimum ($T = 152$~ns, mode $T_1 = 30\,\mu$s) the formula gives $\varepsilon_\text{loss} = 1.11\times10^{-3}$, within $2\%$ of the simulated photon-loss infidelity of $1.13\times10^{-3}$. This linear loss competes with the coherent error, which is largest at short gate durations, to produce the interior optimum of Fig.~\ref{fig:gate_fidelity}, where photon loss accounts for $81\%$ of the total infidelity. Longer mode lifetimes weaken the linear term and move the optimum to longer gates.

\section{Gate Errors due to Dual-Rail Erasure \label{sec:erasure_errors}}
Excitation loss on a DRT, detectable as an erasure~\cite{kubica2023erasure}, has two effects. On a DRT participating in a gate it dephases the partner DRT, through the same which-trajectory record as a photon lost from the mode~\cite{SM}. On a spectator DRT, one of the $N_q-2$ coupled to the chain but idle during the gate, it drops the pair from its one-excitation logical subspace to the ground state. The two logical states pull the driven mode equally to order $1/\Delta^2$, leaving the encoded coupling $\chi_L\sim1/\Delta^4$~\cite{SM}, but the ground state pulls it differently, by the bare-transmon dispersive shift $\chi\sim1/\Delta^2$. The mode frequency therefore shifts by $\chi$ at a random time $t_0$ during the gate of duration $T$, drawn from $p(t_0) = \Gamma_e e^{-\Gamma_e t_0}$ with $\Gamma_e = 1/T_1$ the transmon relaxation rate. For a spectator near the chain edge, $|\chi| \approx 2g^2\eta/[(N+1)\Delta(\Delta+\eta)]$, with $g$, $\Delta$, and $\eta$ as in Eq.~\eqref{eq:drt_ham}, where the factor $1/(N+1)$ is the spectator's participation in the driven mode, at most twice as large at the chain center~\cite{SM}.

The shifted detuning distorts the phase-space loop of the gate. We denote the unshifted detuning $\epsilon_0$, the $\epsilon$ of the main text. The evolution remains an exact displacement times an $\tilde X_1\tilde X_2$-dependent phase~\cite{SM}, and an erasure leaves a residual displacement $\propto(\chi/\epsilon_0)^2$ and a phase deficit $\propto\chi/\epsilon_0$, both vanishing for an erasure at the gate end and growing with the loop phase $\theta=\epsilon_0(T-t_0)$ still ahead at the jump. Averaging the resulting infidelity over $p(t_0)$ with $\Gamma_e T\ll1$ gives, per spectator DRT,
\begin{equation}
\langle 1-F \rangle \approx \left(\frac{7\pi^2}{30} + \frac{5}{8}\right)\Gamma_e T \left(\frac{\chi}{\epsilon_0}\right)^{2} \approx 0.074\,\Gamma_e \chi^2 T^3.
\label{eq:erasure_scaling}
\end{equation}
This is the probability $\Gamma_e T$ that an erasure interrupts the gate, weighted by the squared fractional shift it imparts. At the operating point of Fig.~\ref{fig:gate_fidelity}, exact diagonalization of the spectator--chain Hamiltonian gives $\chi/2\pi = 0.56$~MHz, so $\chi/\epsilon_0 = 0.085$, and numerically averaging the exact expressions gives $\langle 1-F \rangle = 2.7\times10^{-5}$ per spectator DRT at transmon $T_1 = 100\,\mu$s, which Eq.~\eqref{eq:erasure_scaling} reproduces to within $20\%$.

The spectator error grows with the array. With one DRT per two modes, $N = 2N_q-1$, the shift dilutes as $\chi \propto 1/(N+1)$. Resolving one mode from its neighbors requires $T \propto (N+1)/J$, and loop closure sets $\epsilon_0 \propto 1/T$, so $\chi/\epsilon_0$ is independent of $N$ and the per-spectator error of Eq.~\eqref{eq:erasure_scaling} grows as $\Gamma_e T \propto N+1$. To leading order in $(N_q-2)\Gamma_e T$ the spectators add, so the summed error grows as $(N+1)^2$. At transmon $T_1 = 100\,\mu$s it reaches $1.4\times10^{-4}$ at $N = 25$, a seventh of the error budget, and lowers the mode-count ceiling from $29$ to $25$ modes~\cite{SM}. The error is correlated, since a single erasure dephases every gate running concurrently, but the erasure detection flags all affected gates for targeted correction.

\clearpage
\onecolumngrid
\begin{center}
\textbf{\large Supplemental Material for ``All-to-All Connectivity in Superconducting Qubits via Metamaterial-Mediated Interactions''}
\end{center}
\bigskip

\setcounter{equation}{0}
\setcounter{figure}{0}
\setcounter{table}{0}
\setcounter{section}{0}
\setcounter{subsection}{0}
\makeatletter\@removefromreset{equation}{section}\let\@hangfrom@section\SMorig@hangfromsection\let\@sectioncntformat\SMorig@sectioncntformat\makeatother
\setcounter{secnumdepth}{2}
\renewcommand{\theequation}{S\arabic{equation}}
\renewcommand{\thefigure}{S\arabic{figure}}
\renewcommand{\thetable}{S\arabic{table}}
\renewcommand{\thesection}{S\arabic{section}}
\renewcommand{\thesubsection}{\thesection.\arabic{subsection}}
\makeatletter\renewcommand{\p@subsection}{}\makeatother
\renewcommand{\theHequation}{S\arabic{equation}}
\renewcommand{\theHfigure}{S\arabic{figure}}
\renewcommand{\theHtable}{S\arabic{table}}
\renewcommand{\theHsection}{S\arabic{section}}
\renewcommand{\theHsubsection}{S\arabic{section}.\arabic{subsection}}

\noindent This Supplemental Material is organized as follows. Section~\ref{sec:dreh} derives the effective cross-resonance Hamiltonian, Eq.~\eqref{eq:effham} of the main text, for a single DRT coupled to a resonator mode by second-order time-dependent Schrieffer-Wolff perturbation theory, together with the drive-induced frequency shifts and the residual dispersive coupling, evaluated to fourth order in the inverse qubit--mode detuning. Section~\ref{sec:slth} develops the corresponding effective Hamiltonian for the spin-locked transmon. Section~\ref{sec:sm_parity} supports the parity-measurement claims of the main text, the single-pair measurement and the decoding and measurement-time requirements of the parallel case. Section~\ref{sec:sm_dephasing} derives the drive-induced dephasing of the dual-rail transmon quoted in the End Matter. Section~\ref{sec:sm_erasure} computes the average gate infidelity from spectator dual-rail erasure. Section~\ref{sec:sm_ceiling} derives the mode-count ceiling set by photon loss and spectral crowding, and the correction from spectator erasure. Section~\ref{sec:sm_simulation} lists the parameters and protocol of the numerical simulations reported in Fig.~\ref{fig:gate_fidelity} of the main text and collects their error budget at the operating point.

\medskip

\section{Derivation of the Effective Hamiltonian}\label{sec:dreh}

We derive the effective cross-resonance Hamiltonian, Eq.~\eqref{eq:effham} of the main text, for a single dual-rail transmon (DRT) coupled to a single resonator mode, using second-order time-dependent Schrieffer-Wolff (TDSW) perturbation theory~\cite{bravyi2011schrieffer, malekakhlaghFirstprinciplesAnalysisCrossresonance2020a, petrescuAccurateMethodsAnalysis2023}. Beyond the effective coupling, the same calculation yields the drive-induced frequency shifts needed to calibrate the drive tones and the residual dispersive coupling.

\subsection{Normal-mode Hamiltonian}
Starting from Eq.~\eqref{eq:drt_ham} of the main text, we use the symmetric and anti-symmetric normal modes $\tilde{a}_\pm = (a \pm b)/\sqrt{2}$. The DRT logical states are the single-excitation states $\ket{0_L} = \tilde{a}_-^\dagger\ket{gg}$ and $\ket{1_L} = \tilde{a}_+^\dagger\ket{gg}$, split by the dual-rail gap $2g_{DR}$. The resonator mode $c$ couples only to $\tilde{a}_+$, while the anti-symmetric drive of Eq.~\eqref{eq:hdrive} of the main text acts only on $\tilde{a}_-$. The two drive tones are applied at the lab-frame frequencies $\omega_\pm^{\mathrm{lab}} = \nu - \epsilon \pm 2g_{DR}$, symmetric about their center frequency $\nu - \epsilon$, where $\epsilon$ is the detuning of the resonator from the drive center ($\epsilon = 0$ for the single-DRT case of Eq.~\eqref{eq:effham} of the main text). In the frame rotating at the drive-center frequency $\nu - \epsilon$, for the transmons and the resonator alike, and after the rotating-wave approximation with respect to the drive frequencies, the Hamiltonian splits into $H_0 + V(t)$ with
\begin{equation}\label{eq:sm_h0_sa}
H_0 = -(\Delta - \epsilon - g_{DR})\, \tilde{a}_+^\dagger \tilde{a}_+ - (\Delta - \epsilon + g_{DR})\, \tilde{a}_-^\dagger \tilde{a}_-
  - \frac{\eta}{4}\bigl(\tilde{a}_+^{\dagger 2}\tilde{a}_+^2 + \tilde{a}_-^{\dagger 2}\tilde{a}_-^2 + 4\tilde{a}_+^\dagger \tilde{a}_+\, \tilde{a}_-^\dagger \tilde{a}_-
  + \tilde{a}_-^{\dagger 2}\tilde{a}_+^2 + \tilde{a}_+^{\dagger 2}\tilde{a}_-^2\bigr)
  + \epsilon\, c^\dagger c,
\end{equation}
\begin{equation}\label{eq:sm_v_sa}
V(t) = \sqrt{2}\,g\,(\tilde{a}_+^\dagger c + c^\dagger \tilde{a}_+)
  + \sum_{\sigma=\pm}\frac{\Omega_\sigma}{\sqrt{2}}\bigl(\tilde{a}_-^\dagger e^{-i\omega_\sigma t} + \tilde{a}_- e^{i\omega_\sigma t}\bigr),
\end{equation}
where $\omega_\pm = \pm 2g_{DR}$ are the drive frequencies measured in this frame; second-order corrections to these frequencies are derived below. The transmon nonlinearity written in the normal-mode basis contains the pair-exchange term $-(\eta/4)(\tilde{a}_+^{\dagger 2}\tilde{a}_-^2 + \tilde{a}_-^{\dagger 2}\tilde{a}_+^2)$, which is the essential ingredient of the scheme.

\subsection{Two-excitation manifold}

The pair-exchange term hybridizes the two-excitation states $\ket{2_+}$ and $\ket{2_-}$, where $\ket{2_\pm}$ denotes the doubly excited $\tilde a_\pm$ state. Diagonalizing this block with mixing angle $\theta = \frac{1}{2}\arctan\!\left(\frac{-\eta}{4g_{DR}}\right)$ gives the hybrids
\begin{equation}
\ket{h_+} = \cos\theta\,\ket{2_+} + \sin\theta\,\ket{2_-}, \qquad
\ket{h_-} = -\sin\theta\,\ket{2_+} + \cos\theta\,\ket{2_-},
\end{equation}
with energies $E_\pm = -2\Delta - \eta/2 \pm \Lambda$, where $\Lambda \equiv \sqrt{4g_{DR}^2 + \eta^2/4}$, while the state $\ket{1_+ 1_-}$, with one excitation in each mode, remains uncoupled at $E = -2\Delta - \eta$. This hybridization opens the second-order flip-flip pathway $\ket{0_L,0_c} \xrightarrow{\tilde{a}_-^\dagger} \ket{h_\pm,0_c} \xrightarrow{c^\dagger\tilde{a}_+} \ket{1_L,1_c}$, in which the drive populates the $\ket{2_-}$ content of a hybrid and the resonator coupling removes an excitation from its $\ket{2_+}$ content. Both hybrids contribute, with amplitudes $\propto \mp\sin\theta\cos\theta$ weighted by their energy denominators. At $\eta = 0$ the two contributions cancel exactly, and the $\mathcal{O}(\eta)$ remainder of this near-cancellation is the effective coupling.

\subsection{Time-dependent Schrieffer-Wolff transformation}

We treat $V(t)$ as a perturbation of $H_0$ and work in the eigenbasis of $H_0$. The computational subspace $P = \{\ket{0_L}, \ket{1_L}\}\otimes\{\ket{0_c}, \ket{1_c}\}$ is spanned by the four $H_0$ eigenstates in which the DRT is in a logical state and the resonator holds at most one photon; these are the states between which the cross-resonance interaction acts. The same calculation in the block $\{\ket{n_c}, \ket{(n+1)_c}\}$ with $n = 0, 1, 2, \ldots$ gives the same energy denominators, since the resonator energy $n_c\epsilon$ enters every state equally and only the photon-number changes of $\pm1$ survive, and matrix elements scaled by the $\sqrt{n+1}$ of $c^\dagger$, so the result holds as the operator $\tilde{X}(c^\dagger + c)$. Since the nonlinear terms of Eq.~\eqref{eq:sm_h0_sa} vanish on the logical DRT states, the energies are $E_{0_L,n_c} = -(\Delta - \epsilon + g_{DR}) + n_c\epsilon$ and $E_{1_L,n_c} = -(\Delta - \epsilon - g_{DR}) + n_c\epsilon$. We decompose the perturbation into Fourier components, $V(t) = \sum_\mu V^{\mu}e^{i\mu t}$ with $\mu \in \{0, \pm\omega_+, \pm\omega_-\}$, reserving $\nu$ for the resonator frequency. The static component $V^0 = \sqrt{2}\,g\,(\tilde{a}_+^\dagger c + c^\dagger \tilde{a}_+)$ is the resonator coupling, and $V^{-\omega_\pm} = (\Omega_\pm/\sqrt{2})\,\tilde{a}_-^\dagger$, $V^{+\omega_\pm} = (\Omega_\pm/\sqrt{2})\,\tilde{a}_-$ are the drive tones. Writing $V^{\mu}_{ab} = \bra{a}V^{\mu}\ket{b}$ for matrix elements between $H_0$ eigenstates with energies $E_a$, $E_b$, the second-order TDSW effective Hamiltonian in $P$ is
\begin{equation}\label{eq:tdsw_formula}
(H_{2})_{mn}(t) = \frac{1}{2}\sum_{k \notin P}\sum_{\mu,\mu'} V^{\mu}_{mk}\, V^{\mu'}_{kn}
\left(\frac{1}{\Delta^{(\mu)}_{mk}} - \frac{1}{\Delta^{(\mu')}_{kn}}\right)
e^{i(\Delta^{(\mu)}_{mk} + \Delta^{(\mu')}_{kn})t},
\qquad \Delta^{(\mu)}_{ab} = E_a - E_b + \mu,
\end{equation}
where $m, n \in P$ and $\Delta^{(\mu)}_{ab}$ is the energy mismatch of the transition between $\ket{b}$ and $\ket{a}$ when the perturbation contributes frequency $\mu$. The sum runs over the virtual states $k \notin P$, the $H_0$ eigenstates reached from $P$ by a single application of $V$. The resonator coupling $V^0$ moves one excitation between $\tilde{a}_+$ and $c$, and the drive components add or remove one $\tilde{a}_-$ excitation. Starting from $P$, these reach the transmon vacuum $\ket{gg}$ and the two-excitation sector, each tensored with the resonator Fock states $\ket{0_c}$, $\ket{1_c}$, and $\ket{2_c}$. In the two-excitation sector we use the $H_0$ eigenstates $\{\ket{h_\pm}, \ket{1_+ 1_-}\}$ rather than the bare states $\ket{2_\pm}$, so that the energy denominators include the pair-exchange hybridization exactly in $\eta$.

The phase in Eq.~\eqref{eq:tdsw_formula} is $(\Delta^{(\mu)}_{mk} + \Delta^{(\mu')}_{kn})t = (E_m - E_n + \mu + \mu')t$. The virtual energy $E_k$ cancels, so whether a term is static depends only on the two computational states it connects and the total frequency $\mu + \mu'$ supplied by the two factors of $V$. Time-independent (DC) contributions arise when $\mu + \mu' = E_n - E_m$. For diagonal elements ($m = n$) this requires $\mu' = -\mu$, i.e., the same component acting twice, as in the undriven $V^0 V^0$ path or the absorption and re-emission of a drive quantum. These terms give the energy shifts of Sec.~\ref{ssec:sm_dressed}. For off-diagonal elements, $\mu + \mu'$ must match the splitting of the two computational states. The relevant splittings are $E_{1_L,1} - E_{0_L,0} = 2g_{DR} + \epsilon = \omega_+ + \epsilon$ and $E_{0_L,1} - E_{1_L,0} = \epsilon - 2g_{DR} = \omega_- + \epsilon$. One drive component combined with the static $V^0$ ($\mu' = 0$) therefore leaves each cross-resonance element oscillating slowly at $\pm\epsilon$, the deliberate detuning of the tones from the two transitions. These slowly rotating off-diagonal elements are the effective cross-resonance couplings of Sec.~\ref{ssec:sm_gcr}, the amplitudes for the transitions $\ket{0_L,0_c}\leftrightarrow\ket{1_L,1_c}$ and $\ket{1_L,0_c}\leftrightarrow\ket{0_L,1_c}$ mediated by the virtual states. They must be retained. Absorbing the $e^{\pm i\epsilon t}$ phases back into the resonator operators returns the block Hamiltonian $\epsilon\,c^\dagger c - g_{CR}\tilde{X}(c^\dagger + c)$, the M{\o}lmer--S{\o}rensen Hamiltonian displayed above Eq.~\eqref{eq:2qprop} of the main text.

All other terms of Eq.~\eqref{eq:tdsw_formula} oscillate at frequencies large compared to $\epsilon$. We discard them, which amounts to a rotating-wave approximation on the effective Hamiltonian, justified when these frequencies greatly exceed the effective couplings, since the terms then average out over the gate time. Two structural properties follow at second order. First, the static and slowly rotating effective Hamiltonian is additive in the two drive tones, with no cross terms between $\Omega_+$ and $\Omega_-$. Second, the only fast terms are the $\Omega_+\Omega_-$ cross terms, which oscillate at $\omega_+ - \omega_- = 4g_{DR}$ and at $4g_{DR} \pm 2\epsilon$. Their amplitude relative to $g_{CR}$ grows linearly with the drive, and their leading Magnus correction, of order their amplitude squared over $4g_{DR}$, as its cube. At $\Omega_+/2\pi = 100$~MHz (balanced $\Omega_-$, $\epsilon = 4g_{CR}$) the amplitudes are $2.2$--$2.8\,g_{CR}$ and the Magnus correction is $\sim 2\times10^{-2}\,g_{CR}$. An exact Floquet simulation of a single DRT coupled to a single mode, with the DRT-1 parameters of Table~\ref{tab:sim_params} and $\Delta/2\pi = 1.013$~GHz, reproduces the effective couplings of Sec.~\ref{ssec:sm_gcr} to within $2\%$ for $\Omega_\pm/2\pi \le 100$~MHz, which sets the validity range of the second-order effective Hamiltonian. Stronger drives, including the minimum-infidelity point of Fig.~\ref{fig:gate_fidelity} of the main text, lie beyond this range; the simulations there integrate the full Hamiltonian and include all higher-order corrections.

\subsection{Effective cross-resonance coupling}\label{ssec:sm_gcr}

The slowly rotating off-diagonal elements of Eq.~\eqref{eq:tdsw_formula} give the two cross-resonance amplitudes. The red tone $\Omega_-$ drives the flip-flop transition $\ket{1_L,0_c}\leftrightarrow\ket{0_L,1_c}$ and the blue tone $\Omega_+$ the flip-flip transition $\ket{0_L,0_c}\leftrightarrow\ket{1_L,1_c}$,
\begin{align}
g_- &= -\frac{\eta g \Omega_-}{2}\left[\frac{1}{D_-(0)} + \frac{1}{D_-(-\epsilon)}\right], \label{eq:sm_gred}\\
g_+ &= -\frac{\eta g \Omega_+}{2}\left[\frac{1}{D_+(0)} + \frac{1}{D_+(-\epsilon)}\right], \label{eq:sm_gblue}
\end{align}
with the denominator functions
\begin{equation}\label{eq:sm_denoms}
D_-(\epsilon') = (\epsilon' + \Delta - g_{DR})(\epsilon' + \Delta - g_{DR} + \eta), \qquad
D_+(\epsilon') = (\epsilon' + \Delta + g_{DR})(\epsilon' + \Delta + g_{DR} + \eta) - 4g_{DR}^2.
\end{equation}
The two halves of each amplitude are the two energy denominators of Eq.~\eqref{eq:tdsw_formula}. In one the virtual step carries the drive frequency, whose detuning from the transition shifts the denominator by $-\epsilon$, while in the other it carries the static resonator coupling and is unshifted. A brute-force evaluation of the second-order sum reproduces Eqs.~\eqref{eq:sm_gred} and~\eqref{eq:sm_gblue} to machine precision for $\epsilon$ up to $g_{DR}/2$.
Both amplitudes are proportional to $\eta$ and vanish identically at $\eta = 0$, since without the pair-exchange hybridization there is no cross-resonance coupling. The asymmetry between the two denominators reflects the virtual states each pathway visits. The flip-flop transition proceeds through the unhybridized states $\ket{gg,1_c}$ and $\ket{1_+1_-,0_c}$, while the flip-flip transition proceeds only through the hybridized pair $\ket{h_\pm,0_c}$, whose $\Lambda$-shifted energies combine into the $-4g_{DR}^2$ term. At leading order in $1/\Delta$, both reduce to $-\eta g \Omega_\pm/\Delta^2$. This is the same scaling as the cross-resonance coupling of a bare transmon, $\sim g\Omega\eta/\Delta^2$~\cite{magesan2020effective}, so the dual-rail encoding suppresses the dispersive coupling without weakening the desired cross-resonance interaction (see Table~\ref{tab:sm_summary}).

Setting $|g_-| = |g_+|$ combines the two transitions into a pure $\tilde{X}(c^\dagger + c)$ coupling, which requires the amplitude ratio
\begin{equation}\label{eq:sm_ratio}
\frac{\Omega_-}{\Omega_+} = r(\epsilon) = \frac{D_+(0)^{-1} + D_+(-\epsilon)^{-1}}{D_-(0)^{-1} + D_-(-\epsilon)^{-1}}.
\end{equation}
When balanced, the interaction is $-g_{CR}\tilde{X}(c^\dagger + c)$ with $g_{CR} = -g_+$; setting $\epsilon = 0$ reproduces Eq.~\eqref{eq:effham} of the main text and the amplitude ratio quoted there. The relative sign of the two tones is fixed by the operator structure of the two amplitudes. The flip-flip and flip-flop terms combine into $\tilde{X}(c^\dagger + c)$ only if the coefficients of $\tilde{\sigma}^+ c^\dagger$ and $\tilde{\sigma}^- c^\dagger$ are equal, while equal moduli with the opposite relative sign produce the orthogonal combination $\tilde{Y}\,i(c^\dagger - c)$. For the dual-rail transmon, both amplitudes [Eqs.~\eqref{eq:sm_gred} and~\eqref{eq:sm_gblue}] carry the same, negative, sign, so the balance condition is satisfied with a positive amplitude ratio $r > 0$. For the spin-locked qubit, the dressed-$Z$ projection [Eq.~\eqref{eq:sl_floquet_H}] gives the two amplitudes opposite signs, so the tones must instead be driven $\pi$ out of phase, $\Omega_1 = -\Omega_2$ (Sec.~\ref{sec:sl_floquet}).

\subsection{Dressed frequencies}\label{ssec:sm_dressed}

The DC diagonal elements of Eq.~\eqref{eq:tdsw_formula} shift the dual-rail gap and the resonator frequency. Writing the dressed energies of the computational states as $E_{1_L,n_c}$ and $E_{0_L,n_c}$, the shift of the dual-rail gap is $\delta\omega_q = \delta E_{1_L,0} - \delta E_{0_L,0}$:
\begin{equation}\label{eq:sm_dwq}
\begin{gathered}
\delta\omega_q = -\frac{2g^2}{\Delta - g_{DR}}
+ \frac{\Omega_-^2}{2(\Delta - \epsilon + \eta - g_{DR})} + \frac{\Omega_+^2}{2(\Delta - \epsilon + \eta + 3g_{DR})} \\
+ \frac{4\left(2\Delta^2 + \Delta\eta - 2g_{DR}^2\right)}{(4\Delta + \eta - 2\Lambda)(4\Delta + \eta + 2\Lambda)}
\left[\frac{\Omega_-^2}{\Delta - \epsilon - g_{DR}} + \frac{\Omega_+^2}{\Delta - \epsilon + 3g_{DR}}\right] \\
- \frac{Q_-}{4\Lambda(4\Delta + \eta - 2\Lambda)}
\left[\frac{\Omega_-^2}{2\Delta + \eta - 2\Lambda - 2\epsilon - 6g_{DR}} + \frac{\Omega_+^2}{2\Delta + \eta - 2\Lambda - 2\epsilon + 2g_{DR}}\right] \\
- \frac{Q_+}{4\Lambda(4\Delta + \eta + 2\Lambda)}
\left[\frac{\Omega_-^2}{2\Delta + \eta + 2\Lambda - 2\epsilon - 6g_{DR}} + \frac{\Omega_+^2}{2\Delta + \eta + 2\Lambda - 2\epsilon + 2g_{DR}}\right],
\end{gathered}
\end{equation}
with $Q_\pm = 4\Lambda\left(4\Delta + \eta + 4g_{DR} \pm 2\Lambda\right) \pm 8g_{DR}(4\Delta + \eta)$. The first term is the difference of the Lamb shifts of the two logical states; like $\chi_L^{(2)}$, it is independent of $\epsilon$. The remaining terms are drive-induced Stark shifts, in which the detuning enters through the drive frequencies $\pm 2g_{DR}$ measured from transmon transitions shifted by $\epsilon$ in this frame, hence the $-\epsilon$ in their denominators. The resonator frequency shift is $\delta\omega_c = \frac{1}{2}(\delta E_{1_L,1} + \delta E_{0_L,1} - \delta E_{1_L,0} - \delta E_{0_L,0})$:
\begin{equation}\label{eq:sm_dwc}
\begin{gathered}
\delta\omega_c = \frac{g^2 \left(2\Lambda - \eta + 4 g_{DR}\right) \left(2\Lambda + \eta - 4 g_{DR}\right)}{\left(\Delta - g_{DR}\right) \left(2\Lambda - \eta - 4 g_{DR}\right) \left(2\Lambda + \eta + 4 g_{DR}\right)}
+ \frac{\eta g^2 \left(2\Lambda + \eta - 4 g_{DR}\right)}{\Lambda \left(2\Lambda + \eta + 4 g_{DR}\right) \left(2\Delta + \eta + 2 g_{DR} + 2\Lambda\right)} \\
- \frac{\eta g^2 \left(2\Lambda - \eta + 4 g_{DR}\right)}{\Lambda \left(2\Lambda - \eta - 4 g_{DR}\right) \left(2\Delta + \eta + 2 g_{DR} - 2\Lambda\right)}
+ \frac{g^2}{\Delta + \eta - g_{DR}},
\end{gathered}
\end{equation}
whose leading behavior is the Lamb shift $\delta\omega_c \approx 2g^2/\Delta$; like $\chi_L^{(2)}$, it is drive-free and hence independent of $\epsilon$. The corrected drive tones must be centered a detuning $\epsilon$ below the \emph{dressed} resonator frequency and split by the \emph{dressed} dual-rail gap. In the lab frame,
\begin{equation}\label{eq:sm_drive_freq}
\omega_\pm^{\mathrm{lab}} = \left(\nu + \delta\omega_c - \epsilon\right) \pm \left(2g_{DR} + \delta\omega_q\right).
\end{equation}
Since $\delta\omega_q$ depends on the drive amplitudes and on $\epsilon$ through the energy denominators, Eq.~\eqref{eq:sm_drive_freq} is an implicit equation, which we solve by iteration, starting from $\omega_\pm^{\mathrm{lab},(0)} = \nu - \epsilon \pm 2g_{DR}$, computing the shifts, updating the frequencies, and repeating until convergence. Because $\delta\omega_q$ is quadratic in the drive amplitudes, a shaped pulse envelope moves the corrected frequencies during the ramps. A constant-frequency tone satisfies Eq.~\eqref{eq:sm_drive_freq} only at one amplitude, so the envelope must be slow on the scale of the detunings incurred, or the tones chirped. The amplitude \emph{ratio} of Eq.~\eqref{eq:sm_ratio}, by contrast, is independent of the amplitudes and holds pointwise through any common envelope. Even with these corrections the two tones are not simultaneously resonant with their transitions. The residual dispersive shift $\chi_L$ (Sec.~\ref{ssec:sm_chi}) splits the flip-flip and flip-flop frequencies by $\chi_L$, and the symmetric choice of Eq.~\eqref{eq:sm_drive_freq} leaves each tone off-resonant by $\mp\chi_L/2$.

\subsection{Residual dispersive coupling}\label{ssec:sm_chi}

The differential shift $\chi = (\delta E_{1_L,1} - \delta E_{0_L,1}) - (\delta E_{1_L,0} - \delta E_{0_L,0})$ is the static dispersive coupling between the DRT and the resonator, $H_\chi = \frac{\chi}{2}\tilde{Z}\,c^\dagger c$. At second order in the coupling $g$, the contributing terms of Eq.~\eqref{eq:tdsw_formula} combine into the single fraction
\begin{equation}\label{eq:sm_cross_kerr}
\chi_L^{(2)} = -\frac{4\eta^2 g_{DR}\, g^2}{D_-(0)\,D_+(0)},
\end{equation}
whose denominator is the product of the two unshifted coupling denominators of Eq.~\eqref{eq:sm_denoms}. Because it is drive-free, $\chi_L^{(2)}$ does not depend on the drive detuning $\epsilon$. In the frame of Eq.~\eqref{eq:sm_h0_sa} the shift of the transmon energies by $\epsilon$ cancels against the $\epsilon\,c^\dagger c$ term in every $g^2$ energy denominator. Expanding for $\Delta \gg \eta, g_{DR}$ gives $\chi_L^{(2)} \approx -4g^2\eta^2 g_{DR}/\Delta^4$. Compared to the bare-transmon dispersive shift $\sim g^2\eta/\Delta^2$, the dual-rail encoding provides a suppression by two additional powers of $1/\Delta$. The $\mathcal{O}(\eta)$ contributions to the energies of the two logical states cancel in the differential shift, leaving a term proportional to both $\eta^2$ and $g_{DR}$. The same parametric improvement occurs for spin-locked transmons (SLTs) at this order; the SLT's fourth-order term is larger than the dual rail's by $\Delta/2\Omega_{sl}$ (Sec.~\ref{ssec:sl_g4}), where $\Omega_{sl}$ is the spin-locking drive amplitude, the analog of $g_{DR}$ for the DRT.

The dispersive shift can also be evaluated beyond second order. In the absence of the drive, the coupling to the mode conserves the total excitation number, so the relevant energies follow from exact finite blocks. $E_{0_L,0}$ is unshifted, $E_{1_L,0}$ follows from the two-dimensional block $\{\ket{1_L,0_c}, \ket{\mathrm{vac},1_c}\}$, and $E_{1_L,1}$, $E_{0_L,1}$ from the six-dimensional two-excitation sector $\{\ket{0_L,1_c}, \ket{1_L,1_c}, \ket{\mathrm{vac},2_c}, \ket{2_+,0_c}, \ket{2_-,0_c}, \ket{1_+ 1_-,0_c}\}$. Expanding the block eigenvalues to fourth order in the coupling yields, at $\epsilon = 0$ and to leading order in $1/\Delta$,
\begin{equation}\label{eq:sm_chi4}
\chi_L^{(4)} = \frac{4\eta g^4}{\Delta^4},
\end{equation}
a drive-independent contribution arising from the photon-assisted virtual transitions $\ket{1_L,1_c}\leftrightarrow\ket{\mathrm{vac},2_c}$ and $\ket{1_L,1_c}\leftrightarrow\ket{2_+,0_c}$. Combining the two contributions gives the logical-subspace dispersive shift quoted in the main text,
\begin{equation}\label{eq:sm_chi_logical}
H_{\chi} = \frac{\chi_L^{(2)} + \chi_L^{(4)}}{2}\,\tilde{Z}\,c^\dagger c \approx \frac{2\eta g^2\left(g^2 - \eta g_{DR}\right)}{\Delta^4}\,\tilde{Z}\,c^\dagger c,
\end{equation}
where $\chi_L \approx \chi_L^{(2)} + \chi_L^{(4)}$ is the dispersive shift of the encoded (dual-rail) logical qubit. At the DRT-1 parameters of Table~\ref{tab:sim_params} ($g/2\pi = 70$~MHz, $\Delta/2\pi = 1.013$~GHz), exact diagonalization of the undriven excitation-number blocks gives $\chi_L/2\pi = -31$~kHz. The perturbative expressions overestimate this value. Eq.~\eqref{eq:sm_cross_kerr} with the full denominators of Eq.~\eqref{eq:sm_denoms}, combined with the leading-order $\chi_L^{(4)}$ of Eq.~\eqref{eq:sm_chi4}, evaluates to $-41$~kHz, and the fully expanded $1/\Delta^4$ combination of Eq.~\eqref{eq:sm_chi_logical} to $-63$~kHz. At this coupling the perturbative error is $30$--$100\%$ (it drops to the $5\%$ level at $g/2\pi = 35$~MHz), and we quote the exact value where it matters.

The two contributions carry opposite signs and cancel at $g^2 = \eta g_{DR}$ at leading order in $1/\Delta$. At $g/2\pi = 70$~MHz this gives $g_{DR}/2\pi \approx 24$~MHz, and exact diagonalization places the null at $g_{DR}/2\pi = 38$~MHz, the difference being the $\mathcal{O}(\eta/\Delta)$ corrections. At the null the remaining undriven shift is of order $1/\Delta^5$, and it changes by $0.5$~kHz per MHz of $g_{DR}$. The dual-rail coupling is thus a design knob for nulling the dispersive coupling of an idle DRT, which is the coupling a spectator presents to a populated mode. The drive-induced contribution below does not cancel and grows as $1/g_{DR}$, so for the driven DRTs the null trades the static shift for a larger drive-induced one.

The drive itself adds a further contribution to $\chi_L$ at order $g^2\Omega^2$, $\approx \eta^2 g^2\Omega^2(1+r^2)/(2g_{DR}\Delta^4)$ at leading order, with $r = \Omega_-/\Omega_+$ the tone ratio. It carries the same $1/\Delta^4$ suppression as the static terms, is opposite in sign to $\chi_L^{(2)}$, and grows quadratically with the drive. At $\Omega/2\pi = 100$~MHz (balanced $\Omega_-$), the largest drive within the validity range of the second-order effective Hamiltonian, it evaluates to $14$~kHz.

The shift $\chi_L$ is the dual-rail-suppressed counterpart of the bare-transmon dispersive shift $\chi^{\text{tr}} \approx 2g^2\eta/[\Delta(\Delta+\eta)] \approx 2g^2\eta/\Delta^2$ [the single-transmon form of Eq.~\eqref{eq:chi-full}]. The dual-rail encoding trades the $1/\Delta^2$ of an unencoded transmon for the $1/\Delta^4$ of $\chi_L$, the parametric protection central to this work ($\chi^{\text{tr}}/2\pi \approx 1.6$~MHz at the parameters above, some fifty times $|\chi_L|$). The same $\chi^{\text{tr}}$ sets the shift a spectator DRT imparts to the mode when an erasure drops it to $\ket{gg}$ (End Matter, with an extra $1/(N+1)$ mode-dilution factor), and the shift of the auxiliary transmon of the parity readout [Eq.~\eqref{eq:chi-full}], which is deliberately left unencoded. Both are therefore $1/\Delta^2$ shifts rather than $\chi_L$.

\subsection{Robustness to rail asymmetry}\label{ssec:sm_rail_asym}

The derivation above assumed identical rails, with equal couplings $g_a = g_b = g$, equal rail frequencies, and a perfectly anti-symmetric drive. We repeat the second-order TDSW analysis in the bare rail basis with three imperfections, couplings $g_{a,b} = g(1 \pm x)$, a rail detuning $\omega_a - \omega_b = 2\delta\omega$, and a drive imbalance $a^\dagger - (1-y)e^{i\varphi}b^\dagger$. A coupling asymmetry leaves the symmetric coupling, and hence $g_{CR}$ and the static effective Hamiltonian, unchanged at first order; its only first-order effect is a spurious drive $\propto x g\Omega/\Delta$ addressing the wrong normal mode, detuned by the dual-rail gap $2g_{DR}$ and therefore suppressed by the ratio of the spurious amplitude to $2g_{DR}$. A rail detuning tilts the normal modes by the mixing angle $\theta = \delta\omega/2g_{DR}$, so the mode acquires a wrong-axis coupling $\approx g_{CR}(\delta\omega/g_{DR})\,\tilde{Z}(c^\dagger + c)$, which is $2\%$ of $g_{CR}$ at $\delta\omega/2\pi = 2$~MHz. A drive amplitude or phase imbalance only rescales $g_{CR}$ by $1 - y/2$ and rotates the conditional-displacement quadrature by $\varphi/2$, both absorbed in calibration. All remaining corrections are off-resonant by $2g_{DR}$ or second order in the imperfections; exact numerics at the parameters of Table~\ref{tab:sim_params} confirm these scalings. This is the quantitative content of the robustness statement in the main text. Every first-order imperfection is either absorbed in calibration or suppressed by the dual-rail gap.

\subsection{Two DRTs sharing a mode}\label{ssec:sm_twodrt}

The two-qubit effective Hamiltonian displayed above Eq.~\eqref{eq:2qprop} of the main text follows from the single-DRT result with no new ingredients. Each DRT $i$ is driven by its own two tones, balanced according to Eq.~\eqref{eq:sm_ratio}, and contributes its own slowly rotating cross-resonance block $-g_{CR,i}\tilde{X}_i(c^\dagger + c)$ with the common detuning $\epsilon$ set by the shared drive-center frequency. At second order the effective Hamiltonian is additive in the perturbations. Cross terms between the drives of different DRTs vanish identically, because each drive operator acts on its own transmon pair and connects no state reached by the other DRT's drive within a single virtual step. Cross terms between one DRT's drive and the shared resonator coupling of the \emph{other} DRT are exactly the mediated interactions quantified in Sec.~\ref{ssec:sm_mediated}, and are negligible at the operating separation of one DRT per two resonators, where neighboring DRTs attach two sites apart with an empty resonator between them. The result is the M{\o}lmer--S{\o}rensen Hamiltonian above Eq.~\eqref{eq:2qprop} of the main text, $H_\text{eff} = \epsilon\,c^\dagger c - \sum_i g_{CR,i}\tilde{X}_i(c^\dagger + c)$. Equal effective couplings $g_{CR,1} = g_{CR,2}$, required for the displacement to depend only on the parity in the parity-measurement protocol, are obtained by fixing the ratio of the overall drive amplitudes $\Omega_{+,1}/\Omega_{+,2}$ to compensate any difference in the two DRTs' parameters. Since $\tilde{X}_1$ and $\tilde{X}_2$ commute with $H_\text{eff}$, which is linear in $c$ and $c^\dagger$ within each joint $\tilde{X}_1\tilde{X}_2$ eigensector, the Magnus expansion terminates at second order and gives the exact propagator of Eq.~\eqref{eq:2qprop} of the main text.

\subsection{Mediated DRT--DRT cross-Kerr}\label{ssec:sm_mediated}

Two DRTs coupled to the same metamaterial acquire a direct cross-Kerr coupling $\zeta_{LL}\,\tilde{Z}_1\tilde{Z}_2$ mediated by the chain, even without a drive. To verify its exponential suppression with distance, we exactly diagonalize two DRTs attached to sites of an open chain at site distance $d$, that is, with $d-1$ resonators between the two attachment sites, at the parameters of Table~\ref{tab:sim_params}. Because the undriven Hamiltonian conserves the total excitation number, the logical two-qubit energies lie in the two-excitation block, which we diagonalize without truncation error. We find $\zeta_{LL}/2\pi = 51$~kHz, $0.48$~kHz, $4.4$~Hz, and $0.04$~Hz for $d = 1$--$4$, a suppression by $\sim 1/100 \approx (J/\Delta)^2$ per unit of site distance, independent of the total chain length. A site distance of $d = 2$ already suppresses the mediated cross-Kerr below $1$~kHz. This is the configuration simulated in Fig.~\ref{fig:gate_fidelity} of the main text (DRTs at sites $1$ and $3$ of the three-resonator chain), where $\zeta_{LL}/2\pi = 0.48$~kHz is small compared to the residual $|\chi_L|/2\pi = 31$~kHz.

\section{Spin-Locking Effective Hamiltonian}
\label{sec:slth}

We derive the effective cross-resonance Hamiltonian for a spin-locked transmon (SLT) coupled to a resonator mode. The spin-locking drive plays the role of the dual-rail coupling $g_{DR}$. It opens a gap that protects the encoded qubit from dephasing noise and suppresses the unwanted dispersive coupling to the resonator. The central result is that a two-frequency cross-resonance drive maintains the desired qubit--resonator interaction at full strength while the dispersive coupling is suppressed. We first work in the two-level approximation to establish the physical picture, then generalize to transmons.

\subsection{Baseline: qubits without spin locking}

We first establish the reference point, the dispersive and cross-resonance couplings of a qubit without spin locking, against which the spin-locking suppression is later measured. We refer to them as the bad and the good coupling. Consider a qubit coupled to a harmonic oscillator and driven cross-resonantly:
\begin{equation}\label{eq:sl_bare_qubit}
H = \frac{\omega}{2}Z + \nu c^\dagger c + g(\sigma^+ c + c^\dagger \sigma^-) + 2\Omega \cos(\omega_d t)\, Y,
\end{equation}
with $\Delta = \nu - \omega$. The $Y$ quadrature is chosen for consistency with the spin-locked case below, where the cross-resonance drive must be orthogonal to the locking drive; for the baseline the choice is a convention.

\paragraph{Dispersive (bad) interaction.}
Neglecting the cross-resonance drive and performing second-order perturbation theory in $g$ yields the quantum Stark shift
\begin{equation}\label{eq:sl_stark_bare}
H_{\mathrm{Stark}} \sim \frac{g^2}{\Delta}\, Z\, c^\dagger c.
\end{equation}
A constant Lamb shift $\sim (g^2/\Delta)\,Z$ also appears; it renormalizes the qubit frequency and is absorbed into $\omega$.

\paragraph{Cross-resonance (good) interaction.}
After the rotating-wave approximation and in the frame rotating at $\omega_d$, the cross-resonance drive dresses the qubit eigenstates via a small rotation about the $x$-axis. To lowest order in $\Omega/|\omega - \omega_d|$, this yields
\begin{equation}\label{eq:sl_good_bare}
H_{\mathrm{eff}} = (\nu - \omega_d)\, c^\dagger c + g\,\frac{\Omega}{\omega - \omega_d}\, Z\, i(c^\dagger - c).
\end{equation}
Since $\omega_d$ is chosen near $\nu$, the effective cross-resonance coupling strength is $\sim g\Omega/\Delta$; the mode quadrature reflects the drive phase, and the redefinition $c \to ic$ recasts the interaction as $Z(c^\dagger + c)$.

\subsection{Dispersive interaction with spin locking (qubit approximation)}
\label{sec:sl_bad_qubit}

Adding the spin-locking drive suppresses the dispersive coupling. We compute the residual shift $J_{Xc}$ within the two-level approximation. The spin-locked qubit--resonator Hamiltonian is
\begin{equation}
H = \frac{\omega}{2}Z + \nu c^\dagger c + \Omega_{sl}\cos(\omega t)\,X + g(\sigma^+ c + c^\dagger \sigma^-) + H_{\mathrm{cr}},
\end{equation}
where $\Omega_{sl}$ is the spin-locking Rabi frequency and $H_{\mathrm{cr}} = 2\Omega\cos(\omega_d t)\,Y$ is the cross-resonance drive, acting on the quadrature orthogonal to the lock. Performing the rotating-wave approximation for $\omega$ and moving to the rotating frame gives
\begin{equation}\label{eq:sl_rot}
H = \Delta\, c^\dagger c + \frac{\Omega_{sl}}{2}\,X + g(\sigma^- c^\dagger + c\,\sigma^+) + H_{\mathrm{cr}}.
\end{equation}

To compute the dispersive coupling, we drop $H_{\mathrm{cr}}$ and work in the $X$-eigenbasis $\ket{\pm}$. Second-order perturbation theory in $g$ gives the energies of the joint eigenstates $\ket{\pm,n}$, with $n$ the photon number, and the dispersive shift, defined as
\begin{equation}
J_{Xc} = E_{+,0} - E_{-,0} - E_{+,1} + E_{-,1},
\end{equation}
is
\begin{equation}\label{eq:sl_JXc}
J_{Xc} = \frac{\Omega_{sl}\, g^2}{\Delta^2 - \Omega_{sl}^2}.
\end{equation}
Compared to the bare dispersive shift $g^2/\Delta$ [Eq.~\eqref{eq:sl_stark_bare}], this is suppressed in two regimes:
\begin{equation}
J_{Xc} \sim \begin{cases} \dfrac{\Omega_{sl}\, g^2}{\Delta^2} & \text{if } \Omega_{sl} \ll \Delta, \\[8pt] \dfrac{g^2}{\Omega_{sl}} & \text{if } \Omega_{sl} \gg \Delta. \end{cases}
\end{equation}
Both regimes yield a suppression factor of $\min(\Omega_{sl}/\Delta,\,\Delta/\Omega_{sl})$ relative to the unsuppressed case. The crossover $\Omega_{sl} \approx \Delta$ is not an optimum but a resonance. The locked gap matches the detuning, the denominator of Eq.~\eqref{eq:sl_JXc} vanishes, and both the perturbation theory and the protection fail. The two suppression regimes are the two sides of this resonance.

\subsection{Failure of a single-frequency drive}
\label{sec:sl_single_freq}

A natural first attempt is a single cross-resonance drive at frequency $\omega_d \approx \Delta$. In the regimes where the dispersive coupling is suppressed ($\Omega_{sl} \ll \Delta$ or $\Omega_{sl} \gg \Delta$), however, the rotating-wave approximation for the cross-resonance drive breaks down, and we use the Floquet formalism. We work in the $X$-eigenbasis, through the cyclic relabeling $X,Y,Z \to Z,X,Y$, under which the spin-locking term becomes diagonal in $Z$ with splitting $\Omega_{sl}$ and the $Y$-quadrature cross-resonance drive becomes $X$. The qubit Hamiltonian is then
\begin{equation}
H = \Omega_{sl}\ket{1}\!\bra{1} + \Omega(e^{i\omega_d t} + e^{-i\omega_d t})\,X,
\end{equation}
with $\Omega$ the drive amplitude of Eq.~\eqref{eq:sl_bare_qubit}. Expanding in Floquet harmonics and truncating to the $\pm\omega_d$ sidebands, the dressed eigenstates are
\begin{equation}
\ket{\tilde{0}} = \ket{0} + \frac{\Omega}{\omega_d - \Omega_{sl}}\ket{1,-} - \frac{\Omega}{\omega_d + \Omega_{sl}}\ket{1,+},
\end{equation}
and similarly for $\ket{\tilde{1}}$ with exchanged labels, where $\ket{1,\mp}$ denotes the excited qubit state dressed with $\mp 1$ Floquet quantum of the drive, carrying the phase $e^{\pm i\omega_d t}$. Including the resonator, the rotating-frame coupling $g\sigma^- c^\dagger + \mathrm{h.c.}$ of Eq.~\eqref{eq:sl_rot} becomes $g(Z - iX)c^\dagger/2 + \mathrm{h.c.}$ under the cyclic relabeling [$\sigma^- = (X - iY)/2 \to (Z - iX)/2$]. Projecting onto the dressed qubit, with $h \equiv Z - iX$ the qubit factor multiplying $c^\dagger$, the effective cross-resonance coupling is $\frac{g}{2}\bigl(\bra{\tilde{0}} h \ket{\tilde{0}} - \bra{\tilde{1}} h \ket{\tilde{1}}\bigr)c^\dagger + \mathrm{h.c.}$, with
\begin{equation}
\bra{\tilde{0}} h \ket{\tilde{0}} - \bra{\tilde{1}} h \ket{\tilde{1}} = -\frac{4i\,\Omega\,\Omega_{sl}}{\omega_d^2 - \Omega_{sl}^2}.
\end{equation}
In both limiting regimes, the desired interaction is suppressed by the \emph{same} factor as the dispersive coupling:
\begin{equation}
\sim \frac{g\Omega\,\Omega_{sl}}{\Delta^2} \quad (\Omega_{sl} \ll \Delta), \qquad \sim \frac{g\Omega}{\Omega_{sl}} \quad (\Omega_{sl} \gg \Delta).
\end{equation}
This is expected, since both the dispersive and the cross-resonance coupling involve the qubit $Z$ operator, and spin locking suppresses its matrix elements uniformly.

\subsection{Two-frequency drive: recovering the full interaction}
\label{sec:sl_two_freq}

The resolution is to drive at two frequencies $\omega_{+} \approx \Omega_{sl} + \Delta$ and $\omega_{-} \approx |\Omega_{sl} - \Delta|$, which activate near-resonant flip-flip ($\ket{0}\ket{0} \to \ket{1}\ket{1}$) and flip-flop ($\ket{0}\ket{1} \to \ket{1}\ket{0}$) transitions between the dressed qubit and the resonator. In these processes the drive first flips the qubit and then the $Z$ operator in the coupling flips it back while creating or destroying a photon, which bypasses the suppression.

\subsubsection{Regime $\Omega_{sl} \gg \Delta$}

Since both drive frequencies are near $\Omega_{sl}$, the rotating-wave approximation is valid. We treat the flip-flip drive ($\omega_d = \Omega_{sl} + \Delta - 2\Omega^2/\Delta - \epsilon$) explicitly; the flip-flop analysis is analogous.

In the frame rotating with $\omega_d\ket{1}\!\bra{1}$, diagonalizing the qubit yields the dressed states
\begin{equation}
\ket{\tilde{0}} \approx \ket{0} - \frac{\Omega}{\Delta}\ket{1}, \quad \ket{\tilde{1}} \approx \ket{1} + \frac{\Omega}{\Delta}\ket{0},
\end{equation}
with an energy difference $E_0 - E_1 = \Delta - \epsilon$ (accounting for the Stark shift). The matrix element $\bra{\tilde{1}} Z \ket{\tilde{0}} \approx -2\Omega/\Delta$ mediates the near-resonant flip-flip process $\ket{\tilde{1}}\!\bra{\tilde{0}}\,c^\dagger$, yielding
\begin{equation}\label{eq:sl_flipflip}
H_{\mathrm{ff}} = \epsilon\, c^\dagger c - \frac{g\Omega}{\Delta}\left(\ket{\tilde{1}}\!\bra{\tilde{0}}\,c^\dagger + \mathrm{h.c.}\right).
\end{equation}
The flip-flop drive ($\omega_d = \Omega_{sl} - \Delta + 2\Omega^2/\Delta + \epsilon$) gives an analogous term. Choosing the two Rabi frequencies $\pi$ out of phase combines the two into
\begin{equation}\label{eq:sl_combined_qubit}
H = \epsilon\, c^\dagger c - \frac{g\Omega}{\Delta}\, X\,(c^\dagger + c),
\end{equation}
which has the full cross-resonance strength $g\Omega/\Delta$, undiminished by spin locking.

\subsubsection{Regime $\Omega_{sl} \ll \Delta$}

The drive frequencies are now $\omega_{\pm} \approx \Delta \pm \Omega_{sl}$, much larger than the qubit splitting $\Omega_{sl}$. Although the rotating-wave approximation does not apply globally, only the near-resonant terms contribute to the desired interaction, and these can be isolated within a Floquet treatment. The Stark shifts differ from the $\Omega_{sl} \gg \Delta$ case, $\sim 2\Omega_{sl}\Omega^2/\Delta^2$ rather than $\sim \Omega^2/\Delta$, but the final effective Hamiltonian is again Eq.~\eqref{eq:sl_combined_qubit} with the same interaction strength $g\Omega/\Delta$.

\subsection{Floquet analysis with both drives}
\label{sec:sl_floquet}

The preceding analysis treated the two drives independently and summed the resulting Hamiltonians. A Floquet treatment with both drives present confirms this result and gives the Stark shifts.

\subsubsection{Regime $\Omega_{sl} \gg \Delta$}

After the rotating-wave approximation, we go to the frame rotating at $\omega_1 = \Omega_{sl} + \Delta - \epsilon$ and expand in Floquet harmonics of the beat frequency $\omega_{21} = \omega_2 - \omega_1 = -2(\Delta - \epsilon)$. The Floquet-dressed eigenstates are
\begin{equation}
\ket{\tilde{0}} = \ket{0} - \frac{\Omega_1}{\Delta}\ket{1} + \frac{\Omega_2}{\Delta}\,e^{-i\omega_{21}t}\ket{1}, \quad
\ket{\tilde{1}} = \ket{1} + \frac{\Omega_1}{\Delta}\ket{0} - \frac{\Omega_2}{\Delta}\,e^{i\omega_{21}t}\ket{0},
\end{equation}
with Stark-shifted energies
\begin{equation}
E_0 = \frac{\Omega_1^2 - \Omega_2^2}{\Delta}, \qquad E_1 = \Omega_{sl} - \omega_1 - \frac{\Omega_1^2 - \Omega_2^2}{\Delta}.
\end{equation}
For equal drive amplitudes $|\Omega_1| = |\Omega_2|$ the Stark shifts cancel, $E_0 = 0$ and $E_1 = \Omega_{sl} - \omega_1$.

Projecting the coupling $Z\,c^\dagger$ onto the dressed-qubit subspace yields both a non-oscillating component $\propto \ket{\tilde{1}}\!\bra{\tilde{0}}$ (flip-flip) and a component oscillating as $e^{i\omega_{21}t} \propto \ket{\tilde{0}}\!\bra{\tilde{1}}$ (flip-flop). In the rotating frame with respect to $(\Delta - \epsilon)(c^\dagger c + \ket{\tilde{0}}\!\bra{\tilde{0}})$, these combine into
\begin{equation}\label{eq:sl_floquet_H}
H = \epsilon\, c^\dagger c + g\!\left(-\frac{\Omega_1}{\Delta}\ket{\tilde{1}}\!\bra{\tilde{0}}\,c^\dagger + \frac{\Omega_2}{\Delta}\ket{\tilde{0}}\!\bra{\tilde{1}}\,c^\dagger + \mathrm{h.c.}\right).
\end{equation}
Choosing $\Omega_1 = -\Omega_2 \equiv \Omega$ simplifies this to
\begin{equation}\label{eq:sl_floquet_final}
H = \epsilon\, c^\dagger c - \frac{g\Omega}{\Delta}\, X\,(c^\dagger + c),
\end{equation}
confirming Eq.~\eqref{eq:sl_combined_qubit}, i.e., $-g_{CR}X(c^\dagger+c)$ with $g_{CR} = g\Omega/\Delta > 0$, the same sign convention as Eq.~\eqref{eq:effham} of the main text.

\subsubsection{Regime $\Omega_{sl} \ll \Delta$}

The analysis proceeds identically after noting that only the near-resonant terms in each drive contribute. The Stark shifts no longer cancel, $E_0 = 2\Omega_{sl}(\Omega_1^2 + \Omega_2^2)/\Delta^2$ and $E_1 = \Omega_{sl} - \omega_1 - E_0$. The drive frequencies $\omega_{1,2}$ are adjusted to absorb these shifts, and the final Hamiltonian is again given by Eqs.~\eqref{eq:sl_floquet_H} and~\eqref{eq:sl_floquet_final}.

In both regimes, the $\mathcal{O}(g^2)$ corrections to the drive frequencies follow by replacing $\Omega_{sl} + \Delta$ and $|\Omega_{sl} - \Delta|$ in $\omega_{1,2}$ with the exact energy differences $E_{+,1} - E_{-,0}$ and $|E_{+,0} - E_{-,1}|$ from the dispersive-shift calculation of Sec.~\ref{sec:sl_bad_qubit}.

\subsection{Extension to transmons}
\label{sec:sl_transmon}

We now extend the analysis from two-level qubits to transmons with anharmonicity $\eta$, in the regime $\Delta \gg \Omega_{sl}$ proposed for operating spin-locked transmons. The expansion is in $g$ and $\Omega$ relative to $\Delta$. The locking amplitude $\Omega_{sl}$ is not an expansion parameter, and the results below hold whenever $\Delta \gg \Omega_{sl}$; dressed states and matrix elements are written to first order in $\Omega_{sl}/\eta$ where explicit. The $\Omega_{sl} \gg \Delta$ regime of the preceding subsections establishes the two-level physical picture and is not accessible to a transmon, whose anharmonicity bounds the useful locking amplitude. Unless noted otherwise, the numerical benchmarks quoted in this section are evaluated at $\Delta = 10\eta$, $\Omega_{sl} = 0.02\eta$, and $\Omega = 0.1\eta$, which for $\eta/2\pi = 205$~MHz is a locking amplitude $\Omega_{sl}/2\pi \approx 4$~MHz.

\subsubsection{Transmons without spin locking}

For a bare transmon coupled to a resonator with Hamiltonian $H = -(\eta/2)\,a^\dagger a^\dagger a a + g(a^\dagger c + c^\dagger a) + \Delta\, c^\dagger c$ (in the rotating frame), the dispersive and cross-resonance couplings are both reduced relative to the qubit case by a factor $\eta/\Delta$:
\begin{equation}\label{eq:sl_bare_transmon}
\text{dispersive:}\;\sim \frac{2g^2\eta}{\Delta^2}, \qquad \text{cross-resonance:}\;\sim \frac{2g\Omega\eta}{\Delta^2}.
\end{equation}
The dispersive result follows from second-order perturbation theory applied to the $\ket{1,1}$ state, in the basis $\ket{n_a, n_c}$ of transmon and resonator excitation numbers, where virtual transitions to $\ket{2,0}$ (with energy denominator $\Delta + \eta$) partially cancel those to $\ket{0,2}$. The cross-resonance result follows from the drive-induced admixture of $\ket{2}$ into $\ket{\tilde{1}}$, with matrix element $\sqrt{2}\,\Omega/(\omega - \omega_d - \eta)$.

\subsubsection{Dispersive interaction for spin-locked transmons}

In the rotating frame, the spin-locked transmon--resonator Hamiltonian is
\begin{equation}\label{eq:sl_transmon_ham}
H = -\frac{\eta}{2}a^\dagger a^\dagger a a - \delta\, a^\dagger a + \frac{\Omega_{sl}}{2}(a + a^\dagger) + \Delta\, c^\dagger c + g(a^\dagger c + c^\dagger a),
\end{equation}
where $\delta$ is the detuning of the spin-locking drive from the bare $0$--$1$ transition. This detuning is not free. The off-resonant coupling of the locking drive to the $1$--$2$ transition shifts the $0$--$1$ frequency by $\Omega_{sl}^2/2\eta$, and $\delta$ must cancel it,
\begin{equation}\label{eq:sl_clock}
\delta = \frac{\Omega_{sl}^2}{2\eta} + \mathcal{O}\!\left(\frac{\Omega_{sl}^4}{\eta^3}\right),
\end{equation}
which is the clock condition of Ref.~\cite{zuk2024robust}. The shift is second order in $\Omega_{sl}$, but the locked gap is itself $\Omega_{sl}$, so an uncancelled shift rotates the locked states off the $x$ axis by the first-order angle $\Omega_{sl}/2\eta$, the same order the expansion below retains.

With Eq.~\eqref{eq:sl_clock} imposed, $\ket{\pm} = (\ket{0} \pm \ket{1})/\sqrt{2}$ are eigenstates of the spin-locking term with energies $E_{\pm} = \pm\Omega_{sl}/2$. The finite anharmonicity dresses these with $\ket{2}$ to first order in $\Omega_{sl}/\eta$:
\begin{equation}\label{eq:sl_dressed_transmon}
\ket{\tilde{\pm}} = \ket{\pm} \pm \frac{\Omega_{sl}}{2\eta}\ket{2}, \qquad
\ket{\tilde{2}} = \ket{2} - \frac{\Omega_{sl}}{2\eta}\ket{+} + \frac{\Omega_{sl}}{2\eta}\ket{-} + \frac{\sqrt{3}\,\Omega_{sl}}{4\eta}\ket{3},
\end{equation}
with energies unchanged to first order, $E_{\pm} = \pm\Omega_{sl}/2$ and $E_2 = -\eta$.

The dispersive shift is computed from the second-order energy corrections to $\ket{\tilde{\pm},n}$ (with $n = 0,1$ photons). The matrix elements of $a^\dagger$ in the dressed $\{\ket{\tilde{-}}, \ket{\tilde{+}}, \ket{\tilde{2}}\}$ basis are
\begin{equation}
a^\dagger = \begin{pmatrix}
-\tfrac{1}{2} + \tfrac{\Omega_{sl}}{2\eta} & -\tfrac{1}{2} - \tfrac{\Omega_{sl}}{2\eta} & 0 \\
\tfrac{1}{2} - \tfrac{\Omega_{sl}}{2\eta} & \tfrac{1}{2} + \tfrac{\Omega_{sl}}{2\eta} & 0 \\
-1 - \tfrac{\Omega_{sl}}{2\eta} & 1 - \tfrac{\Omega_{sl}}{2\eta} & -\tfrac{\Omega_{sl}}{4\eta}
\end{pmatrix}.
\end{equation} Computing the energies $E_{\pm,0}$ and $E_{\pm,1}$ via second-order perturbation theory in $g$ (keeping terms to first order in $\Omega_{sl}$), one obtains
\begin{align}
E_{+,0} &= \frac{\Omega_{sl}}{2} - \frac{g^2(2 + \Omega_{sl}/\Delta)}{4\Delta}, \label{eq:sl_Ep0t}\\
E_{-,0} &= -\frac{\Omega_{sl}}{2} - \frac{g^2(2 - \Omega_{sl}/\Delta)}{4\Delta}, \label{eq:sl_Em0t}\\
E_{+,1} &= \frac{\Omega_{sl}}{2} + \Delta + \frac{g^2(-2 + 4\Omega_{sl}/\eta - 3\Omega_{sl}/\Delta)}{4\Delta} + \frac{g^2(1 - \Omega_{sl}/\eta - \Omega_{sl}/[2(\eta+\Delta)])}{\eta + \Delta}, \label{eq:sl_Ep1t}\\
E_{-,1} &= -\frac{\Omega_{sl}}{2} + \Delta + \frac{g^2(-2 - 4\Omega_{sl}/\eta + 3\Omega_{sl}/\Delta)}{4\Delta} + \frac{g^2(1 + \Omega_{sl}/\eta + \Omega_{sl}/[2(\eta+\Delta)])}{\eta + \Delta}. \label{eq:sl_Em1t}
\end{align}

The dispersive shift is then
\begin{equation}\label{eq:sl_bad_transmon}
E_{+,0} - E_{-,0} - E_{+,1} + E_{-,1} = \frac{g^2\eta^2\Omega_{sl}}{\Delta^2(\Delta + \eta)^2} \approx \frac{g^2\eta^2\Omega_{sl}}{\Delta^4},
\end{equation}
where the last approximation holds for $\Delta \gg \eta$. Compared to the bare-transmon dispersive shift $\sim g^2\eta/\Delta^2$ [Eq.~\eqref{eq:sl_bare_transmon}], the spin-locked result is suppressed by two additional factors, $\eta/\Delta$ from the transmon structure and $\Omega_{sl}/\Delta$ from the spin locking. The additional $\eta/\Delta$ arises because the $\mathcal{O}(\eta^1)$ contributions to $E_{\pm,1}$ cancel in the combination $-E_{+,1} + E_{-,1}$. The terms linear in $\eta/\Delta$ are independent of $\Omega_{sl}$ at zeroth order and hence drop out of the dispersive shift.

Both the two extra powers of $1/\Delta$ and the factor $\Omega_{sl}$ rely on the two locked states $\ket{\pm}$ being the $\pm x$ eigenstates on the Bloch sphere of the $0$--$1$ subspace, that is, on Eq.~\eqref{eq:sl_clock}. Driving the lock at the bare transmon frequency instead tilts them by $\Omega_{sl}/2\eta$, and the tilt re-exposes the bare-transmon dispersive shift weighted by $\langle Z\rangle_{\tilde{+}} - \langle Z\rangle_{\tilde{-}} = -\Omega_{sl}/\eta$, leaving
\begin{equation}\label{eq:sl_bad_noclock}
E_{+,0} - E_{-,0} - E_{+,1} + E_{-,1}\big|_{\delta = 0} = \frac{g^2\,\Omega_{sl}}{\Delta(\Delta + \eta)},
\end{equation}
the unsuppressed two-level value of Eq.~\eqref{eq:sl_JXc} up to the factor $\Delta/(\Delta+\eta)$. This is larger than Eq.~\eqref{eq:sl_bad_transmon} by $(\Delta/\eta)^2(\Delta+\eta)/\Delta$, two orders of magnitude at $\Delta/\eta = 10$. Exact diagonalization of Eq.~\eqref{eq:sl_transmon_ham} reproduces Eq.~\eqref{eq:sl_bad_transmon} to better than $2\%$ when Eq.~\eqref{eq:sl_clock} holds, and Eq.~\eqref{eq:sl_bad_noclock} to $1\%$ when it does not. Satisfying the clock condition is therefore a requirement of the scheme, not a refinement of it.

\subsubsection{Cross-resonance interaction for spin-locked transmons}

The full Hamiltonian including the cross-resonance drive is
\begin{equation}
H = -\frac{\eta}{2}a^\dagger a^\dagger a a + \frac{\Omega_{sl}}{2}(a + a^\dagger) + \Delta\,c^\dagger c + g(a^\dagger c + c^\dagger a) + 2\Omega\cos(\omega_d t)\,i(a - a^\dagger).
\end{equation}
The drive operator in the dressed transmon basis is
\begin{equation}
i(a - a^\dagger) = -i\begin{pmatrix}
0 & -1 & 1 + \tfrac{\Omega_{sl}}{2\eta} \\
1 & 0 & -1 + \tfrac{\Omega_{sl}}{2\eta} \\
-1 - \tfrac{\Omega_{sl}}{2\eta} & 1 - \tfrac{\Omega_{sl}}{2\eta} & 0
\end{pmatrix}.
\end{equation}

We compute the flip-flip coupling $\ket{\tilde{-},0} \leftrightarrow \ket{\tilde{+},1}$ (at drive frequency $\omega_d \approx \Delta + \Omega_{sl}$) by summing three contributing second-order processes with intermediate states $\ket{\tilde{+},0}$, $\ket{\tilde{2},0}$, and $\ket{\tilde{-},1}$. The result is
\begin{equation}\label{eq:sl_good_transmon}
-ig\Omega\left[-\frac{\tfrac{1}{2} + \tfrac{\Omega_{sl}}{2\eta}}{\Delta} + \frac{(1 - \tfrac{\Omega_{sl}}{2\eta})(1 + \tfrac{\Omega_{sl}}{2\eta})}{\Delta + \eta + \tfrac{\Omega_{sl}}{2}} - \frac{\tfrac{1}{2} - \tfrac{\Omega_{sl}}{2\eta}}{\Delta}\right].
\end{equation}
Expanding to first order in $\Omega_{sl}$:
\begin{equation}
\frac{ig\Omega\eta}{\Delta(\Delta + \eta)} + \frac{ig\Omega\,\Omega_{sl}}{2(\Delta + \eta)^2} \approx \frac{ig\Omega\eta}{\Delta^2},
\end{equation}
the second term being smaller than the first by $\Omega_{sl}/2\eta$. The leading term is the same as the bare-transmon result [Eq.~\eqref{eq:sl_bare_transmon}], confirming that the cross-resonance interaction strength is unaffected by spin locking to leading order. The flip-flop interaction ($\ket{\tilde{-},1} \leftrightarrow \ket{\tilde{+},0}$ at $\omega_d \approx \Delta - \Omega_{sl}$) yields an identical amplitude; both follow from the minimum splitting of the corresponding avoided crossing of the exact Floquet quasi-energies, which reproduces $g\Omega\eta/[\Delta(\Delta+\eta)]$ to $2\%$ and is independent of $\Omega_{sl}$.

The cross-resonance drives shift the locked gap. Because $\omega_d \simeq \Delta \gg \Omega_{sl}$, both drive sidebands contribute at the same order, as in the $\Omega_{sl} \ll \Delta$ regime of Sec.~\ref{sec:sl_floquet}, and the gap becomes
\begin{equation}\label{eq:sl_stark_transmon}
\tilde{\Omega}_{sl} = \Omega_{sl} - \frac{4\Omega^2\Omega_{sl}}{\omega_d^2 - \Omega_{sl}^2}
+ 2\Omega^2\left[\frac{\left(1 - \frac{\Omega_{sl}}{2\eta}\right)^2\left(\eta + \frac{\Omega_{sl}}{2}\right)}{\left(\eta + \frac{\Omega_{sl}}{2}\right)^2 - \omega_d^2} - \frac{\left(1 + \frac{\Omega_{sl}}{2\eta}\right)^2\left(\eta - \frac{\Omega_{sl}}{2}\right)}{\left(\eta - \frac{\Omega_{sl}}{2}\right)^2 - \omega_d^2}\right] \simeq \Omega_{sl}\left(1 - \frac{2\Omega^2}{\Delta^2}\right).
\end{equation}
The first term is the two-level Floquet shift of Sec.~\ref{sec:sl_floquet}; the $\ket{\tilde{2}}$ terms cancel half of it. The two tones shift the gap independently, so with both active $\tilde{\Omega}_{sl} = \Omega_{sl}[1 - 2(\Omega_1^2 + \Omega_2^2)/\Delta^2]$, and the drive frequencies follow from $\tilde{\Omega}_{sl}$. Comparison with exact Floquet quasi-energies confirms Eq.~\eqref{eq:sl_stark_transmon} to $2\times10^{-6}\,\eta$ at $\Omega = 0.1\,\eta$. Keeping instead a single sideband per intermediate state gives $\Omega_{sl} - 2\Omega^2/\Delta - \Omega^2(1 - \Omega_{sl}/2\eta)^2/(\Delta - \eta + \Omega_{sl}/2) - \Omega^2(1 + \Omega_{sl}/2\eta)^2/(\Delta + \eta + \Omega_{sl}/2)$, which errs by $4\times10^{-3}\,\eta$ at the same drive amplitude, twenty times $g_{CR}$.

One further term accompanies the two-tone drive. The tones sit at $\omega_d = \Delta \pm \Omega_{sl}$, detuned from the mode by $\Omega_{sl}$, and reach it through $g$ with an effective amplitude $\sim g\Omega/\Delta$, displacing it by $\sim g\Omega/(\Delta\,\Omega_{sl})$. The displacement is common to both locked states and does not entangle the qubit with the mode; it is small whenever the locked gap exceeds the effective drive amplitude, $\Omega_{sl} \gg g\Omega/\Delta$, which holds at the benchmark parameters of this section.

As in the qubit case, the $\mathcal{O}(g^2)$ corrections to the drive frequencies follow by replacing $\Omega_{sl} + \Delta$ in the flip-flip frequency with $E_{+,1} - E_{-,0}$ [from Eqs.~\eqref{eq:sl_Ep1t} and~\eqref{eq:sl_Em0t}], and $\Delta - \Omega_{sl}$ in the flip-flop frequency with $E_{-,1} - E_{+,0}$ [from Eqs.~\eqref{eq:sl_Ep0t} and~\eqref{eq:sl_Em1t}].

As an end-to-end benchmark, we simulate the full two-tone cross-resonance gate between two spin-locked transmons coupled through a shared resonator, each transmon driven by its locking tone and a cross-resonance tone, with the gate duration optimized. At $\omega/2\pi = 4.6$~GHz, $\nu/2\pi = 6.0$~GHz, $g/2\pi = 100$~MHz, $\eta/2\pi = 300$~MHz, $\Omega_{sl} = 0.5\eta$, and $\Omega = 0.7\Omega_{sl}$, the gate reaches an infidelity of $7.6\times10^{-6}$ with a leakage of $3\times10^{-4}$.

\subsection{Fourth-order dispersive shift and the optimal locking amplitude}\label{ssec:sl_g4}

Beyond second order in $g$, the spin-locked transmon acquires a drive-free dispersive shift whose dual-rail counterpart is suppressed by the larger gap. The second-order shifts of the Fock states, $\delta E_{0,n} = g^2 n/\Delta$ and $\delta E_{1,n} = -g^2(n+1)/\Delta + 2g^2 n/(\Delta+\eta)$, are diagonal in the transmon basis and therefore act \emph{off-diagonally} on the locked $\ket{\pm}$ doublet; second-order perturbation theory across the locked gap $\Omega_{sl}$ yields
\begin{equation}\label{eq:sl_chi4}
\chi^{(4)} = \left[\left(\frac{3}{\Delta} - \frac{2}{\Delta+\eta}\right)^2 - \frac{1}{\Delta^2}\right]\frac{g^4}{2\Omega_{sl}} \approx \frac{2\eta g^4}{\Delta^3\,\Omega_{sl}},
\end{equation}
which exact diagonalization reproduces to within $2\%$.

In the dual rail, the gap $g_{DR} \gg g^2/\Delta$ suppresses the analogous doublet-mediated process, leaving $\chi_L^{(4)} = 4\eta g^4/\Delta^4$ [Eq.~\eqref{eq:sm_chi4}]; the ratio of the two fourth-order terms is $\chi^{(4)}/\chi_L^{(4)} = \Delta/2\Omega_{sl}$, so the comparison between the encodings is set by the locking amplitude, a design knob.

The fourth-order term overtakes the second-order one at $g \sim \Omega_{sl}\sqrt{\eta/2\Delta}$. Conversely, at fixed coupling the total $\chi^{(2)} + \chi^{(4)}$ is minimized at $\Omega_{sl}^* = g\sqrt{2\Delta/\eta}$. Unlike the dual-rail pair, which cancel at $g^2 = \eta g_{DR}$ at leading order, the two terms here add, so the minimum is a compromise rather than a null. At $g/2\pi = 70$~MHz the optimum is $\Omega_{sl}^*/2\pi = 220$~MHz, close to the dual-rail splitting $2g_{DR}/2\pi = 210$~MHz of Table~\ref{tab:sim_params}, so the two encodings favor the same gap. At this matched gap, exact diagonalization gives $\chi/2\pi = 67$~kHz at $g/2\pi = 70$~MHz, within $10\%$ of the closed forms ($\chi^{(2)}/2\pi + \chi^{(4)}/2\pi = 28 + 44$~kHz) and about twice the DRT's exact $|\chi_L|/2\pi = 31$~kHz. Both are $20$--$50\times$ below the bare-transmon shift $\chi^{\text{tr}}/2\pi \approx 1.6$~MHz.

Two caveats accompany this evaluation. First, the clock detuning must be calibrated exactly, since the leading-order Eq.~\eqref{eq:sl_clock} is insufficient at $\Omega_{sl} \sim \eta$ (using it, the exact-diagonalization shift grows to $0.45$~MHz). Second, at small locking amplitudes the fourth-order term dominates instead. At $\Omega_{sl}/2\pi = 20$~MHz it evaluates to $0.43$~MHz at $g/2\pi = 70$~MHz, only $\sim 4\times$ below $\chi^{\text{tr}}$.

\subsection{Summary}

For spin-locked transmons whose locking drive satisfies the clock condition Eq.~\eqref{eq:sl_clock}, the dispersive coupling $\sim g^2\eta^2\Omega_{sl}/\Delta^4$ is suppressed relative to the bare-transmon value $2g^2\eta/\Delta^2$ by $\eta\Omega_{sl}/2\Delta^2$. The two-frequency cross-resonance interaction retains its full strength $\sim g\Omega\eta/\Delta^2$, identical to the bare-transmon result. The ratio of unwanted to desired coupling therefore improves by $\eta\Omega_{sl}/2\Delta^2$ under spin locking, the same parametric improvement as for dual-rail transmons in Sec.~\ref{sec:dreh}. Without the clock condition the desired coupling is unchanged but the dispersive coupling reverts to Eq.~\eqref{eq:sl_bad_noclock}, and the improvement is only the two-level factor $\Omega_{sl}/\Delta$. These statements hold at second order in $g$; at stronger couplings the fourth-order term of Eq.~\eqref{eq:sl_chi4} contributes, and the locking amplitude minimizing the total dispersive shift matches the dual-rail splitting, $\Omega_{sl}^* = g\sqrt{2\Delta/\eta} \approx 2g_{DR}$ (Sec.~\ref{ssec:sl_g4}).

Table~\ref{tab:sm_summary} collects the asymptotic scalings of the bad (dispersive) and good (cross-resonance) couplings at large $\Delta$, for bare, dual-rail, and spin-locked qubits and transmons. The bad/good ratio improves by an additional factor of $\eta/\Delta$ in going from qubits to transmons in both the dual-rail and spin-locked cases, whereas in the bare case the ratio is unchanged. In this sense transmons outperform qubits under either encoding, though not outright, since the overall strength of the good interaction also decreases in going from qubits to transmons.

\begin{table}[t]
\caption{Asymptotic scalings at large $\Delta$ of the bad (dispersive) and good (cross-resonance) couplings for the three encodings, in the two-level (qubit) approximation and for transmons. Dual-rail and spin-locked entries assume balanced two-tone driving (and, for spin locking, the clock condition). Numerical prefactors of order unity are omitted. The last column quotes the second-order suppression of the dispersive coupling relative to the corresponding bare case, exhibiting the common mechanism $\sim \eta G/\Delta^2$ with gap $G = g_{DR}$ or $\Omega_{sl}$; the transmon encodings additionally carry the fourth-order terms shown, discussed in Secs.~\ref{ssec:sm_chi} and~\ref{ssec:sl_g4}.}
\label{tab:sm_summary}
\setlength{\tabcolsep}{6pt}
\begin{ruledtabular}
\begin{tabular}{lccc}
 & Bad (dispersive) & Good (cross-resonance) & Bad, relative to bare \\
\hline
Bare qubit & $g^2/\Delta$ & $g\Omega/\Delta$ & $1$ \\
Bare transmon & $g^2\eta/\Delta^2$ & $g\Omega\eta/\Delta^2$ & $1$ \\
Dual-rail qubit & $g^2 g_{DR}/\Delta^2$ & $g\Omega/\Delta$ & $g_{DR}/\Delta$ \\
Dual-rail transmon & $g^2\eta\left(g^2 - \eta g_{DR}\right)/\Delta^4$ & $g\Omega\eta/\Delta^2$ & $\eta g_{DR}/\Delta^2$ \\
Spin-locked qubit & $g^2\Omega_{sl}/\Delta^2$ & $g\Omega/\Delta$ & $\Omega_{sl}/\Delta$ \\
Spin-locked transmon & $g^2\eta\left(\eta\Omega_{sl} + 2g^2\Delta/\Omega_{sl}\right)/\Delta^4$ & $g\Omega\eta/\Delta^2$ & $\eta\Omega_{sl}/\Delta^2$ \\
\end{tabular}
\end{ruledtabular}
\end{table}

\section{Parity Measurements: Backaction, Decoding, and Measurement Time}\label{sec:sm_parity}

This section supports the parity-measurement claims of the main text. We first justify three properties of the single-pair measurement, its backaction on the DRTs, the $(-1)^n$ phase within the even sector, and its measurement time. We then treat the parallel case. Several auxiliary transmons coupled to the chain are probed with frequency combs, and the photon numbers of all populated modes are decoded jointly from the reflected signals. We choose the auxiliary transmons, and the number of pairs measured at once, to minimize the measurement time per pair at a fixed error per pair.

\subsection{Single-pair parity measurement}\label{ssec:sm_single_parity}

We start with the single-pair parity measurement. During the parity operation, the balanced cross-resonance drive displaces the mode conditioned on $K = \tilde{X}_1 + \tilde{X}_2$. After a time $t$, the even-sector states $\ket{\tilde{+}\tilde{+}}$ and $\ket{\tilde{-}\tilde{-}}$ carry coherent amplitudes $\pm\alpha$ with $|\alpha| = 2g_{CR}t$, while the odd sector leaves the mode in vacuum. The drive is then switched off and the mode photon number is read out through a dispersively coupled auxiliary transmon, $H_{\text{disp}} = (\chi/2)\,\sigma_z^{\text{anc}}\,c^\dagger c$~\cite{essig_multiplexed_2021}.

Three properties follow. \emph{(i) Backaction on the DRTs.} The measured observable is the mode photon number $n$; with the drive off, the mode couples to the DRTs only through the residual dispersive shift $\chi_L$ of Eq.~\eqref{eq:sm_chi_logical}, so the measurement dephases the DRTs at the suppressed rate $\sim \chi_L \bar{n}$, negligible over the measurement time. Within each parity sector, the two eigenstates produce the same $\bar{n} = |\alpha|^2$ (amplitudes $\pm\alpha$ for even, vacuum for odd), so the photon number reveals the parity and nothing else. \emph{(ii) The $(-1)^n$ phase.} The number measurement projects the mode onto a Fock state. Using $\langle n | {-\alpha} \rangle = (-1)^n \langle n | \alpha \rangle$, the even-sector superposition $u\ket{\tilde{+}\tilde{+}}\ket{\alpha} + v\ket{\tilde{-}\tilde{-}}\ket{-\alpha}$ collapses to $u\ket{\tilde{+}\tilde{+}} + (-1)^n v\ket{\tilde{-}\tilde{-}}$. The outcome-dependent phase is a $\tilde{Z}_1$ (equivalently $\tilde{Z}_2$) rotation by $n\pi$, corrected in software once $n$ is known. This is why the measurement must resolve the photon number, not merely its presence. \emph{(iii) Measurement time.} The auxiliary transmon is coupled to its own transmission line at a rate $\Gamma$ and probed in reflection with a frequency comb~\cite{essig_multiplexed_2021}. A mode holding $n$ photons shifts the transmon by about $n\chi$, onto the comb tooth placed there, and the transmon scatters that tooth. Telling neighboring photon numbers apart requires $\Gamma \lesssim \chi$, and a transmon scatters a resonant tone elastically at no more than about $\Gamma/8$ photons per unit time before it saturates (Sec.~\ref{ssec:sm_comb}). The measurement therefore takes several tens of $1/\chi$, the $\mathcal{O}(1/\chi)$ scaling quoted in the main text.

\subsection{Model Hamiltonian}

We consider an auxiliary transmon coupled to site $j_0$ (typically $j_0=1$) of a one-dimensional tight-binding chain of $N$ resonators, the metamaterial. The Hamiltonian is ($\hbar=1$)
\begin{equation}\label{eq:H}
H = H_{\text{chain}} + H_{\text{anc}} + H_{\text{int}},
\end{equation}
where
\begin{align}
H_{\text{chain}} &= \nu \sum_{j=1}^{N} c_j^\dagger c_j + J \sum_{j=1}^{N-1}(c_j^\dagger c_{j+1} + \text{h.c.}), \label{eq:Hchain} \\
H_{\text{anc}} &= \omega\, d^\dagger d - \frac{\eta}{2} d^\dagger d(d^\dagger d - 1), \label{eq:Htransmon} \\
H_{\text{int}} &= g(c_{j_0}^\dagger d + c_{j_0} d^\dagger). \label{eq:Hint}
\end{align}
Here $\nu$ is the bare resonator frequency (as in the main text), $\omega$ the auxiliary-transmon $|g\rangle \to |e\rangle$ frequency, $\eta > 0$ its anharmonicity (so the $|e\rangle \to |f\rangle$ gap is $\omega - \eta$), and $g$ the vacuum Rabi coupling to site $j_0$. The auxiliary transmon operator $d$ is distinct from the DRT rail operators $a,b$ of the main text.

\subsection{Chain diagonalization}

The tight-binding chain with open boundary conditions is diagonalized by the discrete sine transform, as quoted in the main text and repeated here to fix the notation of this section,
\begin{equation}\label{eq:DST}
\tilde{c}_k = \sqrt{\frac{2}{N+1}} \sum_{j=1}^{N} \sin\!\left(\frac{jk\pi}{N+1}\right) c_j, \quad k = 1, \ldots, N,
\end{equation}
which have normal-mode frequencies
\begin{equation}\label{eq:dispersion}
\nu_k = \nu + 2J\cos\!\left(\frac{\pi k}{N+1}\right), \quad k = 1, \ldots, N.
\end{equation}
This is the tight-binding dispersion relation. The mode index $k$ plays the role of a discrete crystal momentum in the first Brillouin zone. The band spans $[\nu - 2J, \nu + 2J]$.

The coupling to site $j_0$ transforms as
\begin{equation}\label{eq:gk}
g_{k} = g \sqrt{\frac{2}{N+1}}\sin\!\left(\frac{j_0 k\pi}{N+1}\right).
\end{equation}
The detuning of mode $k$ from the transmon is
\begin{equation}\label{eq:Deltak}
\Delta_k \equiv \nu_k - \omega = \Delta + 2J\cos\!\left(\frac{\pi k}{N+1}\right),
\end{equation}
where $\Delta \equiv \nu - \omega$ is the bare detuning.
We work in the dispersive regime, with the auxiliary transmon detuned below the whole band, $\omega < \nu - 2J$, so that $\Delta_k > 0$ for every mode, as for the DRTs of the main text. Because the auxiliary transmon hybridizes with all $N$ modes, the dispersive regime is set by its total admixture into the chain, $\sum_k g_k^2/\Delta_k^2$, and not by any single diluted coupling $g_k$. Completeness of the discrete sine transform gives $\sum_k g_k^2 = g^2$, and every detuning lies within $2J$ of $\Delta$, so
\begin{equation}\label{eq:admixture}
\frac{g^2}{(\Delta+2J)^2} \le \sum_k \frac{g_k^2}{\Delta_k^2} \le \frac{g^2}{(\Delta-2J)^2}.
\end{equation}
For $\Delta \gg J$ the admixture reduces to $(g/\Delta)^2$, the site coupling over the bare detuning, and $g \ll \Delta - 2J \le \min_k \Delta_k$ guarantees the dispersive regime. The designs of Table~\ref{tab:sm_readout} have $\Delta/2J = 3$--$15$. If the transmon is instead tuned close to a single mode, $\Delta_k \lesssim J$, that mode dominates the sum and the relevant ratio becomes the diluted $g_k/\Delta_k$. The dispersive shift per mode is
\begin{equation}\label{eq:chi-full}
\chi_k = -\frac{2 g_k^2 \eta}{\Delta_k(\Delta_k + \eta)}.
\end{equation}
Placing the transmon below the band keeps both factors of the denominator positive, so every mode shifts the transmon in the same direction. It also keeps the transmon out of the straddling regime, $-\eta < \Delta_k < 0$. There the shift changes sign, and at $\Delta_k = -\eta$ it diverges, because the $\ket{e}\to\ket{f}$ transition of the transmon becomes resonant with mode $k$. Exact diagonalization of Eqs.~\eqref{eq:H}--\eqref{eq:Hint} confirms Eq.~\eqref{eq:chi-full} per mode to better than $1\%$ deep in the dispersive regime, with relative errors growing as $(g/\Delta)^2$. The optimized designs below sit where these errors reach tens of percent, so the readout model uses exact shifts (Sec.~\ref{ssec:sm_shifts}). This is the bare-transmon dispersive shift $\chi^{\text{tr}}$ discussed below Eq.~\eqref{eq:sm_chi_logical}; here the auxiliary transmon is deliberately left unencoded so that the shift is large enough to read out.

The mirror symmetry of the sine profile bounds how many modes are independently resolvable. Since $\sin^2\!\big(\frac{j(N+1-k)\pi}{N+1}\big) = \sin^2\!\big(\frac{jk\pi}{N+1}\big)$, the columns $k$ and $N+1-k$ of the coupling-weight matrix are identical. An auxiliary transmon at any site therefore sees a mode and its mirror partner with the same spatial weight, and their shifts per photon differ only through the detunings $\Delta_k$. This difference does not separate the two modes. Every auxiliary transmon sees the same $\Delta_k$, so the difference rescales whole columns, $\chi_{\alpha,N+1-k} = r_k\,\chi_{\alpha,k}$ for every $\alpha$, with $r_k = \Delta_k(\Delta_k+\eta)/[\Delta_{N+1-k}(\Delta_{N+1-k}+\eta)]$ between $0.5$ and $2$ at the parameters used below.

The condition number of $\chi$ measures what this proportionality costs. The readout infers the photon numbers $\mathbf{n}$ from the shifts $\chi\mathbf{n}$, so every pattern of photon numbers must move the shifts by more than the readout noise. The largest and smallest singular values of $\chi$, $\sigma_\text{max}$ and $\sigma_\text{min}$, bound how far a pattern of unit norm can move them, and the condition number $\sigma_\text{max}/\sigma_\text{min}$ compares the most and least visible patterns. A condition number near $1$ means that all patterns are about equally visible. A large one means that some pattern is nearly hidden by the noise, and an infinite one that some pattern produces no shift at all. Every mirror pair hides such a pattern, the vector with $r_k$ in entry $k$ and $-1$ in entry $N+1-k$. For identical auxiliary transmons, one at each site, $\chi$ therefore has rank $\lfloor(N+1)/2\rfloor$ rather than $N$, and its condition number is infinite. Computed in double precision it is $(0.7$--$2)\times10^{16}$ for $N = 5$--$20$, the inverse of the machine epsilon $2.2\times10^{-16}$, because rounding leaves $\sigma_\text{min}$ at that fraction of $\sigma_\text{max}$ instead of at zero. Giving the auxiliary transmons distinct frequencies makes $r_k$ differ from transmon to transmon and restores rank $N$, but the matrix stays ill-conditioned. For frequencies spread uniformly over $\pm20$~MHz, the median condition number over $200$ draws grows from $4\times10^{3}$ at $N = 5$ to $6\times10^{4}$ at $N = 20$, so the mirror partners remain unresolvable in practice. Each mirror pair therefore contributes a single measurable combination, $n_k + r_k\,n_{N+1-k}$. The maximum number of simultaneously readable channels is
\begin{equation}\label{eq:Mmax}
M_{\max} = \left\lfloor \frac{N+1}{2} \right\rfloor,
\end{equation}
one per mirror pair; a photon in an unselected mode aliases onto its mirror channel, rescaled by the ratio of their dispersive shifts, rather than adding independent noise.

\subsection{Several auxiliary transmons}\label{sec:multi-transmon}

To read $M$ modes in parallel, we couple $M$ auxiliary transmons to distinct sites $j_1, \ldots, j_M$,
\begin{equation}\label{eq:H-multi}
H = H_{\text{chain}} + \sum_{\alpha=1}^{M} \bigl(H_{\text{anc}}^{(\alpha)} + g\,(c_{j_\alpha}^\dagger d_\alpha + \text{h.c.})\bigr).
\end{equation}
The chain couples the transmons to each other in two ways. The first is coherent exchange, which at second order is $J_{\alpha\beta} = \tfrac{1}{2}\sum_k g_{\alpha k}g_{\beta k}\,(\Delta_{\alpha k}^{-1} + \Delta_{\beta k}^{-1})$, a few MHz between neighbors. Identical transmons would hybridize into delocalized modes. We therefore stagger their frequencies by $40$~MHz, as frequency multiplexing requires anyway, and require $|J_{\alpha\beta}| \le 0.1\,|\omega_\alpha - \omega_\beta|$. The second is a cross-Kerr term between transmons~\cite{majerCouplingSuperconductingQubits2007},
\begin{equation}\label{eq:cross-kerr}
H_{\text{cross}} = \sum_{\alpha < \beta} \zeta_{\alpha\beta}\,\frac{\sigma_z^{(\alpha)}}{2}\frac{\sigma_z^{(\beta)}}{2}, \quad \zeta_{\alpha\beta} = \sum_k \frac{g_{\alpha,k}^2 g_{\beta,k}^2 \eta}{\Delta_{\alpha k}^3(\Delta_{\alpha k} + \eta)},
\end{equation}
which makes each transmon's frequency depend on the states of the others. Its comparator is the readout linewidth $\Gamma$ below. In the optimized designs of Table~\ref{tab:sm_readout}, the second-order estimates give $|\zeta_{\alpha\beta}|/2\pi \le 78$~kHz, at most $4\%$ of $\Gamma$, and an exchange of at most $3\%$ of the frequency separation. We neglect both.

\subsection{Photon-number-dependent shifts}\label{ssec:sm_shifts}

To decode the parallel parity measurement, we need the frequency of each auxiliary transmon as a function of the photon numbers $\mathbf{n} = (n_1, \ldots, n_M)$ of the populated modes, up to $n_\text{max} = 7$ photons per mode. At the couplings of interest this function is not linear in $\mathbf{n}$. We first diagonalize the harmonic part, the transmon as an oscillator coupled to the chain, which gives a transmon-like normal mode $t$ and dressed chain modes $k$. The remaining term is the Kerr nonlinearity $-\tfrac{\eta}{2}A^{\dagger 2}A^2$, with $A = \sum_m u_m b_m$ the bare transmon operator in the normal modes $b_m$. We write the shift of transmon $\alpha$ as
\begin{equation}\label{eq:sm_shift_model}
\delta\omega_\alpha(\mathbf{n}) = \sum_k f_{\alpha k}(n_k) + \sum_{j<k} P_{\alpha jk}(n_j, n_k).
\end{equation}
The single-mode curve $f_{\alpha k}$ comes from exact diagonalization of the transmon with mode $k$, with its slope at $n_k = 0$ set to the exact multi-mode shift per photon. The two-mode term $P_{\alpha jk}$ comes from exact diagonalization with modes $j$ and $k$. Three-mode terms are neglected.

In this basis an eigenstate is labelled by the normal-mode Fock state that dominates it, and its weight on that state measures how dispersive the configuration is. When the transmon is excited, the Kerr term mixes photon configurations of two populated modes, $\ket{e, n_j, n_k} \leftrightarrow \ket{e, n_j + 1, n_k - 1}$, with an amplitude of order $2\sqrt{2}\,\eta\,u_t^2|u_ju_k|$ that grows with the photon numbers. The comb keeps the transmon partly excited, so strong mixing would move photons between parity channels while they are being read. We count a configuration as non-dispersive if any of its states has less than $80\%$ weight on its label, and reject a design if such configurations carry more than $10^{-3}$ of the prior weight, a tenth of the error budget below. This constraint, not the number of available channels, limits how many modes a set of transmons can read at strong coupling.

\subsection{Comb readout of an auxiliary transmon}\label{ssec:sm_comb}

Each auxiliary transmon couples to its own transmission line at radiative rate $\Gamma$ and is read out in reflection with a phase-coherent frequency comb, as in Ref.~\cite{essig_multiplexed_2021}. We treat it as a two-level system whose frequency is fixed by the photon numbers during the readout. The teeth have equal Rabi rates $\Omega$, are spaced by $\delta_c$, and cover every frequency the transmon can take. Tooth $t$ is reflected with the coherent amplitude $\beta r_t$, where $\beta$ is its input amplitude, $\Omega = 2\sqrt{\Gamma}\,|\beta|$, and
\begin{equation}\label{eq:sm_reflection}
r_t = 1 + \sqrt{\Gamma}\,\langle\sigma_-\rangle_t/\beta .
\end{equation}
Here $\langle\sigma_-\rangle_t$ is the Fourier component at the frequency of tooth $t$ of the periodic steady state of the Bloch equations under the full comb, which we obtain by harmonic balance. For a single tone detuned by $\delta$ from the transmon, Eq.~\eqref{eq:sm_reflection} reduces to
\begin{equation}\label{eq:sm_reflection_single}
r(\delta) = 1 - \frac{\Gamma(\Gamma/2 - i\delta)}{\Gamma^2/4 + \delta^2 + \Omega^2/2},
\end{equation}
a $\pi$ phase flip on resonance for $\Omega \ll \Gamma$ and no change far from it. Elastic scattering of this kind, and its saturation with drive power, has been observed for a single superconducting two-level system in an open transmission line~\cite{astafievResonanceFluorescenceSingle2010}.

Each tooth is recorded by heterodyne detection at efficiency $\eta_\text{meas}$ over a time $T$. The demodulated outcome is
\begin{equation}\label{eq:sm_record}
y_t = \sqrt{\eta_\text{meas} T}\,\beta\, r_t + w_t ,
\end{equation}
with $w_t$ complex Gaussian noise, independent across teeth and transmons. Near saturation the transmon also scatters inelastically. This raises the variance $\mathbb{E}|w_t|^2$ above its vacuum value $1$ at the teeth closest to the transmon frequency, and makes the noise phase-sensitive, $\mathbb{E}[w_t^2] \ne 0$. We take both moments from the quantum regression theorem, tooth by tooth. A stochastic-master-equation simulation of the heterodyne record agrees with them within its statistical error for a single tone, and within $3\%$ per tooth for a comb with $\delta_c \ge 2\Gamma$, the regime we use.

Saturation bounds the signal. A transmon on resonance with a tooth changes that tooth's reflected amplitude, relative to a far-detuned transmon, by at most $|\beta(r(0) - 1)|^2 \le \Gamma/8$, reached at $\Omega^2 = \Gamma^2/2$. A stronger comb does not help. A dense comb at this power saturates the transmon further, and most of its scattering becomes inelastic, which carries no coherent signal. The readout rate of one transmon is therefore of order $\eta_\text{meas}\Gamma$, and resolving the photon numbers keeps $\Gamma$ below the shifts they produce.

\subsection{Decoding and error per pair}\label{ssec:sm_decoding}

The decoder estimates $\mathbf{n}$ from the records of all auxiliary transmons, by maximum a posteriori estimation with the likelihood of Eq.~\eqref{eq:sm_record} and a Poisson prior of mean $\bar{n} = 2$ per mode, truncated at $n_\text{max}$. For $N = 5$ we find the estimate by enumeration. For larger $N$ we use branch-and-bound, which is exact up to the resolution of the grid on which its bounds are tabulated, and which agrees with enumeration where both are feasible.

A pair needs two bits from its mode, whether the mode is empty, which gives the parity of the pair, and whether $n$ is odd, which fixes the $(-1)^n$ phase of the even sector. Its error is the probability that either bit is wrong. We define $T_\text{meas}$ as the shortest $T$ at which every pair of the round has error at most $1\%$, so rounds measuring different numbers of pairs are compared at equal error per pair. We estimate it by Monte Carlo over $(\mathbf{n}, \mathbf{y})$. A union bound over the short vectors of the shift lattice guides the search and agrees with the Monte Carlo to within $15\%$. A wrong estimate usually changes several photon numbers at once, so the errors of different pairs are correlated, and the probability that some pair of a round is wrong stays between $1\%$ and $2\%$.

\subsection{Optimized parallel readout}\label{ssec:sm_tmeas}

The auxiliary transmons of a chain have the same coupling $g$ and anharmonicity $\eta$, and frequencies staggered from a detuning $\Delta$ below the band. We keep each of them dispersive up to $n_\text{max}$ photons, $n_\text{max} \le n_\text{crit} = \Delta_k^2/4g_k^2$ for every mode, together with the admixture bound of Eq.~\eqref{eq:admixture} at $0.06$, $g/2\pi \le 200$~MHz, the constraints of Secs.~\ref{sec:multi-transmon} and~\ref{ssec:sm_shifts}, and a total shift below $\eta$. Within these bounds we choose $\Delta$, $\eta \le 350$~MHz, $g$, $\Gamma$, the sites and the populated modes, and the number $M$ of pairs measured in one round, up to $\lfloor(N+1)/2\rfloor$. The comb has tooth spacing $2\Gamma$ and Rabi rate $0.5\,\Gamma$.

Table~\ref{tab:sm_readout_m} gives the measurement time per pair, $T_\text{meas}/M$, and Table~\ref{tab:sm_readout} the optimal designs. At unit efficiency a pair costs $3$ to $5\,\mu$s, for chains of $N = 5$ to $20$. For $N = 5$ measuring one pair at a time is best. For $N = 10$ and $15$ a round of three or four pairs costs about a third less per pair than single pairs, and for $N = 20$ the gain is about $15\%$. Beyond this the photon-hopping constraint forces weaker coupling and smaller shifts, so a round slows down faster than it adds pairs. Once the designs reach this weak-coupling regime at $N = 15$, the round time saturates near $30\,\mu$s, which gives a second, comparable minimum at $M = 8$.

\begin{table}[t]
\caption{Measurement time per pair, $T_\text{meas}/M$ ($\mu$s), at an error of at most $1\%$ per pair, $\eta_\text{meas} = 1$, and optimized auxiliary transmons, for $M$ pairs measured in one round. Entries in parentheses are from the union bound alone, all others from Monte Carlo with exact maximum a posteriori decoding. Bold entries are the optimum for each $N$.}
\label{tab:sm_readout_m}
\setlength{\tabcolsep}{3pt}
\begin{ruledtabular}
\begin{tabular}{ccccccccc}
$N$ & $M=1$ & 2 & 3 & 4 & 5 & 6 & 8 & 10 \\
\hline
5 & \textbf{3.8} & 4.5 & 4.2 & & & & & \\
10 & 5.1 & 3.4 & \textbf{3.3} & (3.6) & 3.8 & & & \\
15 & 5.5 & 5.1 & (4.4) & \textbf{3.5} & (6.0) & 5.7 & 3.9 & \\
20 & 5.7 & 5.2 & (6.6) & \textbf{4.9} & (5.6) & 7.4 & 7.7 & (7.5) \\
\end{tabular}
\end{ruledtabular}
\end{table}

\begin{table}[t]
\caption{Optimal designs of Table~\ref{tab:sm_readout_m}. Auxiliary transmons a detuning $\Delta$ below the bare resonator frequency, staggered by $40$~MHz, with anharmonicity $\eta$, coupling $g$ and linewidth $\Gamma$; $\chi_\text{max}$ is the largest shift per photon and $T_\text{meas}$ the time of one round.}
\label{tab:sm_readout}
\setlength{\tabcolsep}{4pt}
\begin{ruledtabular}
\begin{tabular}{cccccccc}
$N$ & $M$ & $\Delta/2\pi$ & $\eta/2\pi$ & $g/2\pi$ & $\Gamma/2\pi$ & $\chi_\text{max}/2\pi$ & $T_\text{meas}$ \\
 & & (MHz) & (MHz) & (MHz) & (MHz) & (MHz) & ($\mu$s) \\
\hline
5 & 1 & 700 & 350 & 169 & 6.7 & 8.9 & 3.8 \\
10 & 3 & 700 & 350 & 169 & 2.7 & 6.2 & 9.9 \\
15 & 4 & 700 & 250 & 169 & 1.2 & 4.6 & 14.0 \\
20 & 4 & 500 & 250 & 119 & 1.3 & 2.5 & 19.6 \\
\end{tabular}
\end{ruledtabular}
\end{table}

These times are for the ideal case. The detection efficiency is $\eta_\text{meas} = 1$, and a realistic chain lengthens them roughly as $1/\eta_\text{meas}$. Each auxiliary line also opens a Purcell decay channel for the chain modes, at a rate $\approx (g_k/\Delta_k)^2\Gamma$. For the designs of Table~\ref{tab:sm_readout} it reaches $0.05$ to $0.6\,\mu\text{s}^{-1}$, so a populated mode would lose its photons within one round. A filter that blocks the band of the chain on every auxiliary line, as in Ref.~\cite{essig_multiplexed_2021}, has to suppress this channel, and we assume such filters. The readout is assumed QND. Ref.~\cite{essig_multiplexed_2021} observes changes of the measured photon number for $\Omega/\chi \gtrsim 0.1$, and the optimal designs have $\Omega/\chi = 0.1$--$0.4$, so this backaction remains to be bounded. The thresholds of Sec.~\ref{ssec:sm_shifts} are choices. They exclude hopping between populated modes rather than simulate it.

The parity drive is off during the readout, so the photon numbers are not replenished. A photon lost from a populated mode changes the parity of its photon number, which spoils the $(-1)^n$ correction, and can empty the mode, which flips the measured parity of the pair. Its probability is $\approx \bar{n}\,T_\text{meas}/T_1^\text{mode}$ per populated mode, where $T_\text{meas}$ is the time of the whole round, not the time per pair. For the designs of Table~\ref{tab:sm_readout} this is $0.05$--$0.26$ at $T_1^\text{mode} = 150\,\mu$s. The protocol also fails for the even-sector outcome $n = 0$, which has probability $e^{-\bar{n}}$ and looks odd. Keeping the parity drive on during the readout would hold the modes in a driven steady state and remove both failure modes, at the cost of dephasing within the even sector by the photons that leave the mode; that variant remains to be studied.

\section{Drive-Induced Dephasing of the Dual-Rail Transmon}\label{sec:sm_dephasing}

This section derives the drive-induced dephasing rate quoted in the End Matter. Frequency noise on the two rails of the DRT enters as $V(t)=\delta_a(t)\,a^\dagger a+\delta_b(t)\,b^\dagger b=\frac{1}{2}\left[\delta_+(t)L_c+\delta_-(t)L_t\right]$, with $\delta_\pm=\delta_a\pm\delta_b$, a common channel $L_c=\tilde{a}_+^\dagger\tilde{a}_++\tilde{a}_-^\dagger\tilde{a}_-$, and a transverse channel $L_t=\tilde{a}_+^\dagger\tilde{a}_-+\tilde{a}_-^\dagger\tilde{a}_+$, where $\tilde{a}_\pm=(a\pm b)/\sqrt{2}$. The transverse channel remains protected under the cross-resonance drive, because the anti-symmetric drive [Eq.~\eqref{eq:hdrive}] and the anharmonicity change the $\tilde{a}_+$ occupation only by even amounts while $L_t$ changes it by one. We therefore consider the common channel.

A common-channel dephasing error requires the two dressed logical states to hold different total excitation numbers, $\ell_0\neq\ell_1$ with $\ell_i=\langle\tilde{i}_L|L_c|\tilde{i}_L\rangle$; for low-frequency noise, the logical qubit then dephases at the bare-transmon rate scaled by $(\ell_0-\ell_1)^2/2$. Consider one drive tone at $\omega_\sigma$, $\sigma=\pm$, detuned from the transmons by $\Delta_\sigma=\omega_\sigma-\omega\approx\Delta\pm2g_{DR}$. To lowest order in $\Omega_\sigma/\Delta_\sigma$, it admixes $|0_L\rangle$ with the vacuum and with two-excitation states, and $|1_L\rangle$ with two-excitation states only, each with a weight $\sim\Omega_\sigma^2/\Delta_\sigma^2$. The vacuum admixture lowers $\ell_0$, and the two-excitation admixtures raise $\ell_0$ and $\ell_1$, each in inverse proportion to the square of the energy denominator of the virtual transition. For harmonic transmons these contributions cancel, because a linear drive on a harmonic mode is a pure displacement and adds the same excitation number to every state. Any difference must therefore come from the energies of the two-excitation states, which differ between the logical states. The two-excitation state reached from $|1_L\rangle$ is not coupled by $g_{DR}$, and its admixture carries the single energy denominator $\Delta_\sigma+g_{DR}+\eta$, while the vacuum admixture carries $\Delta_\sigma+g_{DR}$. The state reached from $|0_L\rangle$ is not an eigenstate, since the pair-exchange part of the anharmonicity hybridizes it with the other doubly excited state, which lies $4g_{DR}$ away. This is the same hybridization that generates the cross-resonance coupling (Sec.~\ref{ssec:sm_gcr}), and it spreads the admixture of $|0_L\rangle$ over the energy denominators of the two hybrids.

Second-order perturbation theory in the drive gives, for each tone, $\ell_0-\ell_1\approx\Omega_\sigma^2\left[\langle D^{-2}\rangle_{0_L}-\langle D^{-2}\rangle_\text{ref}\right]$, where $\langle\cdot\rangle_{0_L}$ averages over the energy denominators $D$ of the hybridized admixture of $|0_L\rangle$, and $\langle\cdot\rangle_\text{ref}$ places equal weights on the vacuum and on the unhybridized two-excitation denominators. Expanding $D^{-2}$ about the shared mean $\bar{D}=\Delta_\sigma+g_{DR}+\eta/2$, the $k$-th central moment of each spectrum enters at order $1/\Delta_\sigma^{k+2}$. The two spectra share their mean and their variance $\eta^2/4$, so all terms through $1/\Delta_\sigma^4$ cancel. They first differ in the third central moment, $-g_{DR}\eta^2$ for the hybridized spectrum and zero for the symmetric two-line reference. The leading term is therefore fifth order, with $\Omega_\sigma^2/\Delta_\sigma^2$ from the admixture, $\eta^2/\Delta_\sigma^2$ from the spread of the hybridized spectrum (at $\eta=0$ there is no hybridization), and $g_{DR}/\Delta_\sigma$ from its asymmetry (at $g_{DR}=0$ the two doubly excited states are degenerate, and the hybrids reproduce the reference). The two tones contribute independently at this order, and replacing $\Delta_\sigma$ by $\Delta$ changes only terms of order $1/\Delta^6$, so
\begin{equation}\label{eq:sm_dephasing}
\begin{gathered}
\ell_0-\ell_1 \approx \frac{4g_{DR}\,\eta^2\,\Omega^2}{\Delta^5}\left(1+r^2\right),\\
\Gamma_\phi^\text{dressed} \approx 8\,\Gamma^\text{transmon}_\phi\, \frac{g_{DR}^2\,\eta^4\,\Omega^4}{\Delta^{10}}\left(1+r^2\right)^2,
\end{gathered}
\end{equation}
with $r=\Omega_-/\Omega_+$ the tone ratio and $\Gamma_\phi^\text{transmon}$ the dephasing rate of a single bare transmon. We confirmed the expansion against exact diagonalization of the driven dual-rail Hamiltonian, one tone at a time.

\section{Spectator Erasure: Average Gate Fidelity}\label{sec:sm_erasure}

This section derives the spectator-erasure infidelity quoted in the End Matter. We first justify the End Matter's statement that an erasure on a \emph{participating} DRT dephases its partner through the same which-trajectory record as a photon lost from the mode. During the gate, the mode displacement is conditioned on $K = \tilde{X}_1 + \tilde{X}_2$; an excitation loss at time $t_0$ removes DRT~1 from its logical subspace, terminating its coupling, and the photon lost to the environment carries a record of the trajectory the mode followed up to $t_0$. Averaging over this record multiplies the coherence of the surviving DRT by the overlap $|\langle\alpha_+(t_0)|\alpha_-(t_0)\rangle| = e^{-|\alpha_+(t_0)-\alpha_-(t_0)|^2/2}$, where $\alpha_\pm(t_0)$ is the coherent-state amplitude of the mode at $t_0$ in the sector $\tilde{X}_1 = \pm1$, at fixed $\tilde{X}_2$. A photon lost from the mode itself at $t_0$ reveals the same which-trajectory information and produces the same overlap factor. The participating-DRT erasure is therefore bounded by the photon-loss analysis of the End Matter, with the mode population $|\alpha|^2$ evaluated along the interrupted loop.

The remainder of this section treats the spectator case, described by the Hamiltonian
\begin{equation}
H(t) = \epsilon(t) c^{\dagger} c - g_{CR} (\tilde{X}_{1} + \tilde{X}_{2}) (c^{\dagger} + c),
\end{equation}
where $\epsilon(t) = \epsilon_0 + \chi\Theta(t-t_0)$, with $\Theta$ the Heaviside step function, and $c$ ($c^\dagger$) is the annihilation (creation) operator for the resonator.

In the interaction picture with respect to $\epsilon(t)c^\dagger c$, the coupling is $H_I(t) = -g_{CR}K(c^\dagger e^{i\Phi(t)} + \text{h.c.})$, where $\Phi(t) = \int_0^t \epsilon(t')dt'$ and $K = \tilde{X}_1 + \tilde{X}_2$. Since $[H_I(t_1), H_I(t_2)] \propto K^2$ commutes with $H_I$, the Magnus expansion terminates at second order, giving exactly $U_I(T) = D(\alpha_K)e^{i\gamma_K}$, with displacement amplitude $\alpha_K = ig_{CR}K\int_0^T e^{i\Phi(t)}dt$ and phase $\gamma_K = \frac{i}{2}\int_0^T dt_1\int_0^{t_1}dt_2\,[H_I(t_1), H_I(t_2)] \propto K^2$. Tracing out the resonator, which starts in the vacuum state $|0\rangle_c$, gives the qubit map $\mathcal{E}(\rho_q) = \sum_n M_n\rho_q M_n^\dagger$ with Kraus operators $M_n = \langle n|_c U_I(T)|0\rangle_c$; the desired evolution, in which the resonator returns to the vacuum, is $M_0 = e^{-|\alpha_K|^2/2}e^{i\gamma_K}$, while $M_{n>0}$ represent leakage. Since $K^2 = 2(I + \tilde{X}_1\tilde{X}_2)$, we can write $|\alpha_K|^2 = |\alpha(t_0)|^2(I + \tilde{X}_1\tilde{X}_2)$ and $\gamma_K = \gamma(t_0)(I + \tilde{X}_1\tilde{X}_2)$, with scalar coefficients $|\alpha(t_0)|^2$ and $\gamma(t_0)$ that depend on the jump time $t_0$ through the piecewise $\epsilon(t)$ and the gate duration $T = \pi/(2g_{CR})$. For the unperturbed closed loop we take $\epsilon_0 = 4g_{CR} > 0$, so that $\epsilon_0 T = 2\pi$ and the Magnus phase is $\gamma(T)|_{\chi=0} = 2g_{CR}^2 T/\epsilon_0 = +\pi/4$, and at $\chi = 0$ the map is exactly $U_\text{ideal}$ with $F = 1$.

The ideal gate is $U_\text{ideal} = e^{+i\frac{\pi}{4}\tilde{X}_1\tilde{X}_2}$ of Eq.~\eqref{eq:2qprop} of the main text, acting on the $d = 4$ dimensional logical space of the two DRTs. We compare it with $M_0$ alone. Each Kraus operator $M_{n>0}$ leaves photons in the resonator. Keeping only $M_0$ counts these outcomes as errors, so the fidelity below is a lower bound on that of the full map $\mathcal{E}$. For a linear operator $V$ on a $d$-dimensional space, not necessarily trace-preserving, the fidelity averaged over pure input states is~\cite{pedersenFidelityQuantumOperations2007}
\begin{equation}
F = \frac{\text{Tr}(V^\dagger V) + |\text{Tr}\,V|^2}{d(d+1)},
\end{equation}
which reduces to the average gate fidelity of Ref.~\cite{nielsen_simple_2002} when $V$ is unitary. With $V = U_\text{ideal}^\dagger M_0(t_0)$ and $C(t_0) = -|\alpha(t_0)|^2/2 + i\gamma(t_0)$, evaluating both traces in the joint eigenbasis of $\tilde{X}_1$ and $\tilde{X}_2$ gives
\begin{equation}
F(t_0) = \frac{1 + e^{-2|\alpha(t_0)|^2}}{10} + \frac{4}{5}\left| e^{C(t_0)} \cosh\left(C(t_0) - i\frac{\pi}{4}\right) \right|^2.
\end{equation}
Both terms come from the vacuum amplitudes of the four eigenstates. The odd-parity states, $K = 0$, are never displaced and return to the vacuum with amplitude $1$. The even-parity states, $K = \pm2$, are displaced by $|\alpha_K|^2 = 2|\alpha(t_0)|^2$ and return with amplitude $e^{2C(t_0)}$, whose modulus $e^{-|\alpha_K|^2/2}$ is the overlap of the displaced even-parity trajectory with the undisplaced odd-parity one. Flipping one $\tilde{X}$ eigenvalue takes $K = \pm2$ to $K = 0$. This modulus is therefore the overlap factor $e^{-|\alpha_+-\alpha_-|^2/2}$ that opened this section, now evaluated at the end of the gate, for the loop distorted by the jump, rather than at the jump itself.
The overall gate infidelity follows by averaging $1-F(t_0)$ over $p(t_0)$; although $p(t_0)$ is normalized on $[0,\infty)$, an erasure at $t_0 > T$ leaves the completed gate untouched, so only the $t_0 \in [0,T]$ branch contributes. Expanding to leading order in the fractional shift $\chi/\epsilon_0$ makes the structure explicit. In terms of the loop phase still ahead at the jump, $\theta = \epsilon_0(T-t_0)$, an erasure at $t_0$ leaves a residual displacement $|\alpha(t_0)|^2 = \frac{1}{8}(\chi/\epsilon_0)^2(\theta^2 - 2\theta\sin\theta + 2 - 2\cos\theta)$ and a phase deficit $\delta\gamma(t_0) = -\frac{1}{4}(\chi/\epsilon_0)(\theta - \sin\theta)$, both vanishing for an erasure at the gate end and growing with the remaining loop phase. Averaging $1-F(t_0) \approx |\alpha(t_0)|^2 + \frac{4}{5}\delta\gamma(t_0)^2$ over $p(t_0)$ with $\Gamma_e T \ll 1$ gives the per-spectator result quoted in the End Matter.

Two further properties of Eq.~\eqref{eq:erasure_scaling} of the main text follow from the dependence of $\chi$ and $\epsilon_0$ on the chain parameters. First, the shift is diluted by the spectator's participation in the driven mode $k$, $\frac{2}{N+1}\sin^2\!\big(\frac{j_s k\pi}{N+1}\big)$ for a spectator attached to resonator $j_s$, which is $1/(N+1)$ for a spectator near the chain edge and at most twice that at the chain center; the End Matter quotes the edge value. Second, the dependence on the qubit--mode coupling $g$ is linear. A stronger coupling raises $g_{CR} \propto g$ and so lowers the erasure probability $\Gamma_e T \propto 1/g$, but it raises the shift as $\chi \propto g^2$ and with it the squared fractional shift $(\chi/\epsilon_0)^2 \propto g^2$, and the two effects only partially cancel.

\section{Mode-Count Ceiling}\label{sec:sm_ceiling}

This section derives the mode-count ceiling quoted in the main text, the number of modes a single metamaterial supports. Two error sources bound the gate duration from opposite sides. Photon loss grows with the duration, and the spectral crowding of the modes requires a pulse long enough to resolve one mode from its neighbors. The erasure of spectator DRTs adds a third term, treated after the two bounds. The two bounds involve only the gate duration, so the mode-count ceiling they set is independent of the qubit--mode coupling and the drive.

\paragraph{Model.} We consider a chain of $N$ modes carrying one DRT per two modes, so that $N + 1 = 2N_q$. The gate addresses the mid-band channel $k = N_q$, whose mode profile $|\sin(jk\pi/(N+1))| = |\sin(j\pi/2)|$ equals $1$ at the odd sites $j$ occupied by the DRTs. The per-mode coupling is then $g\sqrt{2/(N+1)}$, and the cross-resonance rate of Eq.~\eqref{eq:effham} of the main text is $g_{CR} = G/\sqrt{N+1}$ with $G = \sqrt{2}\,\eta g\Omega/D_+(0)$ [Eq.~\eqref{eq:sm_denoms}]. Near mid-band the mode spacing is $\delta\nu \simeq 2\pi J/(N+1)$, from the dispersion of Eq.~\eqref{eq:dispersion}. As in Fig.~\ref{fig:gate_fidelity} of the main text, the drive envelope $f(t)$ is a flat top of duration $T_\text{flat}$ between Nuttall ramps of duration $\beta T_\text{flat}$ each; $T_\text{flat} = \pi/2g_{CR}$ closes the loop at $\epsilon = 4g_{CR}$. The total duration is $T = (1+2\beta)\,T_\text{flat}$. We assign no entangling phase to the ramps, which makes the bound conservative.

\paragraph{Photon loss bounds the duration from above.} The End Matter gives the loss infidelity $\varepsilon_\text{loss} = A\,T/T_1^\text{mode}$ with $A = \frac{4}{5}\times0.273 = 0.218$, independent of the drive strength. Within a loss budget $\varepsilon_\text{loss}$,
\begin{equation}\label{eq:sm_tmax}
T \le T_\text{max} \equiv \frac{\varepsilon_\text{loss} T_1^\text{mode}}{A}.
\end{equation}

\paragraph{Spectral crowding bounds the duration from below.} The two modes adjacent to the target lie $\delta\nu \mp \epsilon = \delta\nu \mp 4g_{CR}$ from the drive center; the smaller of the two is the $\Delta_\text{spec}$ of the End Matter. Each is driven off-resonantly by the same force $g_{CR}(\tilde{X}_1 + \tilde{X}_2)$ and, at detuning $w$, is left with the residual displacement $\alpha_K = i g_{CR} K\,\mathcal{F}(w)$, with $K = \tilde{X}_1 + \tilde{X}_2$ and $\mathcal{F}(w) = \int_0^T f(t)\,e^{iwt}\,dt$. To leading order the average gate infidelity is $|\alpha|^2$ in the notation of Sec.~\ref{sec:sm_erasure}, so
\begin{equation}\label{eq:sm_epsx}
\varepsilon_X = 2g_{CR}^2\sum_\pm\left|\mathcal{F}(\delta\nu \mp 4g_{CR})\right|^2.
\end{equation}
The static phase that the adjacent modes add to the gate is absorbed in calibration. In units of $1/g_{CR}$ the envelope depends only on $\beta$, so $\varepsilon_X$ depends only on $\beta$ and $\delta\nu/g_{CR}$. Requiring $\varepsilon_X$ to stay within its budget for every spacing larger than $\delta\nu$, so that the bound does not rely on zeros of $\mathcal{F}$, gives $g_{CR} \le \delta\nu/c$, with $c(\beta, \varepsilon_X)$ computed numerically from Eq.~\eqref{eq:sm_epsx}. With $T = (1+2\beta)\pi/2g_{CR}$ and $\delta\nu = 2\pi J/(N+1)$,
\begin{equation}\label{eq:sm_tmin}
T \ge T_\text{min}(N) \equiv \frac{c\,(1+2\beta)}{4J}\,(N+1).
\end{equation}

\paragraph{The ceiling.} A gate meeting both budgets exists only while $T_\text{min}(N) \le T_\text{max}$. Since $T_\text{min}$ grows linearly with $N + 1$ while $T_\text{max}$ is fixed, this condition bounds the chain length,
\begin{equation}\label{eq:sm_nmax}
N + 1 \le \frac{4\,\varepsilon_\text{loss}\,J\,T_1^\text{mode}}{A\,(1+2\beta)\,c}.
\end{equation}
The coupling $g$, the drive $\Omega$, the anharmonicity $\eta$, and the detuning $\Delta$ do not appear in either duration bound, and therefore not in this one. Longer ramps lower $c$ but lengthen the gate through $1+2\beta$. Table~\ref{tab:sm_ceiling} lists the optimum over the split of the $10^{-3}$ budget between $\varepsilon_\text{loss}$ and $\varepsilon_X$. The ceiling is highest at $\beta \approx 0.7$, where $\varepsilon_X \approx 1.6\times10^{-4}$, and stays within $10\%$ of that value for $0.4 \le \beta \le 1.5$:
\begin{equation}\label{eq:sm_nmax_num}
N + 1 \le 4.2\times10^{-4}\,J\,T_1^\text{mode} = 2.6\times10^{-3}\,(J/2\pi)\,T_1^\text{mode}.
\end{equation}
Since $N + 1 = 2N_q$, the largest array that meets this bound holds
\begin{equation}\label{eq:sm_nqmax}
N_{q,\text{max}} = \left\lfloor 1.3\times10^{-3}\,(J/2\pi)\,T_1^\text{mode} \right\rfloor
\end{equation}
DRTs on $N_\text{max} = 2N_{q,\text{max}} - 1$ modes. At $J/2\pi = 80$~MHz and $T_1^\text{mode} = 150\,\mu$s the bound is $N + 1 \le 31.5$, so $N_{q,\text{max}} = 15$ and $N_\text{max} = 29$. At $T_1^\text{mode} = 30\,\mu$s it is $N + 1 \le 6.3$, so $N_{q,\text{max}} = 3$ and $N_\text{max} = 5$.

\paragraph{Spectator erasure.} The erasure of a spectator DRT (End Matter) adds a third term to the budget. On the crowding-limited line $T = T_\text{min}(N)$, a spectator at an odd site shifts the driven mode by $\chi = 2\chi_1/(N+1)$, with $\chi_1 = 2g^2\eta/[\Delta(\Delta+\eta)]$ the full-participation shift, and loop closure gives $\epsilon_0 = 4g_{CR} = 8\pi J/[c(N+1)]$, so the fractional shift $\chi/\epsilon_0 = c\chi_1/(4\pi J)$ is independent of $N$. Summing Eq.~\eqref{eq:erasure_scaling} of the main text over the $N_q-2$ spectators, to leading order in their total erasure probability, gives
\begin{equation}\label{eq:sm_spec}
\varepsilon_\text{spec} = \left(N_q-2\right)\left(\frac{7\pi^2}{30}+\frac{5}{8}\right)\frac{T_\text{min}(N)}{T_1^\text{tr}}\left(\frac{c\,\chi_1}{4\pi J}\right)^2,
\end{equation}
with $T_1^\text{tr}$ the transmon relaxation time, and the budget $\varepsilon_\text{loss} + \varepsilon_X + \varepsilon_\text{spec} \le 10^{-3}$ becomes quadratic in $N + 1$. At the parameters of Table~\ref{tab:sim_params} with the mid-band detuning $\Delta/2\pi = 900$~MHz, $\chi_1/2\pi = 2.0$~MHz, and $\chi/\epsilon_0 = 0.030$. At $\beta = 0.7$, $T_1^\text{mode} = 150\,\mu$s, and $T_1^\text{tr} = 100\,\mu$s the budget holds for $N + 1 \le 26.7$, so $N_{q,\text{max}} = 13$ DRTs on $N_\text{max} = 25$ modes, with $\varepsilon_\text{spec} = 1.4\times10^{-4}$. Re-optimizing $\beta$ does not raise the ceiling, since the spectator term grows as $c^3(1+2\beta)$ and favors the same ramps. Eq.~\eqref{eq:sm_nmax_num} is recovered as $T_1^\text{tr}\to\infty$. The spectator term equals the loss term at $T_1^\text{tr} \approx 25\,\mu$s for $N + 1 = 27$, and for shorter transmon lifetimes the bound on $N + 1$ grows as $\sqrt{J T_1^\text{tr}}/(\chi/\epsilon_0)$ rather than as $J T_1^\text{mode}$. At $T_1^\text{tr} = 30$, $150$, and $300\,\mu$s the bound is $N + 1 \le 21.2$, $28.1$, and $29.9$, so $N_{q,\text{max}} = 10$, $14$, and $14$. At $T_1^\text{mode} = 30\,\mu$s the gate is short and the spectator term negligible, leaving $N + 1 \le 6.4$ and $N_{q,\text{max}} = 3$. Unlike the loss and crowding bounds, this term depends on the qubit--mode coupling, the anharmonicity, and the detuning, through $\chi_1$.

\begin{table}[t]
\caption{Mode-count ceiling of Eq.~\eqref{eq:sm_nmax}, as the bound on $(N+1)/(J T_1^\text{mode})$, for the Nuttall flat-top envelope, optimized over the split of a $10^{-3}$ budget, with the corresponding crowding coefficient $c$ of Eq.~\eqref{eq:sm_tmin} and crosstalk share $\varepsilon_X/10^{-3}$.}
\label{tab:sm_ceiling}
\setlength{\tabcolsep}{6pt}
\begin{ruledtabular}
\begin{tabular}{cccc}
$\beta$ & bound on $(N+1)/(J T_1^\text{mode})$ & $c$ & $\varepsilon_X/10^{-3}$ \\
\hline
$0.1$ & $2.1\times10^{-4}$ & $63$ & $0.14$ \\
$0.3$ & $3.5\times10^{-4}$ & $27$ & $0.16$ \\
$0.5$ & $4.0\times10^{-4}$ & $19$ & $0.16$ \\
$0.7$ & $4.2\times10^{-4}$ & $15$ & $0.16$ \\
$1.0$ & $3.9\times10^{-4}$ & $15$ & $0.04$ \\
$1.5$ & $3.8\times10^{-4}$ & $11$ & $0.06$ \\
\end{tabular}
\end{ruledtabular}
\end{table}

\paragraph{Role of the drive.} The hardware parameters set the gate duration achieved, $T(N) = (1+2\beta)\pi\sqrt{N+1}/2G$, which grows more slowly than $T_\text{min}(N)$. At fixed $\beta$, both bounds hold only for
\begin{equation}\label{eq:sm_drive_branches}
N + 1 \le \min\left\{\left(\frac{2GT_c}{\pi(1+2\beta)}\right)^2, \left(\frac{2\pi J}{cG}\right)^2\right\},
\end{equation}
where $T_c$ is the value of $T_\text{min}$ at the ceiling, equal to $T_\text{max}$ without the spectator term. The first branch requires the gate to fit within $T_c$, and the second requires it to resolve the neighboring modes. The two branches meet at $G^* = \pi\sqrt{(1+2\beta)J/(cT_c)}$. A stronger drive can always be lowered to $G^*$, so only the weak-drive branch limits the array. The ceiling is reached at $G = G^*$, which requires $g_{CR}^* = \pi(1+2\beta)/2T_c$ at the ceiling, that is, $g_{CR}^*/2\pi = 1.3$~MHz at $T_1^\text{mode} = 150\,\mu$s and $5.2$~MHz at $T_1^\text{mode} = 30\,\mu$s. At the DRT-1 parameters of Table~\ref{tab:sim_params} with the mid-band detuning $\Delta/2\pi = 900$~MHz, the leading-order coupling requires $\Omega^*/2\pi = 370$~MHz and $750$~MHz, respectively, beyond the $\Omega/2\pi \le 100$~MHz validity range of the second-order effective Hamiltonian (Sec.~\ref{sec:dreh}). For a weaker drive the array is drive-limited, on the branch $N + 1 \le [2GT_c/\pi(1+2\beta)]^2 \propto (G\,T_c)^2$. The coupling dilution then sets the capacity, which grows quadratically with $T_1^\text{mode}$ and does not depend on $J$.

\paragraph{Assumptions.} The bound neglects the coherent error, which accounts for $19\%$ of the infidelity at the optimum of Fig.~\ref{fig:gate_fidelity} of the main text. It uses the analytic loss coefficient $A$; the simulated photon loss at that optimum gives $A = 0.222$, within $2\%$ of it. It keeps only the two adjacent modes, at unit mode profile. For gates run in parallel on channels away from mid-band, the spacing is reduced by $\sin(\pi k/(N+1))$. The bound assumes single-mode addressing with a smooth flat-top envelope; amplitude-, phase-, and multi-tone-modulated pulses that close the loops of several modes~\cite{zhuArbitraryspeedQuantumGates2006, roosIonTrapQuantum2008, haddadfarshiHighFidelityQuantum2016, shapiraRobustEntanglementGates2018, webbResilientEntanglingGates2018, milnePhaseModulatedEntanglingGates2020} can raise the prefactor of Eq.~\eqref{eq:sm_nmax_num}.

\section{Numerical Simulation Details}\label{sec:sm_simulation}

We numerically simulate Eq.~\eqref{eq:drt_mm_ham} of the main text using \texttt{QuantumToolbox.jl}~\cite{mercurioQuantumToolboxjlEfficientJulia2025a}. The system consists of two DRTs coupled symmetrically to opposite ends of a three-resonator chain. The two transmons of each DRT couple equally to the chain, $g_{na}=g_{nb}\equiv g_n$. The parameters are listed in Table~\ref{tab:sim_params}.

\begin{table}[t]
\caption{Simulation parameters. Paired values correspond to the endpoint
DRTs at sites $(n=1,n=3)$.}
\label{tab:sim_params}
\setlength{\tabcolsep}{6pt}
\begin{ruledtabular}
\begin{tabular}{lcc}
Parameter & Symbol & Value \\
\hline
Resonator frequency & $\nu/2\pi$ & $4.500$~GHz \\
Resonator coupling & $J/2\pi$ & $80$~MHz \\
Transmon frequencies & $(\omega_1,\omega_3)/2\pi$ & $(3.600,3.750)$~GHz \\
Anharmonicities & $(\eta_1,\eta_3)/2\pi$ & $(205,212)$~MHz \\
DRT couplings & $(g_{DR,1},g_{DR,3})/2\pi$ & $(105,110)$~MHz \\
DRT-metamaterial couplings & $(g_1,g_3)/2\pi$ & $(70,65)$~MHz \\
\end{tabular}
\end{ruledtabular}
\end{table}

The gate addresses mode $c_1$, with $\nu_1/2\pi=4.613$~GHz and $(\Delta_1,\Delta_3)/2\pi=(1.013,0.863)$~GHz. This is the band-edge channel $k = 1$. The mode-count ceiling of Sec.~\ref{sec:sm_ceiling} does not depend on this operating point; the mid-band channels it assumes carry larger mode profiles at the qubit sites and smaller detunings $\Delta_k$, which lower the drive needed to reach it. One red and one blue sideband tone are applied to each DRT, with their frequencies, relative amplitudes and phases, and the gate duration optimized at each drive strength. We report logical-subspace fidelity to the maximally entangling $\tilde{X}\tilde{X}$ gate up to local rotations, counting leakage as error. Mode loss is included for $T_1^\text{mode}=30$--$150~\mu$s ($\kappa/2\pi = 5.3$--$1.1$~kHz, $Q=(0.85$--$4.2)\times10^{6}$); both band edges in Fig.~\ref{fig:gate_fidelity} of the main text are simulated separately, with the solid curve at $T_1^\text{mode}=30\,\mu$s. We treat mode loss to first order in $\kappa$. The channel is the no-jump evolution $U$ under $H-\frac{i}{2}\sum_k\kappa_k c_k^\dagger c_k$ plus the single-jump Kraus operators $\sqrt{\kappa_k}\,P\,U(T,t)\,c_k\,U(t,0)\,P$, integrated over the jump time $t$, with $P$ the projector onto the logical subspace. Since the gate returns the modes to vacuum, a lost photon dephases the logical state rather than removing it from the logical subspace, which gives the $\frac{4}{5}$ coefficient in the Resonator Photon Loss section of the End Matter; on a reduced Hilbert space this treatment agrees with the full Lindblad master equation to $0.03\%$. Transmon relaxation is reported separately as a heralded erasure. For the bare-transmon gate with the flowers sequence (green triangles in Fig.~\ref{fig:gate_fidelity} of the main text), transmon relaxation stays in the logical subspace and is not heralded, and we add both loss channels to the coherent error, $1-F = 1-F_\text{coh} + \frac{4}{5}\int_0^T\big[\langle n_\text{tr}\rangle/T_1^\text{tr} + \langle n_\text{mode}\rangle/T_1^\text{mode}\big]\,dt$, with $\langle n_\text{tr}\rangle$ and $\langle n_\text{mode}\rangle$ the total transmon and mode populations. Its loss term agrees with a full Lindblad simulation to within $1.2\%$.

The coherent error of the optimized gate is almost entirely leakage out of the logical subspace. Between $T = 226$ and $199$~ns ($\Omega_+/2\pi = 115$ and $130$~MHz) it rises fivefold, from $4.3\times10^{-5}$ to $2.2\times10^{-4}$, and the rise has one dominant source. $88\%$ of the increase is population left in the upper two-excitation hybrid $\ket{h_+}$ of the site-3 DRT, with the modes empty, starting from the logical states in which that DRT is in $\ket{0_L}$. The same channel carries $70$--$75\%$ of the leakage over the plateau at $\Omega_+/2\pi = 145$--$190$~MHz, evaluated at a smaller photon-number cutoff of the untargeted modes. Rotating the phase of one tone by $\varphi$ shifts the phase of the leaked amplitude by $m\varphi$, with $m$ the net number of photons that tone contributes; the measured $m$ show that the process absorbs two red-tone photons and emits one blue-tone photon. This three-photon process stays $\delta/2\pi \approx -27$~MHz, four to six pulse linewidths, from resonance for $\Omega_+/2\pi = 115$--$205$~MHz; the rise is therefore not a resonance crossing. Mid-gate, $\ket{h_+}$ holds $0.5$--$2.5\%$ of the population, mostly fast dressing whose population returns to the logical states by the end of the pulse. The final leakage is $10^2$ to $5\times10^3$ times smaller and is the residue of several nearly cancelling paths.

The process detuning is a near-cancellation. With the red and blue tones at $\mp 2g_{DR}$ from the drive center, it is $\delta \simeq 7g_{DR} + \Lambda - \Delta - \eta/2$, plus the shifts that the coupling to the modes imparts to $\ket{h_+}$, $\ket{1_L}$, and the calibrated tones. These shifts total $\approx -60$~MHz. At the site-3 parameters the qubit--mode detuning, $\approx 870$~MHz including the drive-center offset, nearly cancels the $g_{DR}$ and $\eta$ terms, $\approx 910$~MHz. Evaluating $\delta$ with the dressed energies of the full static Hamiltonian gives $\partial\delta/\partial g_{DR} \approx 8.7$, $\partial\delta/\partial\eta \approx -0.27$, and $\partial\delta/\partial\Delta = -1$. Raising $g_{DR,3}$ by $3$~MHz would therefore bring the process onto resonance, while $\eta$ barely moves it. For the site-1 DRT, whose detuning is larger and whose $g_{DR}$ is smaller, the same process lies $\approx -215$~MHz from resonance, and only the site-3 DRT leaks.

\subsection{Error budget at the operating point}\label{ssec:sm_budget}

Table~\ref{tab:sm_budget} collects the contributions to the gate infidelity at the operating point of Fig.~\ref{fig:gate_fidelity} of the main text, $T = 152$~ns, $\Omega_+/2\pi = 190$~MHz, and mode $T_1 = 30\,\mu$s. The simulated Hamiltonian contains the full number-conserving coupling of each transmon to all three modes, and the logical states are its dressed eigenstates, so the residual dispersive coupling of Sec.~\ref{ssec:sm_chi} and the mediated cross-Kerr of Sec.~\ref{ssec:sm_mediated} enter the coherent error without approximation. We split the coherent error $1-F = 1-fu$ into leakage $1-u$, the population that leaves the logical subspace with the modes empty, and the in-subspace error $1-f$, the distance of the projected process from the target Weyl class.

The coherent error is leakage. Of its $2.68\times10^{-4}$, the in-subspace error accounts for $7.3\times10^{-7}$. Every coherent channel that acts within the logical subspace is bounded by this value, the residual dispersive coupling among them. At the operating point a single photon shifts the calibrated sideband tones by $18$--$33$~kHz, the drive-dressed counterpart of $\chi_L$, and the static logical $\tilde Z\tilde Z$ coupling is $-0.49$~kHz. The dispersive coupling could also act by distorting the phase-space loop, which would leave photons in the mode at the end of the gate and so appear as leakage. The residual displacement is $|\Delta\alpha|^2 \le 5\times10^{-6}$ in every logical state, so this route is also negligible. The residual dispersive coupling therefore contributes below $10^{-5}$, less than $1\%$ of the budget. The leakage itself is the off-resonant three-photon process of Sec.~\ref{sec:sm_simulation}, which populates a two-excitation state of the site-3 DRT.

\begin{table}[t]
\caption{Gate error budget at the operating point of Fig.~\ref{fig:gate_fidelity} of the main text ($T = 152$~ns, $\Omega_+/2\pi = 190$~MHz, mode $T_1 = 30\,\mu$s, transmon $T_1 = 100\,\mu$s, $T_\phi = 10\,\mu$s). All entries are from the full simulation of Sec.~\ref{sec:sm_simulation}. Transmon relaxation leaves the logical subspace of the DRT and is heralded as an erasure, so it is listed as a probability and excluded from $1-F$.}
\label{tab:sm_budget}
\setlength{\tabcolsep}{4pt}
\begin{ruledtabular}
\begin{tabular}{lcc}
Channel & $1-F$ & Share \\
\hline
Leakage (two-excitation state of the site-3 DRT) & $2.68\times10^{-4}$ & $19\%$ \\
In-subspace coherent error & $7.3\times10^{-7}$ & $<0.1\%$ \\
\quad residual dispersive coupling and $\tilde Z\tilde Z$ & $<7.3\times10^{-7}$ & \\
\quad residual mode displacement & $\le 5\times10^{-6}$ & \\
Resonator photon loss & $1.13\times10^{-3}$ & $81\%$ \\
Quasi-static transmon dephasing & $\sim1\times10^{-7}$ & $<0.01\%$ \\
\hline
Total & $1.40\times10^{-3}$ & \\
\hline
Heralded erasure (probability) & $3.2\times10^{-3}$ & \\
\end{tabular}
\end{ruledtabular}
\end{table}

\end{document}